\documentclass[useAMS,usenatbib]{mn2e}
\usepackage{graphicx}
\usepackage{amssymb}
\usepackage{times}

\newcommand{\hompc}{\,h\,{\rm Mpc}^{-1}}
\newcommand{\mpcoh}{\,h^{-1}\,{\rm Mpc}}

\newcommand{\Omegam}{\ensuremath{\Omega_{\mathrm{m}}}}
\newcommand{\Omegak}{\ensuremath{\Omega_{\mathrm{K}}}}

\newcommand{\Omegab}{\Omega_{b}}
\newcommand{\avg}[1]{\langle#1\rangle}

\def\simgt{\stackrel{>}{{}_\sim}}
\newcommand\nodata{ ~$\cdots$~ }%
\newcommand{\snl}{\Sigma_{nl}}
\newcommand{\Veff}{\ensuremath{V_{\mathrm{eff}}}}

\begin{document}

\title[BAO in SDSS-III BOSS DR9 galaxies] {The clustering of galaxies
  in the SDSS-III Baryon Oscillation Spectroscopic Survey: Baryon
  Acoustic Oscillations in the Data Release 9 Spectroscopic Galaxy
  Sample}

\author[L. Anderson et al.]{\parbox{\textwidth}{\Large
Lauren Anderson$^{1}$,
Eric Aubourg$^{2}$,
Stephen Bailey$^{3}$,
Dmitry Bizyaev$^{4}$,
Michael Blanton$^{5}$,
Adam S. Bolton$^{6}$,
J. Brinkmann$^{4}$,
Joel R. Brownstein$^{6}$,
Angela Burden$^{7}$,
Antonio J. Cuesta$^{8}$,
Luiz N. A. da Costa$^{9,10}$,
Kyle S. Dawson$^{6}$,
Roland de Putter$^{11,12}$,
Daniel J. Eisenstein$^{13}$,
James E. Gunn$^{14}$,
Hong Guo$^{15}$,
Jean-Christophe Hamilton$^{2}$,
Paul Harding$^{15}$,
Shirley Ho$^{3,14}$,
Klaus Honscheid$^{16}$,
Eyal Kazin$^{17}$,
D. Kirkby$^{18}$,
Jean-Paul Kneib$^{19}$,
Antione Labatie$^{20}$,
Craig Loomis$^{21}$,
Robert H. Lupton$^{14}$,
Elena Malanushenko$^{4}$,
Viktor Malanushenko$^{4}$,
Rachel Mandelbaum$^{14,21}$,
Marc Manera$^{7}$,
Claudia Maraston$^{7}$,
Cameron K. McBride$^{13}$,
Kushal T. Mehta$^{22}$,
Olga Mena$^{11}$,
Francesco Montesano$^{23}$,
Demetri Muna$^{5}$,
Robert C. Nichol$^{7}$,
Sebasti\'an E. Nuza$^{24}$,
Matthew D. Olmstead$^{6}$,
Daniel Oravetz$^{4}$,
Nikhil Padmanabhan$^{8}$,
Nathalie Palanque-Delabrouille$^{25}$,
Kaike Pan$^{4}$,
John Parejko$^{8}$,
Isabelle P\^aris$^{26}$,
Will J. Percival$^{7}$,
Patrick Petitjean$^{26}$,
Francisco Prada$^{27,28,29}$,
Beth Reid$^{3,30}$,
Natalie A. Roe$^{3}$,
Ashley J. Ross$^{7}$,
Nicholas P. Ross$^{3}$,
Lado Samushia$^{7,31}$,
Ariel G. S\'anchez$^{23}$,
David J. Schlegel\thanks{BOSS PI: djschlegel@lbl.gov}$^{3}$,
Donald P. Schneider$^{32,33}$,
Claudia G. Sc\'occola$^{34,35}$,
Hee-Jong Seo$^{36}$,
Erin S. Sheldon$^{37}$,
Audrey Simmons$^{4}$,
Ramin A. Skibba$^{22}$,
Michael A. Strauss$^{21}$,
Molly E. C. Swanson$^{13}$,
Daniel Thomas$^{7}$,
Jeremy L. Tinker$^{5}$,
Rita Tojeiro$^{7}$,
Mariana Vargas Maga\~na$^{2}$,
Licia Verde$^{38}$,
Christian Wagner$^{12}$,
David A. Wake$^{39}$,
Benjamin A. Weaver$^{5}$,
David H. Weinberg$^{40}$,
Martin White$^{3,41,42}$,
Xiaoying Xu$^{22}$,
Christophe Y\`{e}che$^{25}$,
Idit Zehavi$^{15}$,
Gong-Bo Zhao$^{7,43}$
 } \vspace*{4pt} \\ 
\scriptsize $^{1}$ Department of Astronomy, University of Washington, Box 351580, Seattle, WA 98195, USA\vspace*{-2pt} \\ 
\scriptsize $^{2}$ APC, Astroparticule et Cosmologie, Universit\'e Paris Diderot, CNRS/IN2P3, CEA/Irfu, Observatoire de Paris, Sorbonne Paris Cit\'e, 10, rue Alice Domon \& L\'eonie Duquet, 75205 Paris Cedex 13, France\vspace*{-2pt} \\ 
\scriptsize $^{3}$ Lawrence Berkeley National Laboratory, 1 Cyclotron Road, Berkeley, CA 94720, USA\vspace*{-2pt} \\ 
\scriptsize $^{4}$ Apache Point Observatory, P.O. Box 59, Sunspot, NM 88349-0059, USA\vspace*{-2pt} \\ 
\scriptsize $^{5}$ Center for Cosmology and Particle Physics, New York University, New York, NY 10003, USA\vspace*{-2pt} \\ 
\scriptsize $^{6}$ Department Physics and Astronomy, University of Utah, UT 84112, USA\vspace*{-2pt} \\ 
\scriptsize $^{7}$ Institute of Cosmology \& Gravitation, Dennis Sciama Building, University of Portsmouth, Portsmouth, PO1 3FX, UK\vspace*{-2pt} \\ 
\scriptsize $^{8}$ Department of Physics, Yale University, 260 Whitney Ave, New Haven, CT 06520, USA\vspace*{-2pt} \\ 
\scriptsize $^{9}$ Observat\'orio Nacional, Rua Gal. Jos\'e Cristino 77, Rio de Janeiro, RJ - 20921-400, Brazil\vspace*{-2pt} \\ 
\scriptsize $^{10}$ Laborat\'orio Interinstitucional de e-Astronomia - LineA, Rua Gal. Jos\'e Cristino 77, Rio de Janeiro, RJ - 20921-400, Brazil\vspace*{-2pt} \\ 
\scriptsize $^{11}$ Instituto de Fisica Corpuscular, Universidad de Valencia-CSIC, Spain\vspace*{-2pt} \\ 
\scriptsize $^{12}$ ICC, University of Barcelona (IEEC-UB), Marti i Franques 1, Barcelona 08028, Spain\vspace*{-2pt} \\ 
\scriptsize $^{13}$ Harvard-Smithsonian Center for Astrophysics, 60 Garden St., Cambridge, MA 02138, USA\vspace*{-2pt} \\ 
\scriptsize $^{14}$ Department of Physics, Carnegie Mellon University, 5000 Forbes Avenue, Pittsburgh, PA 15213, USA\vspace*{-2pt} \\ 
\scriptsize $^{15}$ Department of Astronomy, Case Western Reserve University, Cleveland, Ohio 44106, USA\vspace*{-2pt} \\ 
\scriptsize $^{16}$ Department of Physics and CCAPP, Ohio State University, Columbus, Ohio 43210, USA\vspace*{-2pt} \\ 
\scriptsize $^{17}$ Centre for Astrophysics and Supercomputing, Swinburne University of Technology, P.O. Box 218, Hawthorn, Victoria 3122, Australia\vspace*{-2pt} \\ 
\scriptsize $^{18}$ Department of Physics and Astronomy, UC Irvine, 4129 Frederick Reines Hall, Irvine, CA 92697, USA\vspace*{-2pt} \\ 
\scriptsize $^{19}$ Laboratoire d'Astrophysique de Marseille, CNRS-Universite Aix-Marseille, 38 rue F. Joliot-Curie 13388 Marseille Cedex 13, France\vspace*{-2pt} \\ 
\scriptsize $^{20}$ Laboratoire Astroparticule et Cosmologie, Universite Paris 7 Denis Diderot, Paris, France\vspace*{-2pt} \\ 
\scriptsize $^{21}$ Department of Astrophysical Sciences, Princeton University, Ivy Lane, Princeton, NJ 08544, USA\vspace*{-2pt} \\ 
\scriptsize $^{22}$ Steward Observatory, University of Arizona, 933 North Cherry Ave., Tucson, AZ 85721, USA\vspace*{-2pt} \\ 
\scriptsize $^{23}$ Max-Planck-Institut f\"ur extraterrestrische Physik, Postfach 1312, Giessenbachstr., 85748 Garching, Germany\vspace*{-2pt} \\ 
\scriptsize $^{24}$ Leibniz-Institut f\"{u}r Astrophysik Potsdam (AIP), An der Sternwarte 16, 14482 Potsdam, Germany\vspace*{-2pt} \\ 
\scriptsize $^{25}$ CEA, Centre de Saclay, IRFU, 91191 Gif-sur-Yvette, France\vspace*{-2pt} \\ 
\scriptsize $^{26}$ Universit\'e Paris 6, Institut d'Astrophysique de Paris, UMR7095-CNRS, 98bis Boulevard Arago, 75014 Paris, France\vspace*{-2pt} \\ 
\scriptsize $^{27}$ Campus of International Excellence UAM+CSIC, Cantoblanco, E-28049 Madrid, Spain\vspace*{-2pt} \\ 
\scriptsize $^{28}$ Instituto de Fisica Teorica (UAM/CSIC), Universidad Autonoma de Madrid, Cantoblanco, E-28049 Madrid, Spain\vspace*{-2pt} \\ 
\scriptsize $^{29}$ Instituto de Astrof\'isica de Andaluc\'ia (CSIC), E-18080 Granada, Spain\vspace*{-2pt} \\ 
\scriptsize $^{30}$ Hubble Fellow\vspace*{-2pt} \\ 
\scriptsize $^{31}$ National Abastumani Astrophysical Observatory, Ilia State University, 2A Kazbegi Ave., GE-1060 Tbilisi, Georgia\vspace*{-2pt} \\ 
\scriptsize $^{32}$ Department of Astronomy and Astrophysics, The Pennsylvania State University, University Park, PA 16802, USA\vspace*{-2pt} \\ 
\scriptsize $^{33}$ Institute for Gravitation and the Cosmos, The Pennsylvania State University, University Park, PA 16802, USA\vspace*{-2pt} \\ 
\scriptsize $^{34}$ Instituto de Astrof{\'\i}sica de Canarias (IAC), C/V{\'\i}a L\'actea, s/n, E-38200, La Laguna, Tenerife, Spain\vspace*{-2pt} \\ 
\scriptsize $^{35}$ Dpto. Astrof{\'\i}sica, Universidad de La Laguna (ULL), E-38206 La Laguna, Tenerife, Spain\vspace*{-2pt} \\ 
\scriptsize $^{36}$ Berkeley Center for Cosmological Physics, LBL and Department of Physics, University of California, Berkeley, CA 94720, USA\vspace*{-2pt} \\ 
\scriptsize $^{37}$ Brookhaven National Laboratory, Bldg 510, Upton, New York 11973, USA\vspace*{-2pt} \\ 
\scriptsize $^{38}$ ICREA \& ICC-UB University of Barcelona, Marti i Franques 1, 08028 Barcelona, Spain\vspace*{-2pt} \\ 
\scriptsize $^{39}$ Yale Center for Astronomy and Astrophysics, Yale University, New Haven, CT 06511, USA\vspace*{-2pt} \\ 
\scriptsize $^{40}$ Department of Astronomy and CCAPP, Ohio State University, Columbus, Ohio, USA\vspace*{-2pt} \\ 
\scriptsize $^{41}$ Department of Physics, University of California, 366 LeConte Hall, Berkeley, CA 94720, USA\vspace*{-2pt} \\ 
\scriptsize $^{42}$ Department of Astronomy, 601 Campbell Hall, University of California at Berkeley, Berkeley, CA 94720, USA\vspace*{-2pt} \\ 
\scriptsize $^{43}$ National Astronomy Observatories, Chinese Academy of Science, Beijing, 100012, P.R. China\vspace*{-2pt} \\ 
}

\date{\today} 
\pagerange{\pageref{firstpage}--\pageref{lastpage}} \pubyear{2011}
\maketitle
\label{firstpage}

\begin{abstract}

  We present measurements of galaxy clustering from the Baryon
  Oscillation Spectroscopic Survey (BOSS), which is part of the Sloan
  Digital Sky Survey III (SDSS-III). These use the Data Release 9
  (DR9) CMASS sample, which contains $264\,283$ massive galaxies
  covering $3275$ square degrees with an effective redshift $z=0.57$
  and redshift range $0.43<z<0.7$. Assuming a concordance $\Lambda$CDM
  cosmological model, this sample covers an effective volume of
  2.2\,Gpc${}^3$, and represents the largest sample of the Universe
  ever surveyed at this density, $\bar{n} \approx 3 \times 10^{-4}
  h^{-3} {\rm Mpc}^3$.  We measure the angle-averaged galaxy
  correlation function and power spectrum, including density-field
  reconstruction of the baryon acoustic oscillation (BAO) feature.
  The acoustic features are detected at a significance of $5\sigma$ in
  both the correlation function and power spectrum. Combining with the
  SDSS-II Luminous Red Galaxy Sample, the detection significance
  increases to $6.7\sigma$.  Fitting for the position of the acoustic
  features measures the distance to $z=0.57$ relative to the sound
  horizon $D_V/r_s = 13.67\pm 0.22$ at $z=0.57$. Assuming a fiducial
  sound horizon of $153.19\,{\rm Mpc}$, which matches cosmic microwave
  background constraints, this corresponds to a distance
  $D_V(z=0.57)=2094\pm 34\,{\rm Mpc}$. At 1.7 per cent, this is the
  most precise distance constraint ever obtained from a galaxy
  survey. We place this result alongside previous BAO measurements in
  a cosmological distance ladder and find excellent agreement with the
  current supernova measurements. We use these distance measurements
  to constrain various cosmological models, finding continuing support
  for a flat Universe with a cosmological constant.

\end{abstract}


\section{Introduction}
\label{sec:intro}
Explaining the late-time acceleration of the expansion rate of the
Universe \citep{riess98, Perl99} is one of the most perplexing problems in modern physics.
All known attempts require exotic ingredients: a new, very small energy
scale in a cosmological constant or low-mass field, a change to
General Relativity to weaken gravity on large scales or at low
densities, or extra dimensions of space-time. Empirical observations
will provide clues as to the cause by providing precision measurements
of the expansion history and the growth of cosmological structure over
time \citep[e.g.][]{DETF}.

One of the key methods for measuring the expansion history is to use
features in the clustering of galaxies within galaxy surveys as a
ruler with which to measure the distance--redshift relation. Obtaining
precision distance measurements is a long-standing challenge in
astronomy, and the Baryon Acoustic Oscillations (BAO) signal in the
two-point clustering of galaxies provides a particularly robust
quantity to measure. The BAO arise because the coupling of baryons and
photons by Thomson scattering in the early universe allows acoustic
oscillations at early times, which in turn leads to a rich structure
in the distribution of matter and the anisotropies of the cosmic
microwave background (CMB) radiation.  The distance that acoustic
waves can propagate in the first million years of the universe becomes
a characteristic comoving scale (\citealt{Pee70,Sun70,Dor78}; a description of
the physics leading to the features can be found in \citealt{Eis98} or
Appendix A of \citealt{MeiWhiPea} and a discussion of the acoustic signal
in configuration space can be found in \citealt{EisSeoWhi07}). 
As the acoustic signature is imprinted on
very large scales ($\sim 150$ Mpc) it is quite insensitive to
astrophysical processing that occurs on smaller scales, thus BAO
experiments are affected by a very low level of systematics induced by
such processes.
Recent 
reviews of BAO as a probe of dark energy may be found in \citet{EisReview} and
\citet{WeinbergReview}.

This acoustic signature has now been detected in many different galaxy
surveys, using a variety of methods to analyse the evolved density
field \citep{percival01,Mill01,Eis05,Col05,Hut06,Pad07,BCBL,Per07,Oku08,
Gaz09,Kaz10,Per10,Rei10,Bla11a,Beutler11,Seo12,Pad12},  and it is already
producing stringent constraints on cosmological models.
Constraints can be obtained from either photometric or spectroscopic
samples, though for the same volume and number of galaxies the
spectroscopic samples provide much stronger constraints. The first BAO
measurements came from the 2dF Galaxy Redshift Survey (2dFGRS;
\citealt{colless03}) and the Sloan Digital Sky Survey (SDSS;
\citealt{Yor00}); when combined, the most recent analyses give a 2.7
per cent measurement of the distance-redshift relation at $z=0.275$
\citep[e.g.][]{Per10}.  Adding to these data, \citet{Bla11a} measured
the BAO feature at $z=0.6$, using the WiggleZ survey
\citep{Drinkwater10}, making a 4 per cent distance measurement from
$132\,509$ galaxies. This result was subsequently improved to provide
distance measurements of accuracy 7.2 per cent, 4.5 per cent and 5.0
per cent in three bins centered at redshifts $z=0.44$, $0.60$, and
$0.73$ respectively, using the full sample of $158\,741$ galaxies from
this survey \citep{Bla11b}. \citet{Beutler11} made a 4.5 per cent
measurement at $z=0.1$ with the 6dF Galaxy Redshift Survey (6dFGRS:
\citealt{Jones09}). Thus the BAO technique has recently provided a
distance--redshift relation at a series of redshifts both higher and
lower than the 2dFGRS and SDSS measurements.

The BAO signal in an evolved galaxy field, such as those analysed in
the papers described above, differs from that predicted in the matter
field by linear theory alone. The dominant difference is caused by
matter flows and peculiar velocities on intermediate scales
($\sim20\mpcoh$), which act to suppress small-scale oscillations in
the galaxy power spectrum and smooth the BAO feature in the
correlation function
\citep{EisSeoWhi07,CroSco08,Mat08a,Mat08b}. \citet{Eis07a} suggested
that this smoothing can be reversed, in effect using the phase
information within the density field to reconstruct linear
behaviour. Although not a new idea \citep[e.g.][]{Pee89,Pee90, Nuss92,
Gram93}, the dramatic effect on BAO recovery had not been previously
realised, and the majority of the benefit was shown to be recovered
from a simple reconstruction prescription.  This reconstruction
technique has been used to sharpen the BAO feature and improve
distance constraints on mock data \citep{PadWhi09,Noh09,Seo10,Meh11},
and it was recently applied to the SDSS-II Luminous Red Galaxy (LRG)
sample \citep{Pad12}. The reconstruction was particularly effective in
this case, providing a 1.9 per cent distance measurement at $z=0.35$,
decreasing the error by a factor of 1.7 compared with the
pre-reconstruction measurement.

This study is the first in a set of papers to describe the clustering
of galaxies at $z\sim 0.6$ from Data Release 9 (DR9) of the Baryon
Oscillation Spectroscopic Survey \citep[BOSS;][]{BOSS}, which is part
of SDSS-III \citep{Eis11}.  We present cosmological results based on
fits to the BAO signature in the clustering of $264\,283$ galaxies in
this paper.  Redshift-Space Distortion (RSD) and Alcock-Paczynski
\citep[][AP]{AP} measurements are presented in \citet{Rei12}, and an
interpretation of these results in terms of dark energy and modified
gravity models is presented in \citet{Sam12}.  \citet{Tojeiro12}
describe a new method for improving RSD measurements.  Further
constraints from fitting models to the full shape of the correlation
function are presented in \citet{Sanchez12}. \citet{Nuza12} have
compared the clustering with the outcome of a large-volume
cosmological simulation. 

Each of these papers focusses on the high redshift galaxy sample
from BOSS, denoted ``CMASS'', where a set of colour-magnitude cuts are
used to select a roughly volume-limited sample of massive, luminous
galaxies from a redshift of $0.43$ to $0.7$.  We describe the
construction of the galaxy catalog and measurement of both the
correlation function and power spectrum of this sample, before and
after applying the reconstruction algorithm of \citet{Eis07a}. We
present the results of two pipelines for the analysis of BAO: one
utilising the correlation function, and one utilising the power
spectrum. While both statistics contain the same information, they
often have different advantages in the usage and application in the
literature. We compare and contrast measurements made on mock
catalogues using both techniques, and apply both to measure and
analyse the BAO distance scale using CMASS data.  In companion papers
we present weights used to correct for artificial density fluctuations
caused by observational effects \citep{Ross12} and a set of mock
catalogues used to estimate statistical errors \citep{Man12}.  An
analysis of a lower redshift sample of galaxies from BOSS will be
presented in \citet{Par12}.  Clustering measurements from a smaller
sample of CMASS galaxies (the first six months of BOSS data) were
presented by \citet{Whi11} and used to constrain halo occupation
distributions, but these measurements did not extend to the BAO scale.

The layout of our paper is as follows. We introduce the data in
Section~\ref{sec:data} and the catalogue used in
Section~\ref{sec:catalog}. Analysis techniques are described in
Section~\ref{sec:analysis}, and correlation function and power
spectrum measurements are described and presented in
Sections~\ref{sec:xi} and ~\ref{sec:pk}, respectively. These are compared
and our final distance measurement presented in
Section~\ref{sec:consensus}. This measurement is placed in a cosmological context
in Sections~\ref{sec:ladder} and ~\ref{sec:params}. A brief discussion
is given in Section~\ref{sec:discuss}. Finally, a series of Appendices
test the validity of various aspects of the methods used.

Throughout the paper we assume a fiducial $\Lambda$CDM+GR\footnote{We
do not consider any modifications to general relativity in this paper.}
flat cosmological model with $\Omega_m=0.274$, $h=0.7$,
$\Omega_bh^2=0.0224$, $n_s=0.95$ and $\sigma_8=0.8$, similar to the
best-fit WMAP 7-year model \citep{komatsu11}. These parameters allow
us to translate angles and redshifts into distances and provide a
reference against which we measure distances.
The BAO measurement allows us to constrain changes in the distance scale relative
to that predicted by this fiducial model.

\section{The Data}
\label{sec:data}
The Sloan Digital Sky Survey \citep[SDSS;][]{Yor00} mapped over one
quarter of the sky using the dedicated 2.5-m Sloan Telescope
\citep{Gun06} located at Apache Point Observatory in New Mexico.  A
drift-scanning mosaic CCD camera \citep{Gun98} imaged the sky in five
photometric bandpasses \citep{Fuk96,Smi02,Doi10} to a limiting
magnitude of $r\simeq 22.5$.  The imaging data were processed through
a series of pipelines that perform astrometric calibration
\citep{Pie03}, photometric reduction \citep{Photo}, and photometric
calibration \citep{Pad08}.  The magnitudes were corrected for Galactic
extinction using the maps of \citet{SFD98}. BOSS, as part of the
SDSS-III survey \citep{Eis11}, has imaged an additional $3100$ square
degrees of sky over that of SDSS-II \citep{Abazajian09} in the South
Galactic sky, in a manner identical to the original SDSS imaging.
This increased the total imaging SDSS footprint to $14\,055$ square
degrees, with $7600$ square degrees at $|b| > 20$ deg in the North
Galactic Cap and $3000$ square degrees at $|b| > 20$ deg in the South
Galactic Cap.  All of the imaging was re-processed as part of SDSS
Data Release 8 \citep{DR8}.

BOSS is primarily a spectroscopic survey, which is designed to obtain
spectra and redshifts for 1.35 million galaxies over an extragalactic
footprint covering $10\,000$ square degrees.  These galaxies are
selected from the SDSS imaging and are being observed together with
$160\,000$ quasars and approximately $100\,000$ ancillary targets.
The targets are assigned to tiles of diameter $3^\circ$ using a tiling
algorithm that is adaptive to the density of targets on the sky
\citep{Tiling}.  Aluminium plates are drilled with $1000$ holes whose
positions correspond to the positions of objects on each tile, which are 
manually plugged with optical fibres that feed a pair of double
spectrographs.  The double-armed BOSS spectrographs are significantly
upgraded from those used by SDSS-I/II, covering the wavelength range
$3600\,$\AA\ to $10\,000\,$\AA\ with a resolving power of $1500$ to
$2600$ \citep{Gun12}.  In addition to expanding the wavelength
coverage from the SDSS-I range of $3850$ to $9220\,$\AA, the
throughputs have been increased with new CCDs, gratings, and improved
optical elements, and the $640$-fibre cartridges with $3\arcsec$
apertures have been replaced with $1000$-fibre cartridges with
$2\arcsec$ apertures. Each observation is performed in a series of
$900$-second exposures, integrating until a minimum signal-to-noise
ratio is achieved for the faint galaxy targets.  This ensures a
homogeneous data set with a high redshift completeness of $>97$ per
cent over the full survey footprint.
A summary of the survey design appears in \citet{Eis11}, and a full
description will be provided in \citet{Daw12}.

    \subsection{Galaxy target selection} 
    \label{sec:target}
    BOSS makes use of luminous galaxies selected from the multi-colour
SDSS imaging to probe large-scale structure at intermediate redshift
($0.2<z<0.7$).  The target selection is an extension of the 
targeting algorithms for the SDSS-II \citep{Eis01} and
2SLAQ \citep{Can06} Luminous Red Galaxies (LRGs), 
targeting fainter and bluer galaxies in order to achieve
the a number density of $3\times10^{-4}\,h^{3}{\rm Mpc}^{-3}$.
The majority of the galaxies have old stellar systems whose prominent
4000\,\AA\ break makes them relatively easy to target using
multi-colour data.  The details of the target selection algorithm will
be presented in \citet{PadTS}; we summarise the details relevant to
this paper below.

The galaxy target selection in BOSS is divided into two classes of
galaxies: LOWZ galaxies ($0.2 < z < 0.43$) and CMASS galaxies ($0.43 <
z < 0.7$), analogous to the Cut-I and II SDSS-II LRGs.
The 4000\,\AA\ break resides primarily in the $g$ and $r$ bands for the
LOWZ and CMASS redshift ranges respectively.
The LOWZ sample in DR9 was somewhat
compromised by a target selection error, now fixed, in the early data,
and regardless it would have fewer objects and a lower effective
volume than the SDSS-II LRGs over the same redshift range. We
therefore restrict our analysis here to the CMASS sample and use the
results from \citet{Pad12} for measurements in the lower redshift
range.  The small scale clustering results of the LOWZ sample is
described in the companion paper of \citet{Par12}. Future BOSS
analyses will use both the LOWZ and CMASS samples.

The CMASS sample was designed to loosely follow a constant stellar
mass cut (hence the name C{\it onstant}MASS) based on the passive
galaxy template of \citet{Mar09}, and was designed to produce a
uniform mass distribution at all redshifts. The distribution of CMASS
stellar masses \citep{Mar12} and velocity dispersions \citep{Thomas12}
in various redshift bins confirm that this goal was achieved. Unlike
SDSS-II LRGs, we do not exclusively target intrinsically red
galaxies with the CMASS cut. In fact, \citet{Mas11} showed that 26 per cent of CMASS
galaxies are massive spirals. Therefore, whereas both the LOWZ and
CMASS samples are colour-selected, CMASS is a significantly more
complete sample than LOWZ at high stellar masses. This issue is
discussed in detail in \citet{Tojeiro12}, which considers the passive
evolution of galaxies between the SDSS-II Luminous Red Galaxies (which
form a subset of the LOWZ sample) and the CMASS sample.  Most CMASS
objects are central galaxies residing in dark matter halos of
$10^{13}\,h^{-1}M_\odot$, but a non-negligible fraction are satellites
that live primarily in halos about 10 times more massive
\citep{Whi11,Nuza12}.  Galaxies in the CMASS sample are highly biased
($b\sim 2$), and bright enough to be used to trace a large
cosmological volume with sufficient number density to ensure that
shot-noise is not a dominant contributor to the statistical error in
BAO measurements. The combination of large volume, high bias, and
reasonable space density makes CMASS galaxies particularly powerful
for probing statistical properties of large-scale structure.

The CMASS target selection makes use of four definitions of flux
computed by the photometric pipeline.
All magnitudes have been photometrically calibrated using the
uber-calibration of \citet{Pad08} and corrected for Galactic extinction \citep{SFD98}.
The model fluxes are computed using either
a PSF-convolved exponential or de Vaucouleurs light profile fit to the $r$-band
only, and are denoted with the ${\it mod}$ subscript.
Cmodel fluxes are computed
using the best-fit linear combination of an exponential and a
de Vaucouleurs light profile fit to each photometric band
independently \citep{Abazajian09}, and are denoted with the subscript ${\it cmod}$.
Point-spread function (PSF) fluxes are computed by fitting a PSF model to
the galaxy, and are denoted with the subscript ${\it psf}$ \citep{Sto02}.
Fibre fluxes are computed within a 2 arcsec aperture after the image
is convolved with a kernel to produce a 2 arcsec FWHM PSF,
and are denoted with the subscript ${\it fib2}$.
Colours are computed using model fluxes.
Magnitude cuts are performed on cmodel and fibre fluxes.

The CMASS algorithm selects luminous galaxies at $z\simgt 0.4$,
extending Cut II of \citet{Eis01} to both fainter and bluer galaxies.
We first select objects classified as galaxies by the imaging pipeline.
These must then pass the following criteria:
\begin{eqnarray}
  17.5 < i_{\rm cmod} &<& 19.9, \\
  r_{\rm mod} - i_{\rm mod} &<& 2.0, \\
  d_\perp &>& 0.55, \\
  i_{\rm fib2} &<& 21.5, \\
  i_{\rm cmod} &<& 19.86 + 1.6(d_{\perp} - 0.8),
\end{eqnarray}
where the auxiliary colour $d_\perp$ is defined as \citep{Can06}
\begin{equation}
  d_\perp = r_{\rm mod} - i_{\rm mod} - (g_{\rm mod} - r_{\rm mod})/8.0.
\end{equation}

CMASS objects must also pass the following star-galaxy separation cuts
\begin{eqnarray}
  i_{\rm psf} - i_{\rm mod} &>& 0.2 + 0.2(20.0 - i_{\rm mod}), 
  \label{eq:sgcut1} \\
  z_{\rm psf} - z_{\rm mod} &>& 9.125 - 0.46z_{\rm mod},
  \label{eq:sgcut2}
\end{eqnarray}
unless they also pass the LOWZ criteria (see \citealt{Ross12, Par12}),
which only uses the default SDSS-II star-galaxy separation
criterion. Stars will have essentially identical model and PSF fluxes,
and this star-galaxy separation cuts on the difference between these
two magnitudes.  The slope with apparent magnitude in
equations~\ref{eq:sgcut1} and \ref{eq:sgcut2}, which is not used in
the standard star-galaxy separator of the photometric pipeline
\citep{Str02}, was set empirically by analysing commissioning
spectroscopic data that relaxed these cuts. Our choices yield a sample
with approximately 3 per cent stellar contamination, and it discards
approximately 1 per cent of genuine galaxy targets, mostly at the
faint end.  The star-galaxy separation is known to fail when the
seeing is poor, as PSF and model magnitudes approach one another for
all object types in poor seeing.  However this has been shown to have
negligible effect on the angular distribution of targets in SDSS
\citep{Ross11}.

    \subsection{Masks}  
    \label{sec:masks}
    We use the {\sc Mangle\/} software \citep{Mangle} to track the areas
covered by the BOSS survey and the angular completeness of those
regions. The mask is constructed of spherical polygons, which form the base unit
for the geometrical decomposition of the sky. The angular
mask of the survey is formed from the intersection of the imaging 
boundaries (expressed as a set of polygons) and spectroscopic sectors
(areas of the sky covered by a unique set of spectroscopic tiles)
\citep[see][]{Tiling,Teg04,DR8}.


\begin{figure}
\begin{center}
\resizebox{3.2in}{!}{\includegraphics{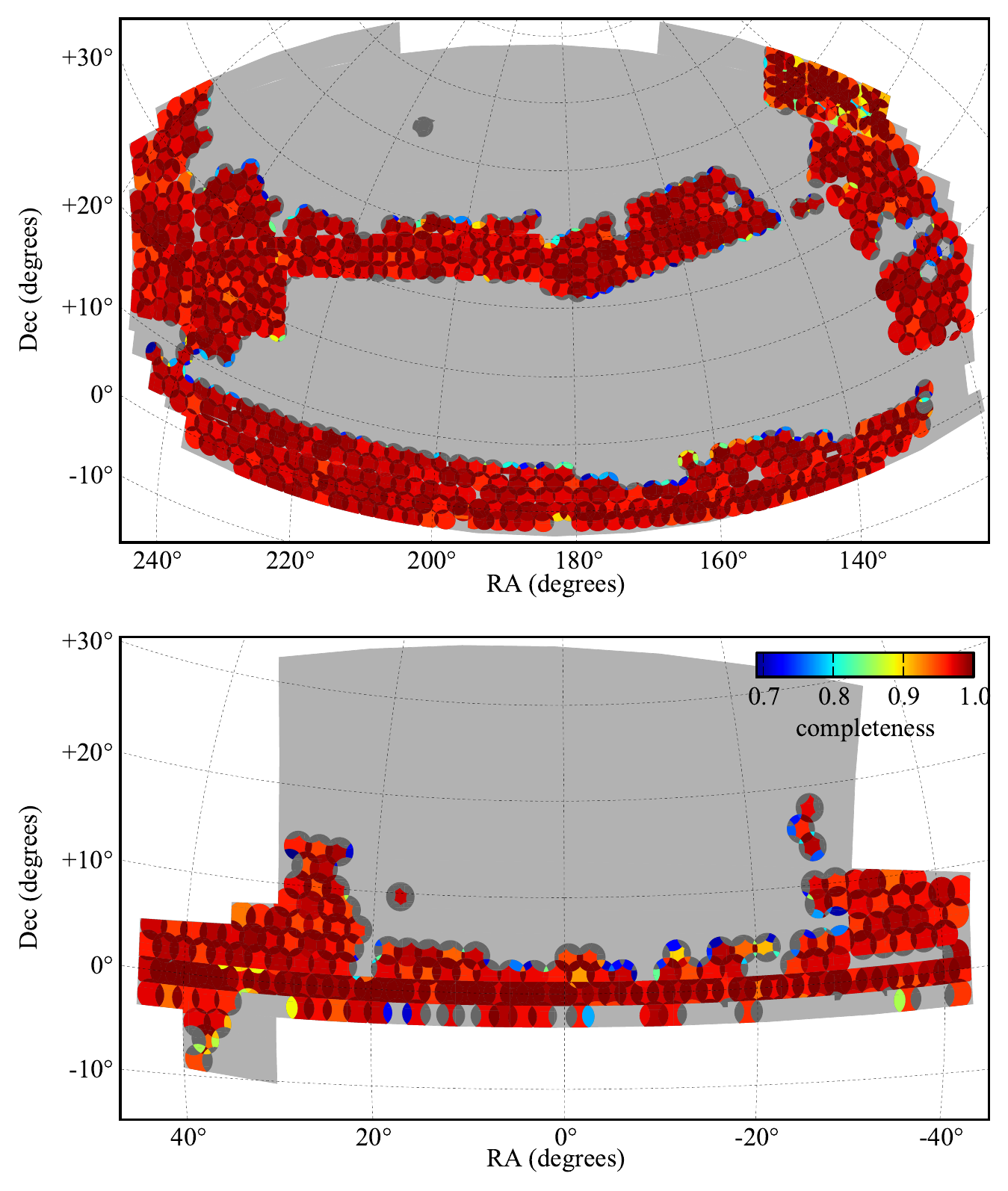}}
\end{center}
\caption{The sky coverage of the galaxies used in this analysis. The
  light grey shaded region shows the expected total footprint of the
  survey, totalling $10\,269$ deg$^2$. The coloured and dark grey
  regions indicate the DR9 spectroscopic coverage of the survey,
  totalling $3792$ deg$^2$.  Colours indicate the completeness within
  each sector used to build the random catalog as defined in
  Eq.~\ref{eq:comp}.  Sectors coloured dark grey are removed from the
  analysis by the cuts described in Section~\ref{sec:targ_summary}.
  The total effective area (accounting for all applied cuts and the
  completeness in every sector included) used in our analysis is
  $3275$ deg$^2$.  The low completeness at many edges is due to
  unobserved tiles that will overlap the current geometry in future
  data releases. }
\label{fig:boss_footprint}
\end{figure}

In addition to tracking the outline of the survey region and the
position of the spectroscopic plates, we apply several ``vetos'' in
constructing the catalog. Regions were masked where the imaging was
unphotometric, the PSF modelling failed, the imaging reduction pipeline
timed out (usually due to too many blended objects in a single field),
or the image was identified as having critical problems. Small regions
around the centre posts of the plates where fibres cannot be placed
due to physical limitations and around bright stars in the Tycho
catalog \citep{tycho2} were also masked.  The mask radius for stars
from the Tycho catalog was
\begin{equation}
  R = (0.0802B^2 - 1.860B + 11.625) {\rm\ arcmin} \,,
\end{equation}
where $B$ is the Tycho $B_T$ magnitude clipped to fall in the range
$[6,11.5]$.  
We also place a mask at the locations of objects with higher priority
(mostly high-$z$ quasars) than galaxies. A galaxy cannot be observed
at a location within the fibre collision radius of these points.
In total we masked $\sim5$ per cent of the area cover by the
set of observed tiles due to our ``veto'' mask. The sky coverage of
the CMASS galaxies is shown in Fig.~\ref{fig:boss_footprint}, and the
basic parameters including areas and galaxy numbers are presented in
Table~\ref{tab:basic_props}.

    \subsection{Measuring galaxy redshifts}  
    \label{sec:redshifts}
    Spectroscopic calibration, extraction, classification, and redshift analysis
were carried out using the \texttt{v5\_4\_45} tag of the \texttt{idlspec2d} 
software package.\footnote{\texttt{http://www.sdss3.org/svn/repo/idlspec2d/tags/v5\_4\_45/}}.
The classification and redshift of each object are determined by
fitting their coadded spectra to a set of galaxy, quasar, and star
eigentemplates.  The fit includes a polynomial background 
(quadratic for galaxies, quasars, and cataclysmic variable stars; cubic for all other stars)
to allow for residual
extinction effects or broadband continua not otherwise described by
the templates.  The reduced $\chi^2$ versus redshift is mapped in
redshift steps corresponding to the logarithmic pixel scale of the
spectra, $\Delta \log_{10}(\lambda) = 0.0001$.  Galaxy templates are
fit from $z=-0.01$ to $1.00$, quasar templates are fit from $z=0.0033$
to $7.00$, and star templates are fit from $z=-0.004$ to $0.004$ ($\pm
1200\,$kms$^{-1}$).  The template fit with the best reduced $\chi^2$ is
selected as the classification and redshift, with warning flags set
for poor wavelength coverage, 
broken/dropped and sky-target fibres, and best fits which are within
$\Delta \chi^2/\rm{dof} = 0.01$ of the next best fit (comparing only
to fits with a velocity difference of less than 1000~kms$^{-1}$).  This
method is the same as used for the SDSS DR8; see \cite{DR8} for
further explanation.

For galaxy targets, a dominant source of false identifications is due to
quasar templates with unphysical fit parameters, {\em e.g.}, large negative
parameters causing a quasar template emission feature to fit a galaxy
absorption feature.  Thus, for galaxy targets, the best classification
and redshift are selected only from the fits to galaxy and star templates, 
and we restrict to fits the pipeline classifies as robust.
\footnote{
These fits are stored in the
``*\_NOQSO'' versions of the Z, Z\_ERR, ZWARNING, CLASS, SUBCLASS,
and RCHI2DIFF fields in the upcoming Data Release 9.
This analysis uses Z\_NOQSO redshifts for targets selected with
CLASS\_NOQSO=``GALAXY'' and ZWARNING\_NOQSO=0.}

\begin{figure}
  \centering
  \includegraphics[width=\linewidth]{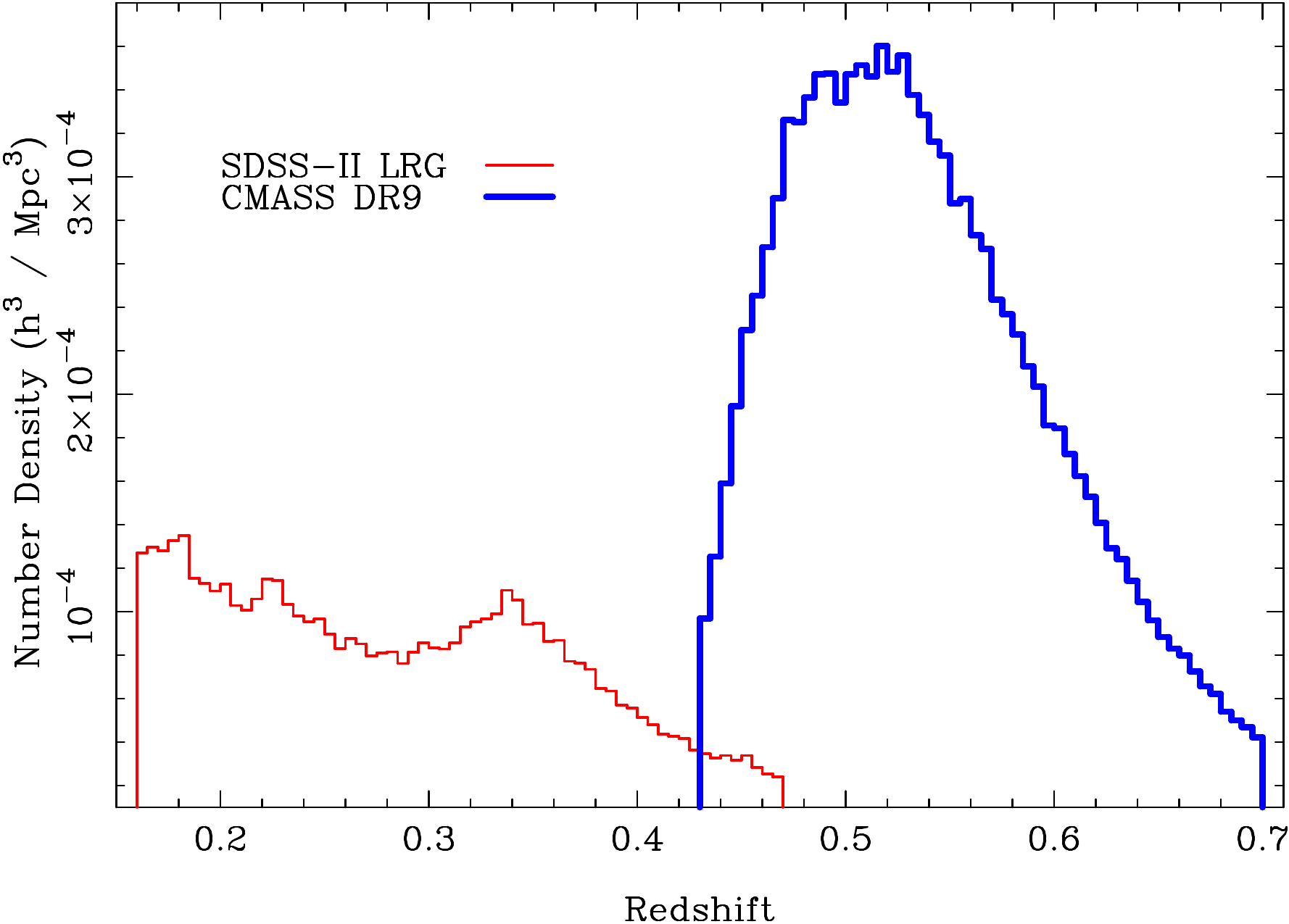}
  \caption{ The galaxy number density as a function of redshift for
    the BOSS DR9 CMASS sample (thick blue line) used in this analysis,
    which ranges in redshift between $0.43<z<0.7$. For comparison, we
    also plot the density for a SDSS-II DR7 LRG sample (thin red line)
    covering $0.16<z<0.47$, which was used in \citet{Pad12}. Note that
    both selections include a small fraction of objects that fall
    outside the redshift cuts shown here.  }
\label{fig:zdist}
\end{figure}

Fig.~\ref{fig:zdist} shows the galaxy number density of the CMASS
sample, compared with the SDSS-II LRG sample.
The CMASS galaxies have approximately three times the density of the 
SDSS-II LRG sample, and sample the underlying density field with 
lower noise and higher fidelity.
Although redshifts are recovered at higher and
lower redshifts, we limit the CMASS redshift range to $0.43<z<0.7$: at
lower redshifts the BOSS LOWZ sample is more dense, and we wish to
remove overlap between samples. At $z>0.7$, the efficiency with which
we recover redshifts decreases, potentially leading to increased 
systematic errors \citep{Ross12}.

\section{Catalog creation}
\label{sec:catalog}

    \subsection{Target photometry}
    \label{sec:photometry}
    Target galaxies are selected as described in Section~\ref{sec:target}
based on the best reduced photometry available {\it at the time of
target selection}. All imaging data used by BOSS is based on
photometry from SDSS Data Release 8 (DR8; \citealt{DR8}).  During the
early phases of BOSS, the final DR8 imaging data were still being
processed, and therefore some of the SDSS imaging (9 per cent) used for
targeting CMASS galaxies is now designated as secondary\footnote{i.e.
there is an overlapping observation with higher quality photometry} in
the DR8 database.  Although the measured object parameters from
different observing runs over the same region agree within the
photometric errors, there can be significant differences between
target samples selected from these different observing runs.  These
differences between our target catalog and that obtained using DR8
primary photometry arise due to the stochastic variations one expects
given the magnitude errors and the different photometry. We have
therefore produced a ``combined photometry target'' sample that uses
the photometry input to each run of the targeting software. Thus,
rather than thinking of the set of BOSS CMASS galaxies as being a
unique sample of galaxies chosen with the properties described in
Section~\ref{sec:target}, we should really consider the stochastic
nature caused by photometric errors: we are simply observing one of
the samples that could have been selected with these properties: using
different imaging data we would find a different sample. We do not
expect this issue to have any impact on the analysis or results
presented here, as it is stochastic in nature. We only include this
description for completeness and to aid future uses of these data.

    \subsection{Close-pair corrections}
    \label{sec:closepair}
    The protective sheath around each spectroscopic fibre and the ferrule
that holds the fiber in the plug plate has a diameter of
62$^{\prime\prime}$ on the focal plane, so no two objects separated by
less than this can be observed on a single plate. This means that
groups and clusters of galaxies with members closer than this apparent
separation will be systematically under-sampled, strongly affecting
the measured small-scale clustering signal if uncorrected. The
targeting algorithm has been designed to place a fibre on as many
objects within tight groups as possible (The selection is random with
respect to galaxy properties other than position on the sky). Where
there are two plates covering a sector we find that approximately $25$
per cent of the pairs separated by $< 62^{\prime\prime}$ only have one
galaxy observed; this fraction reduces to $<7$ per cent when a sector
is covered by three or more plates. An algorithm that corrects for
these effects on small scales is presented and tested in
\citet{guo2012}. On large scales, this procedure is equivalent to
upweighting the galaxies nearest to each unobserved galaxy, and we
adopt this procedure in the analysis presented here. An alternative
would have been to upweight all galaxies within the sector to
compensate, which would have better shot-noise properties. However,
the lost galaxies will predominantly be in groups, and thus may have
different clustering properties than the average galaxy. We therefore
accept the subsequent slightly increased shot-noise contribution. This
close-pair correction weight is denoted $w_{\rm cp}$ throughout this
paper (see also Section~\ref{sec:final_weights}). For each target we
set $w_{\rm cp}=1$, and add one to this for each CMASS target within
$62^{\prime\prime}$ that failed to get a fiber allocated. This
correction affects $\sim 5$ per cent of all the galaxies, with most of
these in pairs with $w_{\rm cp} = 2$.

    \subsection{SDSS-II redshifts} 
    \label{sec:known_redshifts}
    Accurate redshifts for a subsample of the target galaxies were previously
obtained within the SDSS-II survey \citep{Abazajian09}.  These
galaxies were not re-observed by BOSS. We have redshift measurements
for 100 per cent of the SDSS-II galaxy subsample by definition, and
although these lie within the BOSS survey region, the angular
distribution will systematically differ from that of the remaining
subsample. We do not try to define a survey mask that amalgamates the
SDSS-II and BOSS observations because the SDSS-II redshifts do not
even form a random subsample of the BOSS targets based on galaxy
properties. Instead, we subsample the SDSS-II galaxies to match the
sector completeness of the galaxies observed within the BOSS program,
where sector completeness $C_{\rm BOSS}$ is defined within each mask
sector as
\begin{equation}  \label{eq:comp}
  C_{\rm BOSS} = \frac{N_{\rm obs}+N_{\rm cp}}{N_{\rm targ}-N_{\rm known}},
\end{equation}
where $N_{\rm obs}$, $N_{\rm targ}$, $N_{\rm cp}$ and~$N_{\rm known}$ are defined in
Section~\ref{sec:targ_summary}. 

We also subsample the galaxies so that including the galaxies with
previously known redshifts does not change the fraction of close pairs
observed in any sector.This task is accomplished by calculating, for
each sector, the fraction of close pairs within the sample of targets
from which the known galaxies have been removed. We then subsample
close pairs introduced when including the known galaxies, randomly
removing either galaxy in a pair until the fraction of the introduced
close pairs matches that of the sample without known galaxies.

Thus we force the known redshifts to have the same statistical
properties as the BOSS galaxies within each sector, such that the
angular distribution of the combined sample follows that of the BOSS
angular mask.

    \subsection{Redshift-failure corrections} 
    \label{sec:zfail_corrections}
    We do not achieve stellar classification or a good redshift
determination for every spectrum taken. The probability of
successfully obtaining an accurate spectrum is dependent on the fibre
used - some fibres have degraded transmission, the optical quality
(resolution) is better for spectra in some regions of the CCD than
others, and the quality of the sky-subtraction is worse where the
resolution is lower. Bundles of fibres are generally allocated to similar regions
on the plates, and thus the failure rate is a strong function of
position in the field-of-view for each observation. This effect is
shown in Fig.~3 of our companion paper \citep{Ross12}, where specific
regions within each field-of-view are demonstrated to have worse-than-average
failure probabilities. Our overall redshift success rate is 98.2 per
cent, so we lack redshifts for a sufficiently small subsample that the
effect on the measured clustering signal is very small.

We correct for this minor issue by upweighting the nearest (based on
an angular search) target object for which a galaxy redshift, or
stellar classification, has been successfully achieved. The redshift
distribution of these nearest neighbours matches that of the full
sample (see Fig. 4 of \citealt{Ross12}), suggesting that the
anisotropic component of the redshift-failure distribution which
should be the difference between the two does not depend on
redshift. Consequently, upweighting the nearest galaxy with a good
redshift should match the true large-scale density and correct for the
spatially dependent redshift failure effects. Further details can be
found in Section 2.3 of \citet{Ross12}. This redshift-failure
correction weight is denoted $w_{\rm rf}$ throughout this paper (see
also Section~\ref{sec:final_weights}). As in the case of $w_{\rm cp}$,
we define it to be unity for all galaxies and then add one if there is
a nearby redshift failure.

    \subsection{Summary of target objects}  
    \label{sec:targ_summary}
    \begin{table}
\begin{center}
\begin{tabular}{lrrr}
Property & NGC & SGC & total\\ \hline
$\bar{N}_{\rm gal}$ & 222\,538 & 60\,792 & 283\,330 \\
$\bar{N}_{\rm known}$ & 3766 & 1810 & 5576 \\
$\bar{N}_{\rm star}$  & 7201 & 1771 & 8972 \\
$\bar{N}_{\rm fail}$ & 3751 & 1122 & 4873 \\
$\bar{N}_{\rm cp}$ & 14\,116 & 3640 & 17\,756 \\
$\bar{N}_{\rm missed}$ & 4931 & 1911 & 6842 \\
&&&\\
$\bar{N}_{\rm used}$ & 207\,246 & 57\,037 & 264\,283 \\
$\bar{N}_{\rm obs}$ & 233\,490 & 63\,685 & 297\,175 \\
$\bar{N}_{\rm targ}$ & 256\,303 & 71\,046 & 327\,349 \\
&&&\\
Total area / deg$^2$ & 2635 & 709 & 3344 \\
Effective area / deg$^2$ & 2584 & 690 & 3275
\end{tabular}
\end{center}
\caption{Basic parameters of the CMASS DR9 sample, when summed over
  all mask sectors (see Section~\ref{sec:masks}). We define
  $\bar{N}_x=\sum_{\rm sectors}N_x$, and the meaning of each
  $N_X$ is given in the
  text. In our clustering analyses, we only consider galaxies with
  $0.43<z<0.7$, which is why $\bar{N}_{\rm used}<(\bar{N}_{\rm gal}+\bar{N}_{\rm  known})$. 
  We split between the Northern Galactic Cap (NGC) and
  Southern Galactic Cap (SGC) regions, for which we calculate separate
  galaxy and random catalogues. The total area is the sum
  of the areas of all mask sectors passing our completeness cut, and the effective area 
  is the sum when we multiply the area of each sector by its completeness, $C_{\rm BOSS}$.}
\label{tab:basic_props}
\end{table}
 
To summarise, the following outcomes are available for BOSS targets
that are covered by the survey:
\begin{enumerate}
  \item galaxies with redshifts from good BOSS spectra (we denote the
    number in each sector by $N_{\rm gal}$),
  \item galaxies with redshifts from SDSS-II spectra ($N_{\rm known}$),
  \item spectroscopically confirmed stars ($N_{\rm star}$),
  \item objects with BOSS spectra from which stellar classification or redshift
determination failed ($N_{\rm fail}$),
  \item objects with no spectra, in a close-pair ($N_{\rm cp}$),
  \item objects with no spectra, or spectra removed following the
    subsampling discussed in Section~\ref{sec:known_redshifts}, not in a
    close-pair ($N_{\rm missed}$).
\end{enumerate}

In the following we define the number of target objects per sector
\begin{equation}
  N_{\rm targ} = N_{\rm star}+N_{\rm gal}+N_{\rm fail}+
                 N_{\rm cp}+N_{\rm missed}+N_{\rm known},
\end{equation}
and the number of targets observed per sector
\begin{equation}
  N_{\rm obs} = N_{\rm star}+N_{\rm gal}+N_{\rm fail}.
\end{equation}
The number of good galaxies used in the analysis per sector $N_{\rm
  used}$, is less than $N_{\rm gal}+N_{\rm known}$ as we only use
galaxies with $0.43<z<0.7$. Table~\ref{tab:basic_props} gives the
total split of galaxies in the CMASS DR9 target sample into these
categories, where we define $\bar{N}_x=\sum_{\rm sectors}N_x$, and the
areas and weighted areas for the CMASS sample in the Northern Galactic
Cap (NGC) and Southern Galactic Cap (SGC), and combined as derived
from the DR9 data.

Considering the numbers of galaxies in each category per sector, we
can define a sector completeness as in Eq.~\ref{eq:comp}, and the
galaxies with previously known redshifts are subsampled to match this
completeness, as well as the BOSS-only close-pair fraction as detailed
in Section~\ref{sec:known_redshifts}. The distribution of sector
completenesses across the BOSS footprint is shown in
Fig.~\ref{fig:boss_footprint}.  To remove sectors that have only been
partially observed, we only retain sectors with $C_{\rm BOSS}$ greater
than 70 per cent.

We also make a cut on the total redshift failure within each
sector. First, we define a redshift completeness by
\begin{equation}  \label{eq:red_comp}
  C_{\rm red} = \frac{N_{\rm gal}}{N_{\rm obs}-N_{\rm star}}.
\end{equation}
Then a sector is removed if it has more than 10 BOSS spectra, but
fewer than 80 per cent of the non-stellar spectra have good redshift
measurements (i.e. we remove sectors with $N_{\rm obs}-N_{\rm cp}>10$
and $C_{\rm red}<0.8$). For these sectors we assume that there was a
serious problem with the observations. Plate 3698 observed MJD 55501
is responsible for many redshift failures; it was poor data
inadvertently included in DR9 with a CMASS failure rate of 23 per
cent.

    \subsection{Systematic weights}
    \label{sec:sysweights}
    \cite{Ross11} have presented a critical examination of the large-scale
angular clustering of CMASS target galaxies. They demonstrated that
the density of stars has a significant effect on the observed density
of galaxies, and this can introduce spurious fluctuations in the
galaxy density field.  This effect arises because stars have a
large-scale power signature in their distribution across the
sky. Additional potential systematics such as Galactic extinction,
seeing, airmass, and sky background have also been investigated, and
all have been found to potentially introduce spurious fluctuations
into the galaxy density field, albeit to varying degrees. These
non-cosmological fluctuations can be corrected for using a weighting
scheme that minimises these fluctuations as a function of a given
systematic effect (see Fig. 4 of \citealt{Ross11}).

\cite{Ross12} investigated systematic effects on the 3D clustering of
the DR9 CMASS sample.  They found that stellar density is the primary
source of systematic error, and that computing a set of weights based
on stellar density and $i_{fib2}$ magnitude alone has a similar effect
as fitting for all five systematic sources simultaneously.
Over-fitting these fluctuations can result in removing cosmological
power, if the weights remove what are in truth statistical
fluctuations. Hence the simplicity of correcting for one systematic
only, with the added dependence on $i_{fib2}$, minimises this
risk. This approach was tested by making use of mock catalogues. We
refer the reader to \cite{Ross12} for a detailed study of the effect
of all weighting schemes, and an analysis of each Galactic hemisphere
separately. Note that \cite{Ross12} explicitly verify that the BAO
scale is insensitive to these systematic effects.

The adopted methodology for computing the angular systematic weights
used throughout this paper is as follows. The weights are defined as
\begin{equation}
w_{sys} (n_s, i_{fib2}) = A + B n_s
\end{equation}
where $n_s$ is the stellar density. $A$ and $B$ are given by
\begin{eqnarray}
  A &=& A_0 i_{fib2} + A_1, \\
  B &=& B_0  i_{fib2} + B_1,
\end{eqnarray}
where the coefficients $A_0=4.41$, $A_1=-0.17$, $B_0=-1.36e-3$ and
$B_1=6.65e-5$ were fitted so as to give a flat relation between galaxy
density and $n_s$.  For $i_{fib2} < 20.45$, $A$ and $B$ were fixed at
the $i_fib2=20.45$ values. These weights were applied to each galaxy
individually, according to the stellar density of the patch of sky in
which it lies, and to its observed $i_{fib2}$. The stellar density map
was computed using a HEALPix \citep{healpix} grid with Nside = 256,
which splits the sky into equal area pixels of 0.0525 deg$^{2}.$ This
pixel size is much smaller than the scale at which the systematic
effect of stars becomes important ($\theta > 1^{\rm o}$), but large
enough that the mean number of stars in a pixel is greater than 80
(implying any shot-noise effects will be small). 
As this is a large scale effect, a relatively coarse mask is sufficient.

\subsection{Final weights and effective volume} 
\label{sec:final_weights}

As described in the previous sections, galaxies are weighted to allow
for close-pair corrections with $w_{\rm cp}$, redshift failures with
$w_{\rm rf}$ and angular systematic weights with $w_{\rm sys}$.  We
also apply weights to optimise our clustering measurements in the face
of shot-noise and cosmic variance \citep{FKP94}, 
\begin{equation}
  w_{\rm FKP} = \frac{1}{1+\bar{n}(z_i) P_{0}} \,,
\end{equation}
where $\bar{n}(z_i)$ is the mean density at redshift $z_i$ and $P_0 =
20\,000 h^{-3} {\rm Mpc}^3$. This ignores the scale dependence of the
power spectrum and chooses a value optimised for the BAO feature, $P_0
\sim P(k=0.1 h {\rm Mpc}^{-1})$.  We make this simplification for
convenience; using the full scale-dependent weights proposed by
\citet{FKP94} does not change our results and errors.  \cite{Ross12}
find this weighting reduces the variance on the CMASS DR9 mock galaxy
sample $\xi(s)$, typically by 20 per cent relative to no
weighting. Combining $w_{\rm FKP}$ with the systematic weights $w_{\rm
  sys}$, the redshift-failure weights $w_{\rm rf}$, and the close-pair
weights $w_{\rm cp}$, the final weights applied to the galaxies are
given by
\begin{equation}
  w_{\rm tot} = w_{\rm FKP} w_{\rm sys} (w_{\rm rf} + w_{\rm cp} -1).
\end{equation}
Here $w_{\rm FKP}$ and $w_{\rm sys}$ are mutliplicative weights
depending on spatial location, while $w_{\rm rf}$ and $w_{\rm cp}$ are
additive weights, with default of unity. Using these weights, we
calculate the effective volume using our fiducial cosmology as
\begin{equation}
  \Veff = \sum\limits_{i} \left(\frac{\bar{n}(z_i) P_0}{1+\bar{n}(z_i) P_{0}}\right)^2 \Delta V(z_i) \,,
\end{equation}
where $\Delta V(z_i)$ is the volume of the shell at $z_i$ (accounting for
the observational area).  We find $\Veff = 2.2\,{\rm Gpc}^3$ for our CMASS sample
which covers the redshift range $0.43<z<0.7$. The weighted mean redshift
of galaxies within the sample is $z_{\rm eff}=0.57$, which we define as
the effective redshift of our clustering measurements.

    \subsection{Random catalog generation}
    \label{sec:random}
    The evaluation of the correlation function and of the power spectrum
requires an estimate of the average galaxy density. To provide such an
estimate, we generate random catalogs of unclustered objects with the
detailed redshift and angular selection functions of the sample,
accounting for the complex survey geometry. In particular, the random
catalogs account for the differences between the Northern and Southern
Galactic Caps. To minimise the shot-noise induced on clustering
measurements, these catalogues have $70$ times more objects than the
corresponding galaxy catalogues.  Numerous tests have confirmed that
the survey selection function can be factorised into angular and
redshift pieces \citep{Ross12}. The redshift selection function can be
taken into account by distributing the objects of the random catalogue
according to the observed redshift distribution of the sample. We use
the ``shuffled'' catalogue as defined in \citet{Ross12}, where the
redshifts are matched to randomly selected galaxies. We do this
separately for the NGC and SGC samples (see
Appendix~\ref{sec:northsouth} for a further discussion of this). The
completeness on the sky is determined from the fraction of target
galaxies in a sector for which we obtained a spectrum, with the
sectors being areas of the sky covered by a unique set of
spectroscopic tiles (see Section~\ref{sec:masks}). We upweight close
pairs and redshift failures in the galaxy catalog as described in
Section~\ref{sec:zfail_corrections}, and therefore include these
targets when calculating the completeness for the random catalog. Thus
the random catalogue was subsampled to the sector completeness as
given by Eq.~\ref{eq:comp}.

\section{Analysis}
\label{sec:analysis}
We analyse the BAO feature and fit for a distance to $z=0.57$ using
both the correlation function (Section~\ref{sec:xi}) and power
spectrum (Section~\ref{sec:pk}) of the 3D galaxy distribution. The
steps in both analyses parallel one another: (i) density-field
reconstruction of the BAO feature, (ii) computation of the two-point
statistics, (iii) estimation of errors on these measurements by
analysing mock catalogs and (iv) extraction of a distance measurement
by fitting the data. This section details these steps. Details
specific to each method as well as the results are discussed in later
sections.

    \subsection{Reconstruction}
    \label{sec:reconstruction}
    As described in Section~\ref{sec:intro}, the statistical sensitivity
of the BAO measurement is limited by non-linear structure
formation. Following \citet{Eis07a} we apply a procedure to {\it
  reconstruct} the linear density field. We emphasise that this
improvement is not a deconvolution of the correlation function, but
uses information encoded in the full density field. In addition to
undoing the smoothing of the BAO feature, reconstruction also removes
the expected bias ($< 0.5$ per cent) in the BAO distance scale that
arises from the same second-order effects that smooth the BAO feature,
which simplifies analyses. Reconstruction has recently been applied to
the SDSS-II DR7 LRG sample at $z=0.35$ \citep{Pad12}, and our
implementation is very similar; we refer the reader there for details
and simply summarise the steps here:
\begin{enumerate}
\item Smooth the observed density field to suppress the effects of
  shot-noise and highly non-linear features. We use a Gaussian of width
  $l=15\mpcoh$, but demonstrate that our results are insensitive to
  this particular choice (see Appendix \ref{sec:robust_recon}).
\item Embed the observed density field into a larger volume with a
  constrained Gaussian realisation. The correlation function of the 
  constrained realisation is chosen to match the observed unreconstructed
  correlations, but we find that our results are insensitive to the 
  details of this choice.
\item Estimate the displacements $\mathbf{q}$ from the galaxy density field
  $\delta_{\rm gal}$ using the continuity equation $\nabla \cdot \mathbf{q} =
  -\delta_{\rm gal}/b_{\rm gal}$ where $b_{\rm gal}$ is the galaxy bias. 
  In detail, the above continuity equation is modified to account for linear redshift
  space distortions although we find our results are insensitive to the 
  details of this prescription (see \citealt{Pad12} for the modified 
  continuity equation). The galaxy bias $b_{\rm gal}$ is set to a value
  estimated from the unreconstructed correlation function; Appendix~\ref{sec:robust_recon}
  demonstrates that our results are insensitive to errors in this choice.
\item Shift the galaxies by $-\mathbf{q}$. Shift the galaxies by an
  additional $-f q_{s} \mathbf{\hat{s}}$ where $\mathbf{\hat{s}}$ is
  the redshift direction, and $f$ is the logarithmic derivative of the
  linear growth rate with respect to the scale factor. This latter
  shift corrects for linear redshift space distortions. We denote this
  density field by $D$.
\item Generate a sample of points, randomly distributed according to
  the selection function of the survey. Shift these points by
  $-\mathbf{q}$. Note that we do not correct the random points for
  redshift space distortions. We denote this density field by $S$.
\item The reconstructed density field is defined by the difference
  between the density fields defined by $D$ and $S$.
\end{enumerate}

Given the large separation between the data in the Northern and Southern 
Galactic Caps, we run reconstruction on these independently.

    \subsection{Covariances}
    \label{sec:covar}
    We estimate the sample covariance matrix for the spherically averaged
correlation function and for the spherically averaged power spectrum
from the distribution of values recovered from 600 galaxy mock
catalogs. The galaxy mock catalogs are detailed in \citet{Man12}, and
were generated using a method similar to PTHalos \citep{ScoShe02},
which was calibrated using a suite of N-body simulations from
LasDamas\footnote{{\tt http://lss.phy.vanderbilt.edu/lasdamas}}
(McBride et al., in preparation), and we were able to recover the
clustering of halos at $\sim 10$ per cent accuracy. A $10$ per cent
shift in the amplitude of the covariance matrix corresponds to a $3$
per cent shift in the error on the measured amplitude, and we would
expect shifts of a similar order in other measured parameters. We
consider that knowing our errors to this level of accuracy is adequate
for our analyses.

The method can be summarised as follows: we first generate 600 matter
fields at $z=0.55$ using second order perturbation theory (2LPT) for
our fiducial cosmology.  We choose a fixed redshift for simplicity,
which assumes that the evolution over the redshift range is small.
Modelling the evolution in the mocks would require a significantly more
complex approach, as both the halo mass and the method of populating
galaxies would both evolve with redshift.  We chose to model $z=0.55$
since this corresponds to the median (unweighted) redshift of the
observed galaxy sample, and the galaxy mocks were created by matching
smaller scale clustering without weighting (see
Section~\ref{sec:sysweights}).  Since we impose the same sampling as
the data, the effective weighted redshift of the mocks is still
$z_{eff} = 0.57$.  We expect the difference in the covariance matrix
between $z=0.55$ and $z=0.57$ to be small and not a significant
contribution to the errors of our current results.

We identify dark matter halos using particles from the mass field, but
we must calibrate the halo mass from these 2LPT halos
\citep[see][]{Man12}. We populate the halos with galaxies using a halo
occupation distribution (HOD) of the form described in \citet{ZCZ07},
with the exact parameter values determined to reproduce $\xi(r)$ from
the CMASS DR9 data on scales of $30-80\mpcoh$. Although the 2LPT field
does not include strong non-linear corrections (important on small
scales), our method produces mocks that match the clustering of CMASS
galaxies well into the quasi-linear regime when compared with mocks
from N-body simulations (which do contain the full non-linear
evolution).

The galaxy mock catalogs are initially constructed in $2400\mpcoh$
boxes and then reshaped to fit the DR9 geometry.  The mock galaxies
include redshift-space distortions, follow the observed sky
completeness, and are down-sampled to correspond to the radial number
density of the observed data.  This was done separately for the NGC
and SGC to model the sampling in the data (see
Appendix~\ref{sec:northsouth}).  These mock catalogs have also been
used in several related studies, such as analysing CMASS DR9
systematics \citep{Ross12}, analysis of the clustering of galaxies
through simulations \citep{Nuza12}, cosmology constraints on redshift
space distortions (RSD, \citealt{Rei12}), and implications of RSD for
non-standard cosmologies \citep{Sam12}. They were also used when
comparing evolution between the CMASS and SDSS-II Luminous Red Galaxy
samples \citep{Tojeiro12}, and making growth measurements from this
comparison \citep{Tojeiro12b}. For this paper, the measurements from
the NGC and SGC mocks were combined to provide a full measurement for
each of 600 mock catalogs, just as is done in the observational
sample. The galaxy mock catalogs and the derived covariance matrices,
which we refer to as sample covariance matrices, will be made publicly
available\footnote{\texttt{http://marcmanera.net/mocks/}}.

The true covariance matrix depends on cosmological parameters as well as 
the treatment of galaxy bias, both of which we neglect.  However, we 
expect this dependence to be relatively weak in the parameter range 
allowed by our data.  For example, \citet{Labatie12} study the effect of 
cosmology dependence in the covariance matrix for the galaxy correlation 
function for determining the acoustic scale; they find that the best-fit 
value undergoes a small shift of $0.3 \sigma$ with a negligible change 
in the error bar.  The effect for the CMASS DR9 data should be even smaller 
as it covers a larger volume and has more constraining power equating to 
less variation in parameters than the survey assumed in \citet{Labatie12}.  
We therefore consider it reasonable to assume no cosmological dependence 
in the covariance matrix, which we calculate from galaxy mock catalogs 
based on a fixed, fiducial cosmology.

Finally, we explicitly test the sample covariance matrix 
using two alternate methods to estimate covariances:
\begin{itemize}
\item a smooth Gaussian model covariance where parameters are fit to
the galaxy mock catalogs \citep{Xeaip}, and 
\item an analytic estimate
generated from the monopole power spectrum \citep{deputteretal12}.
\end{itemize}
Both of these methods have the advantage of being smooth estimates of
the covariance, which eliminates noise that may complicate the use of
the estimated sample covariance matrix. We find that all three methods
give consistent results within our quoted errors, where we tested by
re-fitting the BAO using these three estimates of the covariance
matrix. Further comparison of methods is provided in
\citet{Man12}. For the rest of this paper, we use the sample
covariance matrices derived directly from the mocks to determine
errors.

Although we have 600 independent mocks for each of the NGC and SGC
samples, these were drawn from the same 600 density fields, and so are
not fully independent when combined. To ensure that this is handled
correctly, we form our covariance matrix for the combined sample by
averaging two covariance matrices, each calculated from 300
independent joint NGC+SGC samples.

    \subsection{Fitting a Distance}
    \label{sec:analysis_fit}
    The distance redshift relation for a (thin) redshift slice may be
characterised by the distance to the mean redshift of the sample and
its derivative. Getting the former wrong dilates all distances in the
survey and shifts the BAO feature in the angle averaged correlation
function, but it retains the underlying isotropy of the clustering
signal. Getting the latter wrong distorts the isotropy of the survey
and induces higher-order moments in the angle dependence of the
clustering. Since this work is limited to the angle-averaged
clustering measurements, we are sensitive primarily to the
former. These measurements constrain the distance combination
\begin{equation}
  D_V \equiv \left[ cz (1+z)^2 D_A^2 H^{-1} \right]^{1/3},
\end{equation}
where $D_A$ is the angular diameter distance and $H$ is the Hubble
parameter, and we assume that the survey extents are much larger than
the scales of interest. We also assume that the enhanced clustering
amplitude along the line-of-sight due to RSDs does not alter the relative
importance of radial and angular modes when calculating spherically
averaged statistics (as for the SDSS-II LRGs, \citealt{Per10}).

We convert our measurements of the power spectrum and correlation
function into a distance measurement by fitting the acoustic feature
to an appropriately dilated template while simultaneously fitting for
a ``nuisance'' broad-band shape. This procedure has formed the basis
of a number of previous analyses
(e.g. \citealt{Eis05,Kaz10,Per10,Beutler11,Bla11a,Bla11b} and
\citealt{Pad12}) and has been explicitly tested to be unbiased even in
the presence of incorrect assumptions about the underlying cosmology
and galaxy bias \citep{Xeaip}. The power spectrum and correlation
function models are dilated according to $P_{t}(k/\alpha)$ and
$\xi_{t}(\alpha r)$ respectively, where $P_t$ and $\xi_t$ are the
template functions for a fiducial cosmology. In order to account for
the effects of non-linear evolution on the BAO feature, the BAO in
these templates is artificially smeared out according to the
prescription given in \citet{EisSeoWhi07}. Any deviation of the true
distance-redshift relation from the fiducial choice is encoded in
$\alpha$ and can be related to the distance to the weighted-mean
redshift of the sample by
\begin{equation}
  D_V/r_s = \alpha \left(D_V/r_s \right)_{\rm fid}.
\label{eqn:alpha_def}
\end{equation}
This distance is relative to the sound horizon $r_s$, for which we
adopt the definition in Eqs. 4 through 6 of \citet{Eis98}. As was
discussed in \citet{Meh12}, the sound horizon in the above is merely a
proxy for the distance information encoded in the BAO features in the
correlation function and power spectrum and is therefore insensitive
to the choice of definition as long as it is consistently used when
estimating cosmological parameters. The effective redshift of the
CMASS DR9 sample is $z=0.57$ (see Section~\ref{sec:sysweights}), and
we assume a fiducial cosmology as described in
Section~\ref{sec:intro}. This yields fiducial values $D_V(z=0.57) =
2026.49\,{\rm Mpc}$, $r_s = 153.19\,{\rm Mpc}$ and $(D_V/r_s)_{\rm fid}
= 13.23$.

\section{The Correlation Function}
\label{sec:xi}

    \subsection{Measuring the correlation function}
    \label{sec:analysis_xi}
    
\begin{figure*}
  \centering
  \resizebox{0.95\columnwidth}{!}{\includegraphics{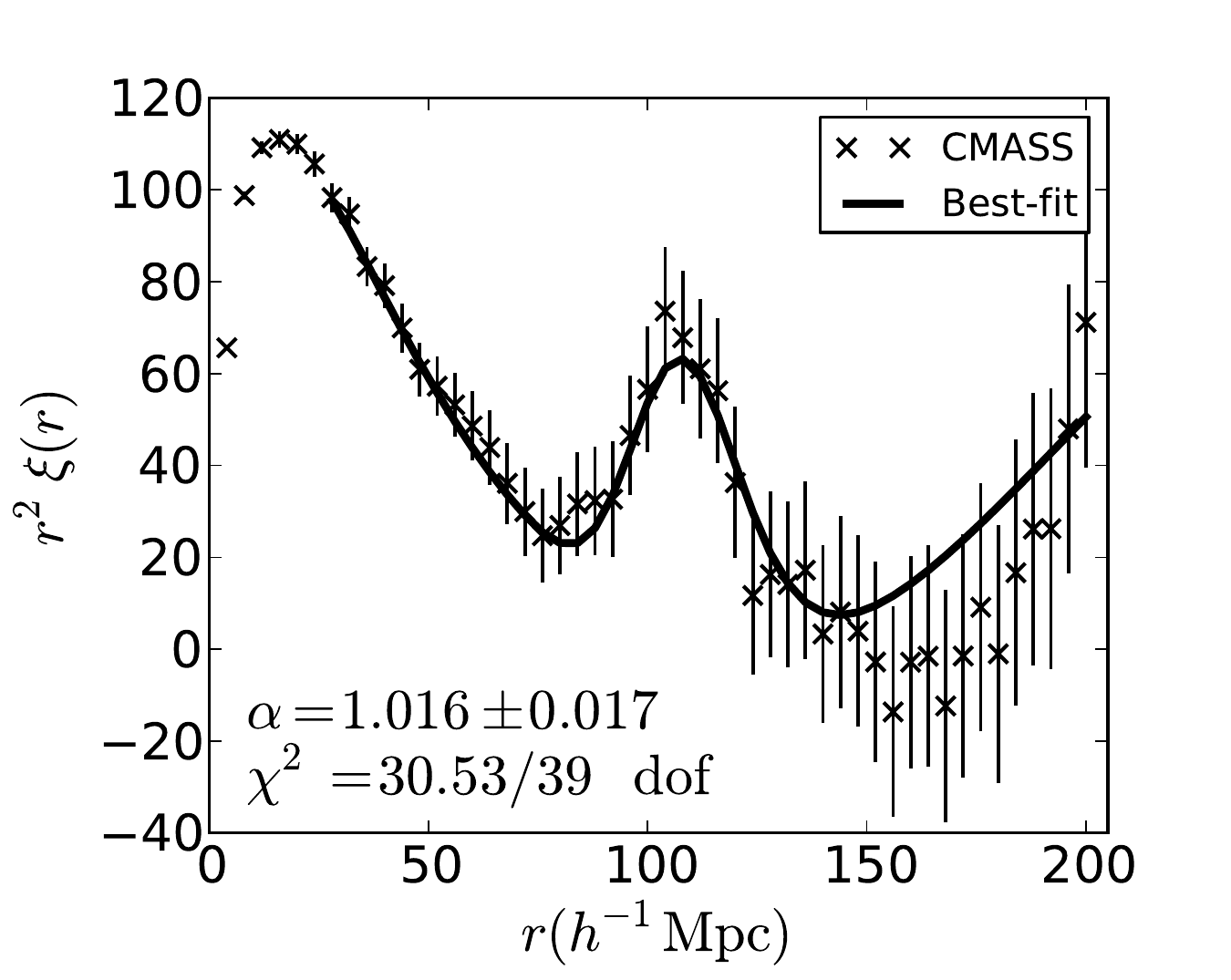}}
  \resizebox{0.95\columnwidth}{!}{\includegraphics{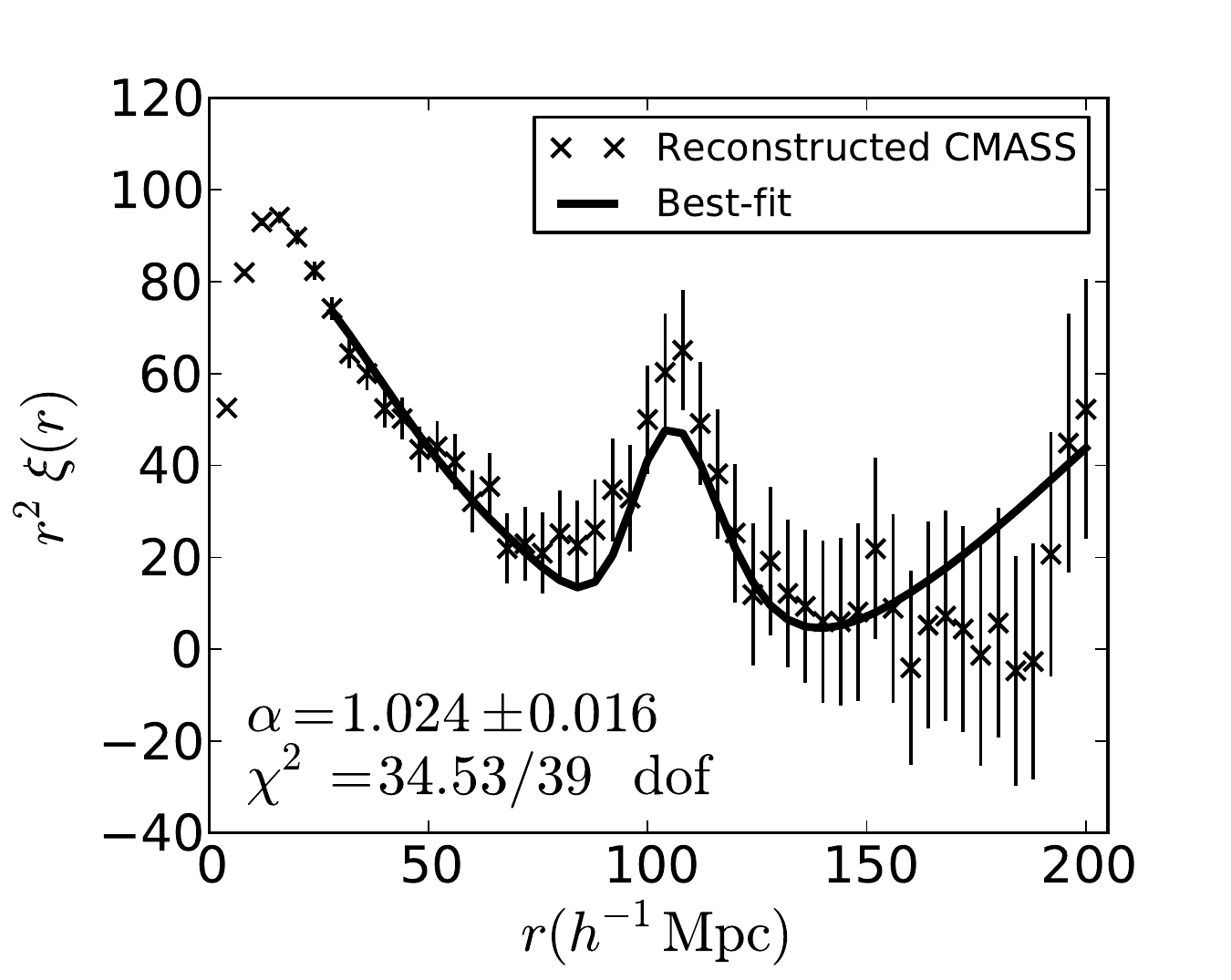}}
  \caption{ The CMASS correlation function before (left) and after
    (right) reconstruction (crosses) with the best-fit models
    overplotted (solid lines). Error bars show the square root of the
    diagonal covariance matrix elements, and data on similar scales
    are also correlated. The BAO feature is clearly evident, and well
    matched to the best-fit model. The best-fit dilation scale is
    given in each plot, with the $\chi^2$ statistic giving goodness of
    fit. }
  \label{fig:DR9xi}
\end{figure*}

\begin{figure}
  \centering
  \resizebox{0.95\columnwidth}{!}{\includegraphics{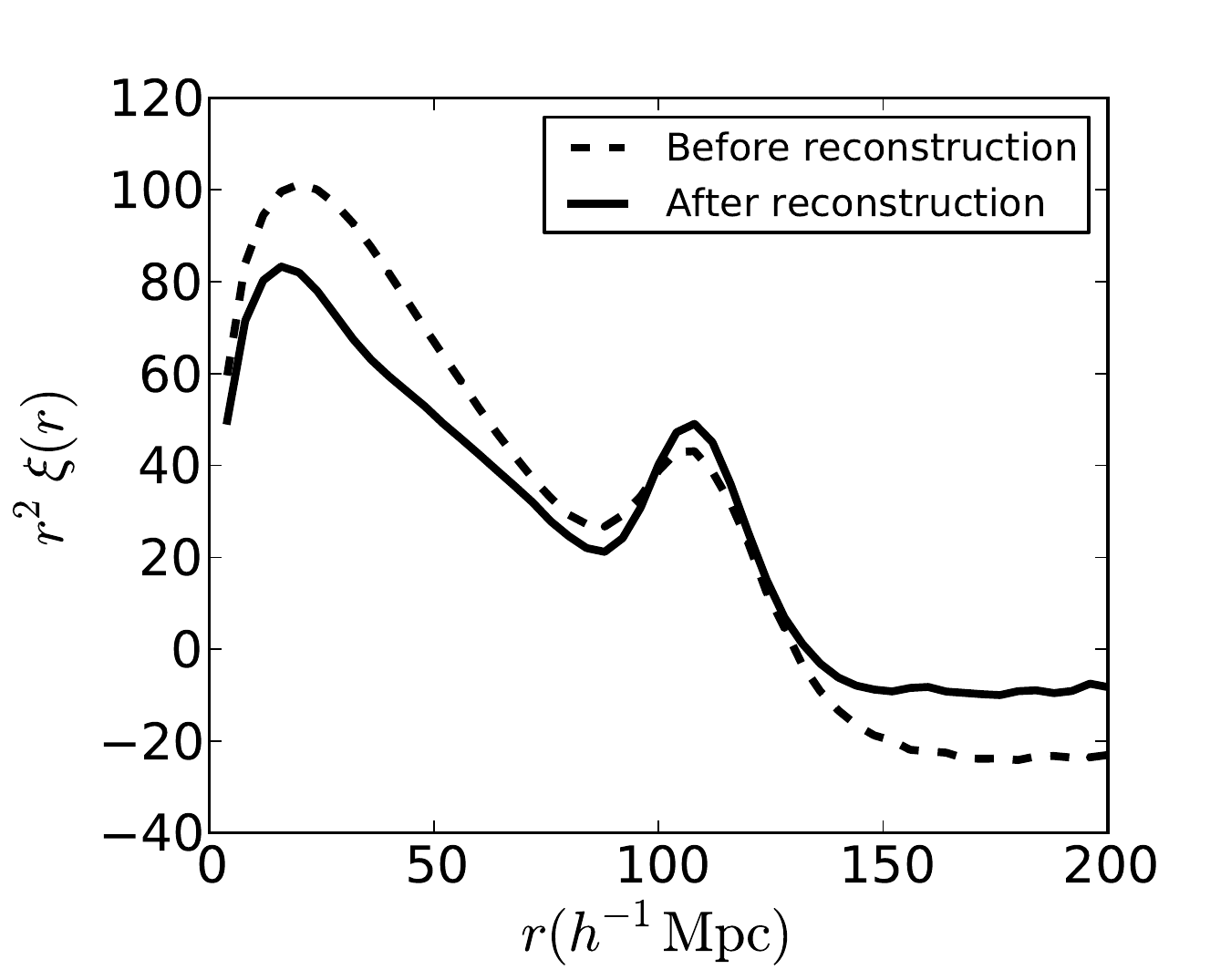}}
  \caption{Average of the mock correlation functions before and after
    reconstruction showing that the average acoustic peak sharpens up
    significantly after reconstruction. This indicates that, on
    average, our reconstruction technique effectively removes some of
    the smearing caused by non-linear structure growth, affording us
    the ability to more precisely centroid the acoustic peak.}
  \label{fig:mock_avg}
\end{figure}

Fig.~\ref{fig:DR9xi} plots the CMASS correlation functions before and after reconstruction, 
with the best-fit model (see below) overplotted. 
We estimate $\xi(r)$ using the \citet{LanSza93} estimator
\begin{equation}
  \xi(r) = \frac{DD-2DR+RR}{RR}
\end{equation}
where $DD$, $DR$ and $RR$ are suitably normalised numbers of
(weighted) data-data, data-random and random-random pairs. For the
case of the reconstructed correlation function, the $DR$ and $RR$ pair
counts in the numerator are replaced by $DS$ and $SS$, where $S$
represents the shifted random particles. The errors are estimated by
applying the same procedure to the mock catalogs and constructing the
sample covariance matrix from the 600 realisations of $\xi(r)$.  The
average correlation function from the 600 mock catalogs is presented
in Fig.~\ref{fig:mock_avg}.  The errors in Fig.~\ref{fig:DR9xi} are
from the diagonal of the covariance matrix. We caution the reader that
these errors are highly covariant, and assessing the significance
requires analysing the full covariance matrix.


    \subsection{Fitting the correlation function}
    \label{sec:fitxi}
    Our correlation function fits are based on the procedure described in
\citet{Xeaip}. We give a brief summary of the techniques here.

Our correlation function model is given by
\begin{equation}
\xi^{\rm fit}(r) = B^2\xi_m(\alpha r)+A(r)
\label{eqn:fform}
\end{equation}
where
\begin{equation} 
\xi_m(r) = \int \frac{k^2dk}{2\pi^2}P_m(k)j_0(kr) e^{-k^2a^2},
\label{eqn:xu_xim}
\end{equation}
and
\begin{equation}
A(r) = \frac{a_1}{r^2} + \frac{a_2}{r} + a_3.
\label{eqn:aform}
\end{equation}
In Eq.~\ref{eqn:xu_xim}, the Gaussian term has been introduced to damp
the oscillatory transform kernel $j_0(kr)$ at high-$k$ to induce
better numerical convergence. The exact damping scale used in this
term is not important, and we set $a=1\mpcoh$, which is significantly
below the scales of interest. The $A(r)$ term is composed of nuisance
parameters $a_{1,2,3}$ that help marginalize over the unmodeled
broadband signal in the correlation function. Such broadband effects
include redshift-space distortions, scale-dependent bias and any
errors made in our assumption of the model cosmology. These effects
may bias our measurement of the acoustic scale if not removed. $B$ is
a multiplicative constant, allowing for an unknown large-scale
bias. We use a template $P_m(k)$ of the form
\begin{equation}
P_m(k) = [P_{\rm lin}(k)-P_{\rm no bao}(k)] e^{-k^2\Sigma_{\rm nl}^2/2}+P_{\rm no bao}(k),
\label{eqn:template}
\end{equation}
as given in \citet{EisSeoWhi07}. Here, $P_{\rm lin}(k)$ is the linear theory
power spectrum and $P_{\rm no bao}(k)$ is the power spectrum with the
BAO feature erased. The $\Sigma_{\rm nl}$ term is used to damp the
acoustic oscillations in the linear theory power spectrum, serving to
model the effects of non-linear structure growth. We fix $\Sigma_{\rm nl}
= 8\mpcoh$ in our fits to the pre-reconstruction correlation functions
and $\Sigma_{\rm nl} = 4\mpcoh$ in our fits to the post-reconstruction
correlation functions. We normalise the template to the observed or
mock correlation function being fit at $r=50\mpcoh$, thereby ensuring
that $B^2\sim1$. These parameters were tuned on our mock catalogs, 
and we explicitly verify that are results are insensitive to 
these particular choices in Appendix~\ref{sec:robust_fit}.

The scale dilation parameter $\alpha$ defined in
Eq. \ref{eqn:alpha_def} captures our distance constraints; $\alpha$
measures the relative position of the acoustic peak in the data versus
the model, thereby characterising any observed shift. If $\alpha>1$,
the acoustic peak is shifted towards smaller scales, and vice versa for
$\alpha<1$.

We obtain the best-fit value of $\alpha$ by computing the $\chi^2$
goodness-of-fit indicator at intervals of $\Delta\alpha=0.001$ in the
range $0.8<\alpha<1.2$, then identify the value of $\alpha$ that gives
the minimum $\chi^2$ and take this as our best-fit value. The $\chi^2$
as a function of $\alpha$ is given by
\begin{equation}
\chi^2(\alpha) = [\vec{d}-\vec{m}(\alpha)]^TC^{-1}[\vec{d}-\vec{m}(\alpha)],
\end{equation}
where $\vec{d}$ is the measured correlation function and
$\vec{m}(\alpha)$ is the best-fit model at each $\alpha$. $C$ is the
sample covariance matrix, and we use a fitting range of
$28<r<200\mpcoh$. We therefore fit over 44 points using 5 parameters,
leaving us with 39 degrees-of-freedom (dof). Assuming a multi-variate
Gaussian distribution for the fitted data (this is tested and shown to
be a good approximation in \citealt{Man12}), the probability
distribution of $\alpha$ is 
\begin{equation}
  p(\alpha)  \propto e^{-\chi^2(\alpha)/2}.
\end{equation}
The normalisation constant is determined by ensuring that the
distribution integrates to 1. In calculating $p(\alpha)$, we also
impose a 15 per cent Gaussian prior on $\log(\alpha)$ to suppress
values of $\alpha \ll 1$ that correspond to the BAO being shifted to
the edge of our fitting range at large scales. The sample variance is
larger at these scales, and the fitting algorithm is afforded some
flexibility to hide the acoustic peak within the larger errors.

The standard deviation of this probability distribution serves as an
error estimate on our distance measurement. The standard deviation
$\sigma_{\alpha}$ for the data and each individual mock catalog can be
calculated as $\sigma_{\alpha}^2 = \langle \alpha^2 \rangle - \langle
\alpha \rangle^2$, where the moments of $\alpha$ are
\begin{equation}
  \langle \alpha^{n} \rangle = \int d\alpha\,p(\alpha) \alpha^n \,\,.
\end{equation}
Note that $\langle\alpha\rangle$ refers to the mean of the $p(\alpha)$
distribution in this equation only.

In reference to the mocks, $\langle\alpha\rangle$ will denote the
ensemble mean of the $\alpha$ values measured from each individual
mock, and $\tilde{\alpha}$ will denote the median. The term
``Quantiles'' will denote the 16$^{\rm th}$/84$^{\rm th}$ percentiles,
which are approximately the 1$\sigma$ level if the distribution is
Gaussian. The scatter predicted by these quantiles suffers less than
the rms from the effects of extreme outliers.

    \subsection{Results}
    \label{sec:results_xi}
    
\begin{figure}
  \centering
  \resizebox{0.95\columnwidth}{!}{\includegraphics{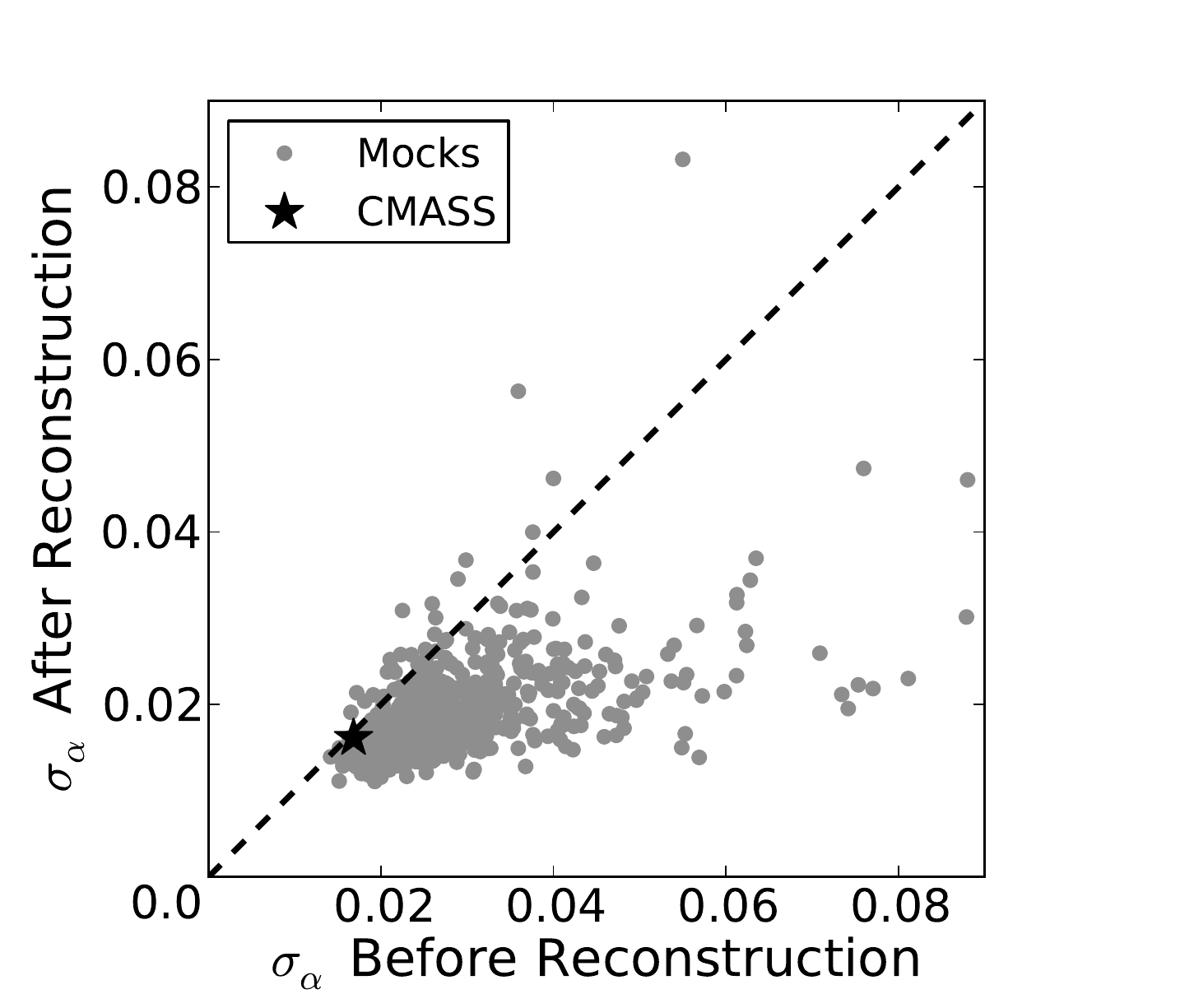}}
  \caption{ Comparisons of $\sigma_\alpha$ errors in mock catalogs before and
    after reconstruction as measured from $\xi(r)$. Reconstruction
    tends to improve our ability to measure $\alpha$; on a mock-by-mock
    basis, the average amount of improvement in $\sigma_\alpha$ is a
    factor of 1.54. However, the amount of improvement varies, and 26
    (out of 600) of the mocks actually see $\sigma_\alpha$ increase
    from pre-reconstruction to post-reconstruction. The CMASS DR9
    point is overplotted as the black star and falls within the locus
    of the mock points. 44 (out of 600) of the mocks have a ratio of
    $\sigma_\alpha$ after reconstruction compared to before
    reconstruction that is greater than the CMASS DR9 value. Hence,
    the fact that the error on $\alpha$ measured from CMASS DR9 does
    not decrease significantly after reconstruction is not unexpected
    in the context of the mocks. One can also see that most of the
    extreme outliers in $\sigma_\alpha$ before reconstruction have
    significantly smaller errors after reconstruction.  }
  \label{fig:comprec}
\end{figure}

Using the procedure described in \S\ref{sec:fitxi}, we measure the
shift in the acoustic scale from the CMASS DR9 data to be $\alpha =
1.016 \pm 0.017$ before reconstruction and $\alpha = 1.024 \pm 0.016$
after reconstruction. The quoted errors are the $\sigma_\alpha$ values
measured from the probability distributions, $p(\alpha)$. Plots of the
data and corresponding best-fit models are shown in
Fig. \ref{fig:DR9xi} for before (left) and after (right)
reconstruction. We see that for CMASS DR9, reconstruction has not
significantly improved our measurement of the acoustic scale. However,
in the context of the mock catalogues, this result is not surprising.

Fig. \ref{fig:comprec} shows the $\sigma_\alpha$ values measured from
the mocks before reconstruction versus those measured after
reconstruction from the correlation function fits. The CMASS DR9 point
is overplotted as the black star and falls within the locus of mock
points. However, we see that before reconstruction, our recovered
$\sigma_\alpha$ for CMASS DR9 is much smaller than the mean expected
from the mocks. 
For typical cases, reconstruction improves errors on $\alpha$, but if one
has a ``lucky'' realisation that yields a low error to begin with, then 
reconstruction does not produce much improvement. The mock catalog
comparison in Figure~\ref{fig:comprec} shows that the BOSS DR9
data volume is just such a ``lucky'' realisation, with a strong and
well defined acoustic peak, and it is therefore unsurprising that 
reconstruction does not reduce the error on $\alpha$.

\begin{figure*}
  \centering
  \resizebox{0.95\columnwidth}{!}{\includegraphics{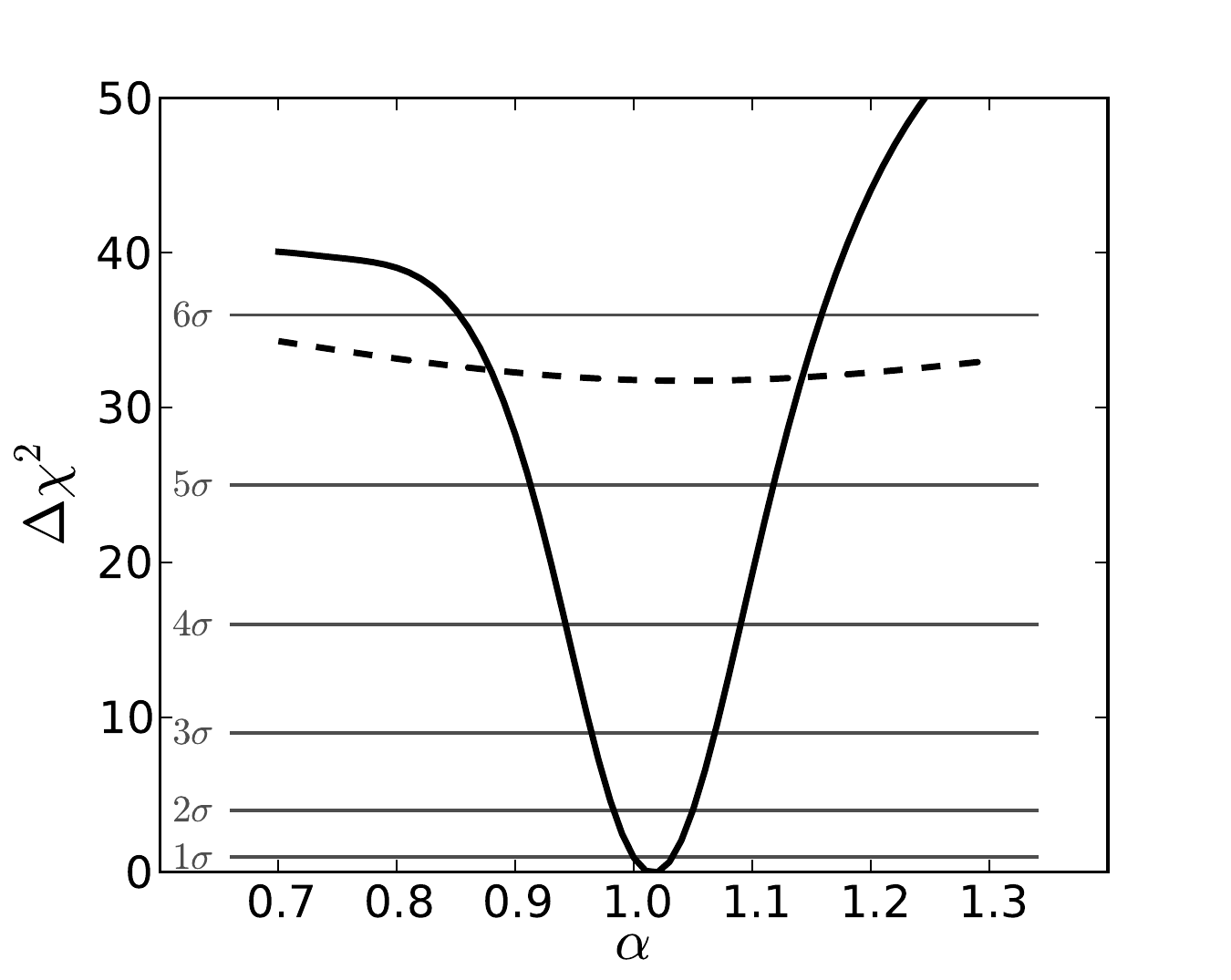}}
  \resizebox{0.95\columnwidth}{!}{\includegraphics{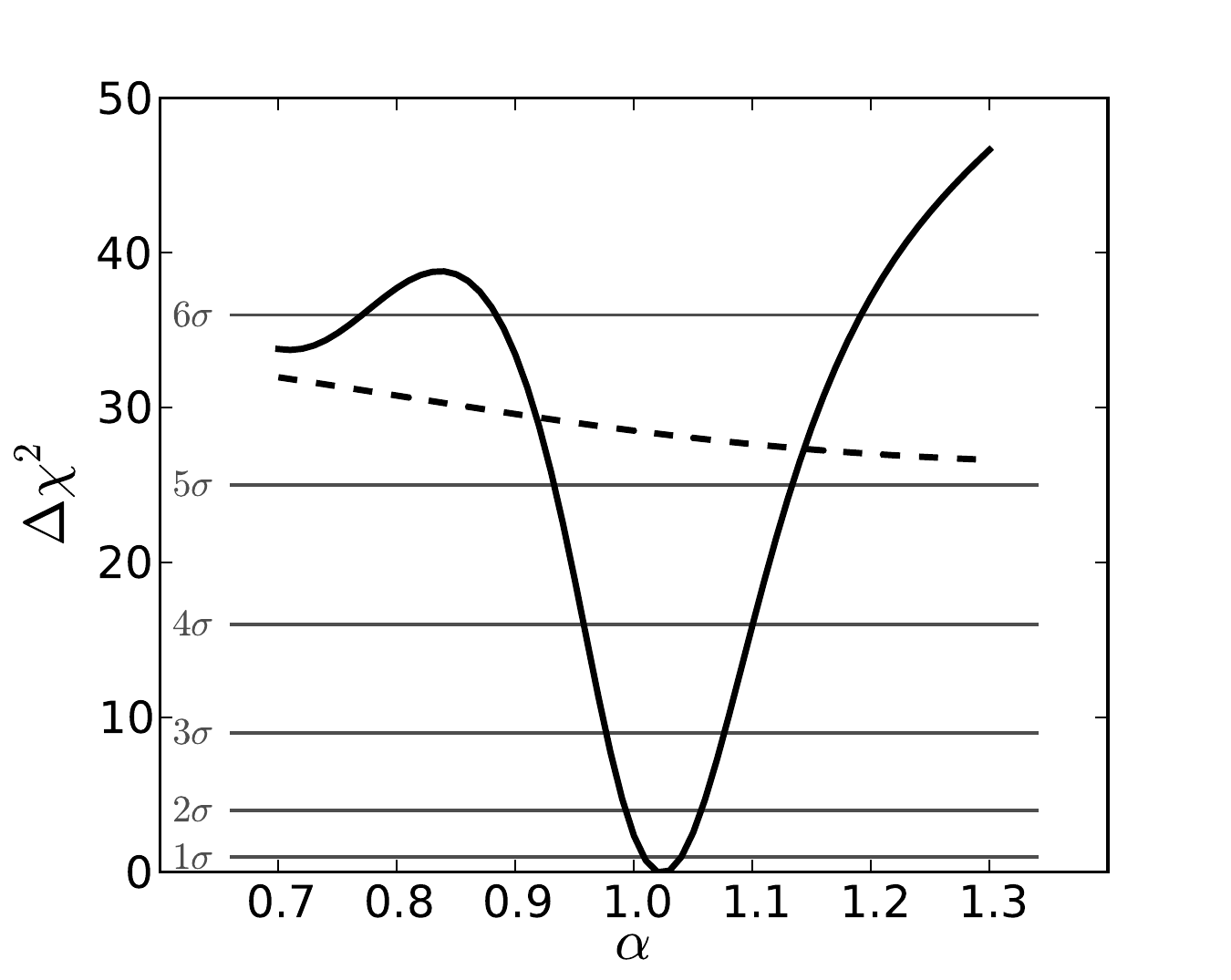}}
  \caption{ Significance of the CMASS DR9 BAO feature before (left)
    and after (right) reconstruction as measured from $\xi(r)$. The
    dashed lines correspond to fits to the data using a model without
    BAO. The quantity $\Delta\chi^2 = \chi^2 - \chi^2_{min}$, where
    $\chi^2_{min}$ corresponds to the minimum $\chi^2$ where the
    best-fit value of $\alpha$ lies. By comparing the minimum of the
    solid curve with the dashed curve, we can quantify how confident
    we are that the BAO can be measured from the CMASS DR9
    sample. Similarly, comparing the minimum and the plateau of the
    solid curve tells us how confident we are that we have measured
    the correct local minima for the acoustic scale. One can see that
    both before and after reconstruction, we detect the BAO at greater
    than $5\sigma$ confidence and the global minimum is itself found
    within a valley that is $6\sigma$ deep.  }
  \label{fig:DR7_9chi2}
\end{figure*}

The BAO detection in the CMASS DR9 data is highly significant as
illustrated in Fig. \ref{fig:DR7_9chi2}. Here we have plotted
$\Delta\chi^2 = \chi^2 - \chi^2_{min}$, where $\chi^2_{min}$ is the
minimum $\chi^2$ that corresponds to the best-fit value of
$\alpha$. The dashed line overplotted corresponds to fits to the data
using a model without a BAO signature. This figure captures two tests
of BAO significance: the first requires a comparison between the solid
and dashed curves, and indicates how confident we are that the BAO
feature exists in the CMASS DR9 data. The second uses the plateau
height of the $\Delta\chi^2$ curve to indicate how confident we are
that we have measured an acoustic feature.

The panel on the left corresponds to our pre-reconstruction results
and the panel on the right corresponds to our post-reconstruction
results. Before reconstruction, the minimum of the solid curve lies
beyond a $\Delta\chi^2$ of 25 from the dashed curve, indicating that
the BAO is detected in CMASS DR9 at greater than $5\sigma$
confidence. Local maxima are seen at greater than $\Delta\chi^2$ of 36
above the minimum, indicating that the data prefer our best-fit value
of $\alpha$ at more than $6\sigma$. We see similar confidence levels
post-reconstruction.

To verify the robustness of our techniques, we also measure the best-fit
$\alpha$ and $\sigma_\alpha$ for our 600 mock catalogues using our
fiducial fitting and reconstruction parameters. We then repeat the same
fitting with slightly altered models as well as on correlation functions
computed from catalogues that were reconstructed using different
parameters (bias, growth factor and smoothing scale). These fitting
results are discussed in more detail in Appendix \ref{sec:robust_tests}
and summarised in Table \ref{tab:mock_alphas}. The values in the
table are computed after discarding the mocks with $\sigma_\alpha >
0.07$. Before reconstruction there were 10 such instances, and after
reconstruction there was only 1 such instance. These large uncertainties
in the measured $\alpha$ indicate that the acoustic signature is weak
in these realisations and is therefore not detected with high fidelity
(see \citealt{Xeaip} for a more detailed description of this approach). We
find that regardless of fitting model parameters or reconstruction
parameters, we always recover consistent measurements of the $\alpha$
and $\sigma_\alpha$. Hence, our fiducial model should be trusted to
return reliable measurements of the acoustic scale.

Before reconstruction our mocks yield $\langle\alpha\rangle=1.004$
with an average error on any single realisation (i.e. the
rms or standard deviation) of $0.027$ and a standard error
on the mean of 0.001. The median is $\tilde{\alpha}=1.004$
with quantiles of $^{+0.026}_{-0.026}$. After reconstruction, we
obtain $\langle\alpha\rangle=1.004$ with average error on any single
realisation of $0.018$ and standard error on the mean of 0.001. The median
is $\tilde{\alpha}=1.004$ with quantiles of $^{+0.017}_{-0.018}$. One
can see that given the error on the mean, we detect a statistically
significant shift in our measured mean from the true acoustic scale
($\alpha=1$) expected in the mocks. This small systematic shift is
discussed in more detail in Section~\ref{sec:consensus}.

Most important, the average error on $\alpha$ recovered from the mocks
has decreased after reconstruction. This is illustrated in
Fig. \ref{fig:comprec}, where an overall improvement in
$\sigma_\alpha$ is evident after reconstruction. The greatest
improvements occur when the pre-reconstruction errors are the
worst. The average decrease in $\sigma_\alpha$ is a factor of 1.54,
which is equivalent to the effects of increasing the survey volume by
a factor of 2.3. Therefore, reconstruction appears to significantly
improve our ability to measure $\alpha$ precisely, on average. This
point is further illustrated in Fig.  \ref{fig:mock_avg}, where we
have plotted the average mock correlation function before and after
reconstruction. One can see the sharpening up of the acoustic peak,
indicating the effectiveness of the reconstruction algorithm in
partially removing the smearing of the BAO caused by non-linear
structure growth. This improvement is what allows a more precise
centroiding of the peak location. In fitting the average mock
correlation function before and after reconstruction, we find
$\Sigma_{nl}$, the damping of the BAO due to non-linear evolution,
decreases from $7.58\mpcoh$ to $3.23\mpcoh$.  Beyond reducing the
distance errors, reconstruction also makes our distance estimates more
robust to parameter choices in our fitting algorithms and reduces the
scatter between the distance estimates from the the correlation
function and the power spectrum. We quantify these improvements
further in following sections.

\begin{figure}
  \centering
  \resizebox{0.95\columnwidth}{!}{\includegraphics{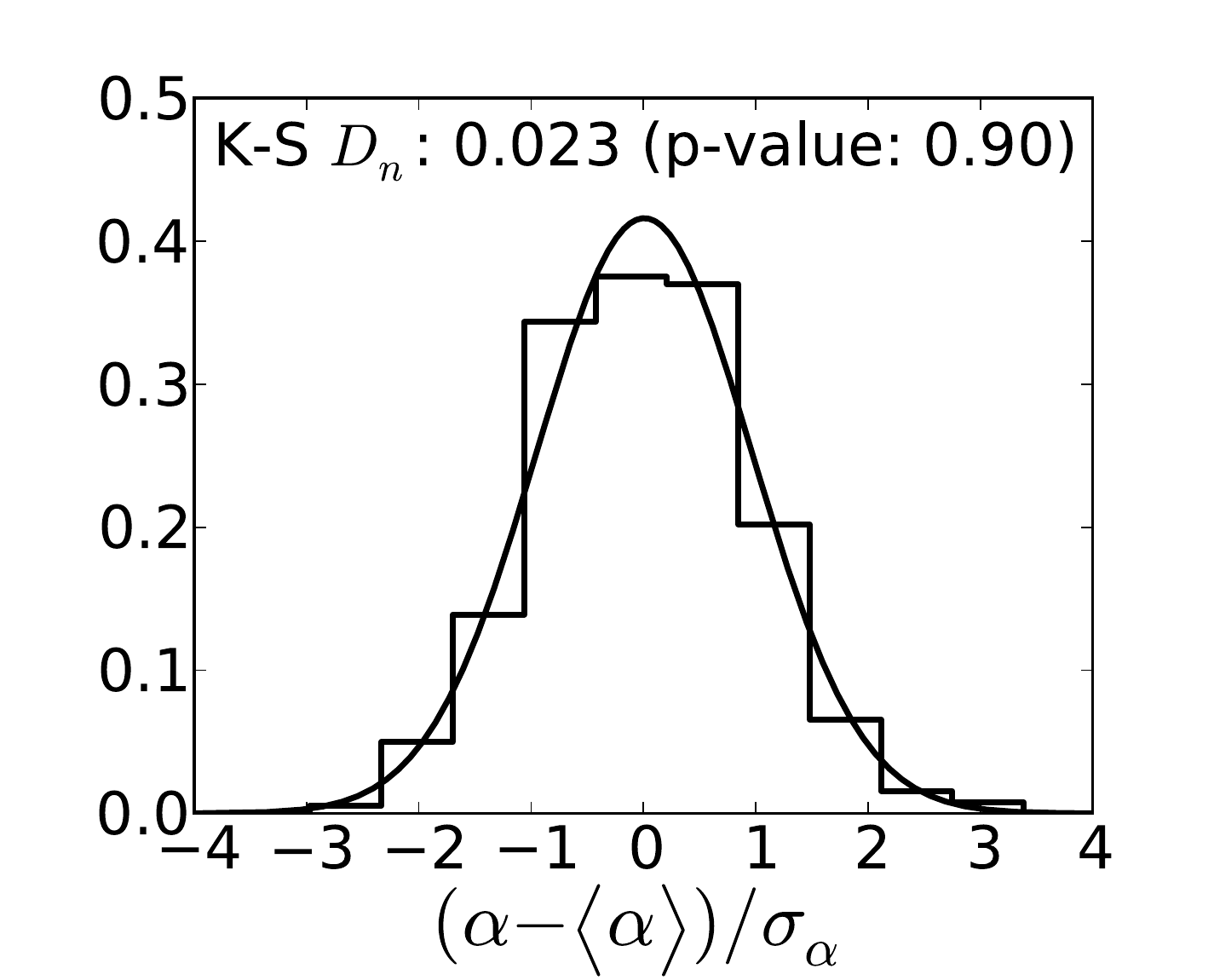}}
  \caption{ Histogram of $(\alpha -
    \langle\alpha\rangle)/\sigma_\alpha$ measured from $\xi(r)$ of the
    post-reconstruction mocks, where $\langle\alpha\rangle$ is the
    mean. This quantity is a proxy for the signal-to-noise ratio of
    our BAO measurement. We see that this distribution is close to
    Gaussian as indicated by the near-zero K-S $D_n$. The
    corresponding p-value indicates that we are 90 per cent certain
    our values are drawn from a Gaussian distribution, indicating that
    the values of $\sigma_\alpha$ we measure from the $\chi^2$
    distribution are reasonable descriptors of the error on $\alpha$
    measured by fitting $\xi(r)$.  }
  \label{fig:snr_xi_rec}
\end{figure}

We next compare the observed scatter in the best-fit $\alpha$ in the
mocks to the $\sigma_\alpha$ estimated in each fit from the
$\chi^2(\alpha)$ curve.  In Fig.~\ref{fig:snr_xi_rec}, we plot a
histogram of $(\alpha-\langle\alpha\rangle)/\sigma_\alpha$ from the
mocks and compare the result to the unit normal distribution.  We find
excellent agreement; a Kolmogorov-Smirnov (K-S) test finds a high
likelihood that the observed distribution is drawn from a unit normal.
Hence the Gaussian probability distribution obtained from the $\chi^2$
statistic is an appropriate characterisation of the error on $\alpha$.

\section{The Power Spectrum}
\label{sec:pk}

    \subsection{Measuring the power spectrum}
    \label{sec:analysis_pk}
\begin{figure*}
  \centering
  \resizebox{0.9\columnwidth}{!}{\includegraphics{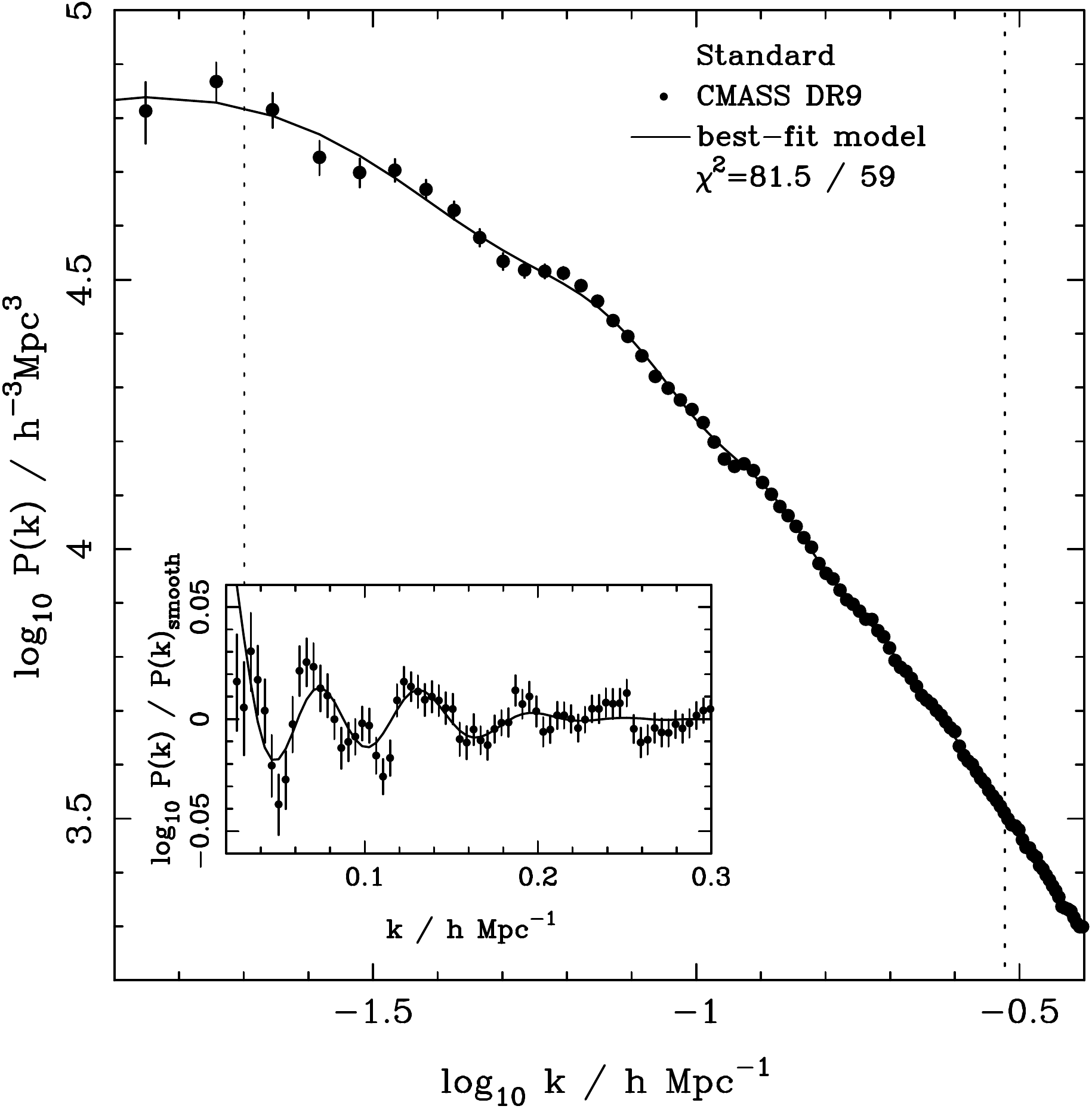}}
  \hspace{0.1\columnwidth}
  \resizebox{0.9\columnwidth}{!}{\includegraphics{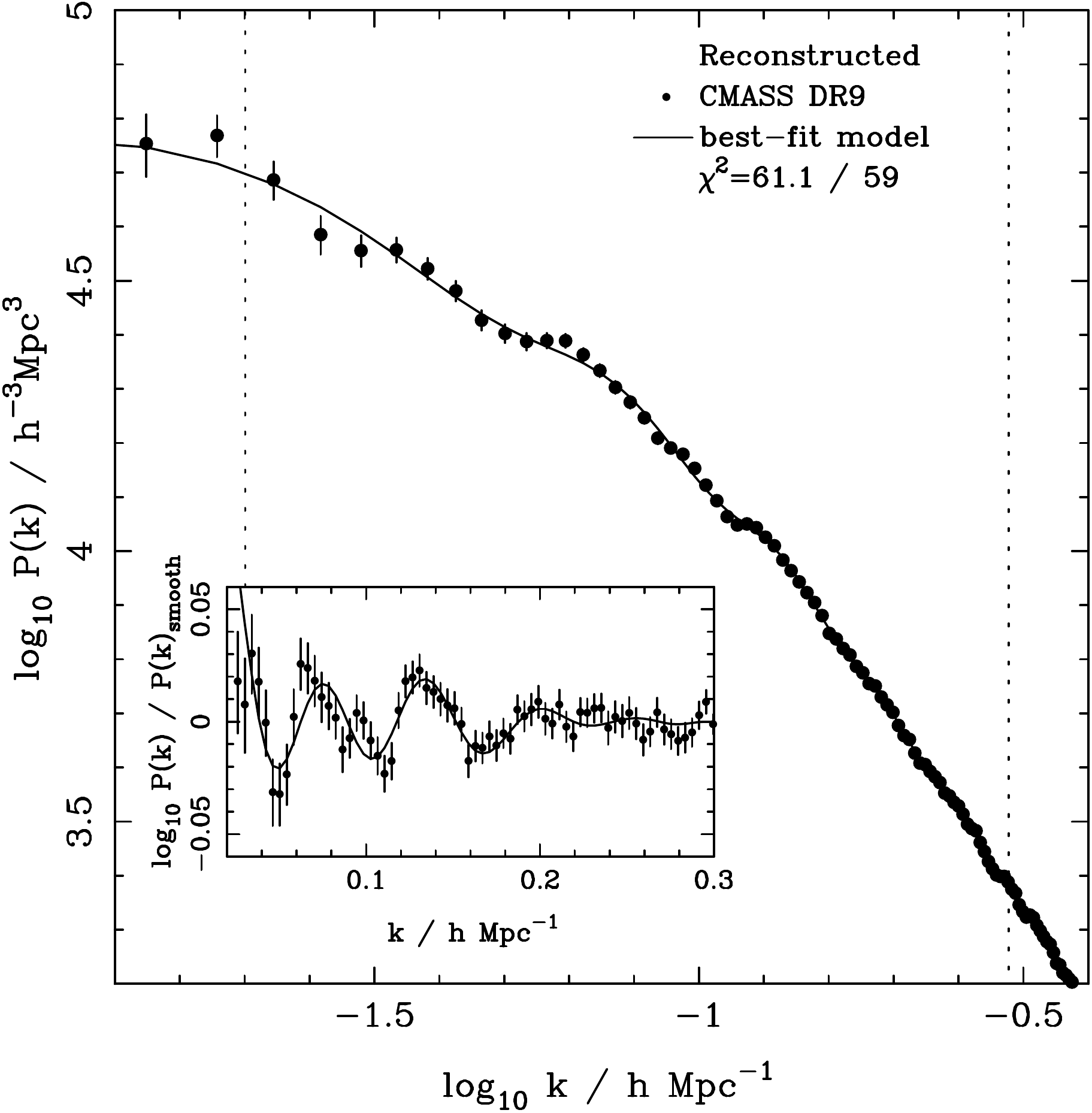}}
  \caption{The CMASS DR9 power spectra before (left) and after (right)
    reconstruction with the best-fit models overplotted. The vertical
    dotted lines show the range of scales fitted ($0.02<k<0.3\hompc$),
    and the inset shows the BAO within this $k$-range, determined by
    dividing both model and data by the best-fit model calculated
    (including window function convolution) with no BAO. Error bars
    indicate $\sqrt{C_{ii}}$ for the power spectrum and the rms error
    calculated from fitting BAO to the 600 mocks in the inset (see
    Section~\ref{sec:covar} for details).}
  \label{fig:DR9pk}
\end{figure*}

    The power spectra recovered from the CMASS DR9 data are shown in
Fig.~\ref{fig:DR9pk} before (left) and after (right)
reconstruction. The inset shows the oscillations in these data,
calculated by dividing by a smooth model (see Section~\ref{sec:fitpk}
for details). The effect of the reconstruction algorithm is clear -
the large-scale power is decreased corresponding to the removal of RSD
effects, with the small-scale power being further reduced by the
reduction in non-linear power. These data represent the most accurate
measurement of a redshift-space galaxy power spectrum ever obtained.

Power spectra were calculated using the Fourier method first developed
by \citet{FKP94}, as described in \citet{percival07b} and \citet{Rei10}. We work in
redshift-space as if observed recession velocities solely arise from
the Hubble expansion. As we focus on measuring angle-averaged Baryon
Acoustic Oscillations, we do not convert from a galaxy density field
to a halo density field as in \cite{Rei10}, or apply corrections for
Finger-of-God effects. Given a weight $w_i$ for galaxy $i$ at location
${\bf r}_i$, the overdensity field can be written
\begin{equation}  \label{eq:field}
  F({\bf r}) = \frac{1}{N} \left[
    \sum_i w_i\delta_D({\bf r}_i-{\bf r}) 
      - \langle w({\bf r})n({\bf r})\rangle \right],
\end{equation}
where $N$ is a normalisation constant
\begin{equation}  \label{eq:N}
  N\equiv\left\{\int d^3r
    \langle w({\bf r})n({\bf r})\rangle^2
  \right\}^{1/2},
\end{equation}
and $\langle w({\bf r})n({\bf r})\rangle$ is the expected weighted
distribution of galaxies at location ${\bf r}$ in the absence of clustering, 
and $n({\bf r})$ is
the galaxy density. The quantity $\delta_D$ is the standard
Dirac-$\delta$ function. We do not apply luminosity-dependent weights
(as applied by \citealt{percival07b,Rei10}), as we are only interested
in the BAO, and not the overall shape of the power spectrum.

We chose to model the expected distribution of galaxies using a random
catalogue with points selected at the mean galaxy density $\langle
n({\bf r})\rangle$, which is then weighted in a similar manner to the
galaxies. The calculation of this catalogue was described in
Section~\ref{sec:random}. The weights in the random catalogue are then
renormalised, and compared with the weights applied to the galaxies so
that $\int F({\bf r})\,dr=0$, thereby matching the total weighted
number density in galaxy and random catalogues.

Power spectra are calculated using a $2048^3$ grid in a cubic box of
length $8000\mpcoh$. This zero-pads the galaxies - the minimum and
maximum galaxy redshifts of the sample correspond to distances of
$1170\mpcoh$ and $1780\mpcoh$, so the galaxies form an angular sector
of a thick shell within this cube. The Nyquist frequency for the
Fourier transform is approximately $0.8\hompc$, which is significantly
larger than the maximum frequency fitted of $0.3\hompc$ (see
Section~\ref{sec:fitpk}). The smoothing effect of the cloud-in-cell
assignment used to locate galaxies on the grid
\citep[e.g.][chap.~5]{he81} is corrected, and shot-noise is 
subtracted following the assumption that galaxies form a Poisson
sampling of the density field (see \citealt{FKP94} for details). The
power spectrum is then spherically averaged, leaving an estimate of
the ``redshift-space'' power, binned into bins in $k$ of width
$0.04\hompc$.

    \subsection{Fitting the power spectrum}
    \label{sec:fitpk}
    We fit the observed redshift-space power spectrum, calculated as
described in Section~\ref{sec:pk}, with a two component model
comprising a smooth cubic spline multiplied by a model for the BAO,
following the procedure developed by
\citet{percival07a,Per07,Per10}. The model power spectrum is given by
\begin{equation}  \label{eq:pk_model}
  P(k)_{\rm m}=P(k)_{\rm smooth} \times B_m(k/\alpha),
\end{equation} 
where $P(k)_{\rm smooth}$ is a smooth model that fits the overall
shape of the power spectrum, and the BAO model $B_{\rm m}(k)$,
calculated for our fiducial cosmology, is scaled by the dilation
parameter $\alpha$ as defined in Eq.~\ref{eqn:alpha_def}. The
calculation of the BAO model is described in detail below. This
scaling of the acoustic signal is identical to that used in
the correlation function fits, although the differing non-linear
prescriptions in (Eqns~\ref{eqn:fform} \&~\ref{eq:pk_model}) means
that the non-linear BAO damping is treated in a subtly different way.

Each power spectrum model to be fitted is convolved with the survey
window function, giving our final model power spectrum to be compared
with the data. The window function for this convolution is the
normalised power in a Fourier transform of the weighted survey
coverage, as defined by the random catalogue, and is calculated using
the same Fourier procedure described in Section~\ref{sec:pk}
(e.g. \citealt{Per07}). This is then fitted to express the window
function as a matrix relating the model power spectrum evaluated at
$1000$ wavenumbers, $k_n$, equally spaced in $0<k<2\hompc$, to the
central wavenumbers of the observed bandpowers $k_i$:
\begin{equation}  \label{eq:defwindow}
  P (k_i)_{\rm fit} =\sum_n W(k_i,k_n) P(k_n)_{\rm m} - W(k_i,0).
\end{equation}
The final term $W(k_i,0)$ arises because we estimate the average
galaxy density from the sample, and is related to the integral
constraint in the correlation function. In fact this term is smooth
(as the power of the window function is smooth), and so can be
absorbed into the smooth component of the fit, and we therefore do not
explicitly include this term in our fits.

To model the overall shape of the galaxy clustering power spectrum we
use a cubic spline \citep{press92}, with nine nodes fixed empirically
at $k=0.001$, and $0.02<k<0.4$ with $\Delta k=0.05$, matching
that adopted in \citet{Per07,Per10}. This model was tested in these
papers, but we show in Section~\ref{sec:robust_pk} that it also
provides an excellent fit to the overall shape of the DR9 CMASS mock
catalogues, and that there is no evidence for deviations for the fits
to the data.

To calculate our fiducial BAO model, we start with a linear matter
power spectrum $P(k)_{\rm lin}$, calculated using {\sc CAMB}
\citep{lewis00}, which numerically solves the Boltzman equation
describing the physical processes in the Universe before the
baryon-drag epoch. We then evolve using the {\sc HALOFIT} prescription
\citep{smith03}, giving an approximation to the evolved power spectrum
at the effective redshift of the survey. To extract the BAO, this
power spectrum is fitted with a model as given by
Eq.~\ref{eq:pk_model}, where we adopt a fixed BAO model ($B_{\rm EH}$)
calculated using the \citet{Eis98} fitting formulae at the same
fiducial cosmology. Dividing $P(k)_{\rm lin}$ by the best-fit smooth
power spectrum component from this fit produces our BAO model, which
we denote $B_{\rm CAMB}$.

We damp the acoustic oscillations to allow for non-linear effects  
\begin{equation}  \label{eq:bao_damp}
  B_m=(B_{\rm CAMB}-1)e^{-k^2\Sigma_{nl}^2/2} + 1,
\end{equation} 
where the damping scale $\Sigma_{nl}$ is a fitted parameter. We assume
a Gaussian prior on $\Sigma_{nl}$ with width $\pm2\mpcoh$, centred on
$8.24\mpcoh$ for pre-reconstruction fits and $4.47\mpcoh$ for
post-reconstruction fits, matching the average recovered values from
fits to the 600 mock catalogs with no prior. The exact width of the
prior is not important, but if we do not include such a prior, then
the fit can become unstable with respect to local minima at extreme
values.

We fit over scales $0.02\hompc<k<0.3\hompc$: these limits are imposed
because the BAO have effectively died out for $k>0.3\hompc$, and
scales $k<0.02\hompc$ are sensitive to observational systematics
\citep{Ross12}. We bin the measured power spectrum in $k$ bins of
width $0.004\hompc$, so 70 data points are included in the fits. The
function $P(k)_{\rm smooth}$ depends on 9 free parameters, the
amplitudes of the spline nodes. Thus, including $\alpha$ and
$\Sigma_{nl}$ we fit to 11 parameters in total, and the fit has 59
degrees-of-freedom. Goodness of fit between model $P(k)_{\rm fit}$ and
data is calculated using the $\chi^2$ statistic. We consider intervals
of $\Delta\alpha=0.002$ in the range $0.7<\alpha<1.3$ and, for each
value of $\alpha$ to be tested, we use the Powell routine
\citep{press92}, starting from a series of widely separated start
points, to find the spline node values and $\Sigma_{nl}$ that result
in the minimum value of $\chi^2$.

For each power spectrum fitted, we have estimated the error on the
best-fit value of $\alpha$ by considering the $\Delta\chi^2=1$
interval and by integrating over the likelihood surface. These
measurements are found to match extremely well for all of the fits,
suggesting that the likelihood is well behaved around the minima. We
also consider the distribution of best-fit $\alpha$ recovered from the
mock catalogues, as discussed in subsequent sections. For the results
presented, in order to be consistent we adopted the procedure described
in Section~\ref{sec:fitxi} to make measurements from the $\chi^2$
surfaces resulting from the power spectrum fits.

    \subsection{Results}
    \label{sec:results_pk}
    We have measured the best-fit $\alpha$ and $\sigma_\alpha$ using the
procedure described in Section~\ref{sec:fitpk} for power spectra
calculated from each of 600 mock catalogues and from the CMASS DR9
data, either before or after applying the reconstruction
algorithm. The maximum likelihood solution for the dilation parameter
from the full CMASS DR9 sample is $\alpha=1.022\pm0.017$ before
reconstruction and $\alpha=1.042\pm0.016$ after reconstruction. Errors
were determined from the moments of $\alpha$ calculated by integrating
the likelihood surface. The CMASS DR9 power spectra before (left) and after
(right) reconstruction are plotted in Fig.~\ref{fig:DR9pk}, compared
with the corresponding best-fit models. Before reconstruction, we
find $\chi^2_{\rm min}=81.5$, while post-reconstruction this reduces
to $\chi^2_{\rm min}=61.1$, with 59 degrees-of-freedom.

\begin{figure}
  \centering
  \resizebox{0.95\columnwidth}{!}{\includegraphics{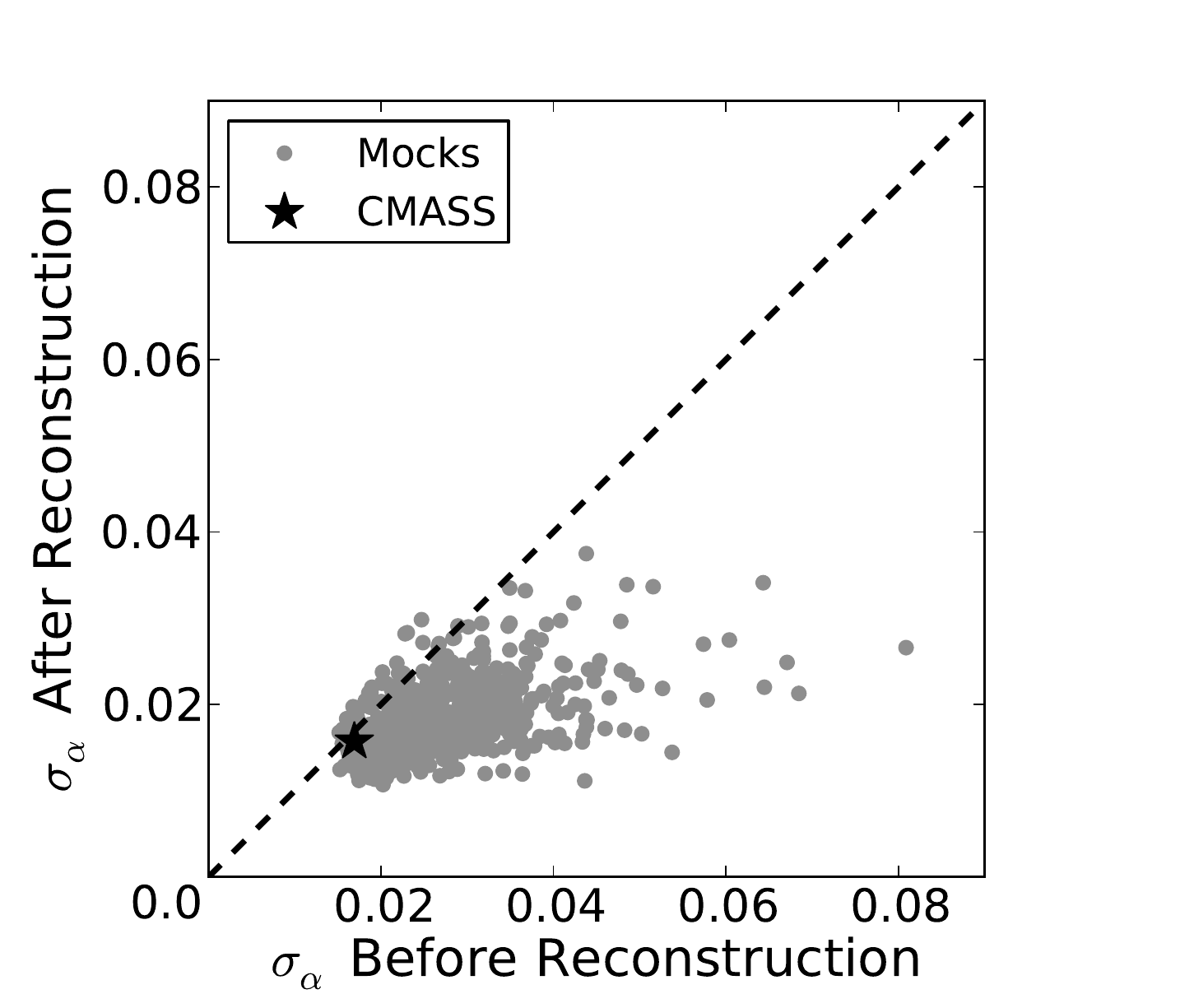}}
  \caption{Comparison of mock $\sigma_\alpha$ errors before and after
    reconstruction as measured from $P(k)$. This plot is analogous to
    Figure \ref{fig:comprec} obtained from our fits to $\xi(r)$. We
    see the same overall average factor of 1.54 decrease in
    $\sigma_\alpha$ as seen for $\xi(r)$. This again indicates that in
    general, reconstruction improves our ability to precisely measure
    $\alpha$. On a mock-by-mock basis, we see a range of results, with
    36 (out of 600) mocks actually having $\sigma_\alpha$ increase
    with reconstruction.  We once again see that reconstruction does
    not appear to improve the error on our CMASS DR9 measurement of
    $\alpha$, and that this is not unexpected in the context of the
    mocks: the CMASS DR9 measurement error lies within the locus of
    values recovered from the mocks, with 61 mocks (out of 600)
    showing a lower fractional improvement than that of the CMASS DR9
    data.}
  \label{fig:comperr_pk}
\end{figure}

As for the correlation function measurements, we find that
reconstruction does not significantly improve the measurement of the
dilation parameter from the CMASS DR9 data. This is consistent with
the results from the mock catalogs: we have measured and fitted power
spectra calculated for all 600 mock catalogs, and
Fig.~\ref{fig:comperr_pk} presents a scatter plot of the recovered
errors before and after reconstruction. Although reconstruction
improves the fit for the majority of the mock catalogs, there are a
small number for which reconstruction increases the recovered
error. Also, we see that improvement is more likely where the
pre-reconstruction error is high, suggesting that variation in the
error recovered from different catalogues is dominated by the
``noise'' that reconstruction is able to remove. In this plot, the
star marks the result from the CMASS DR9 data, showing that the
pre-reconstruction error recovered is significantly smaller than the
mean expected from the mocks. Given this result, we should not be
surprised that reconstruction only has a small effect on these data.

\begin{figure*}
  \centering
  \resizebox{0.95\columnwidth}{!}{\includegraphics{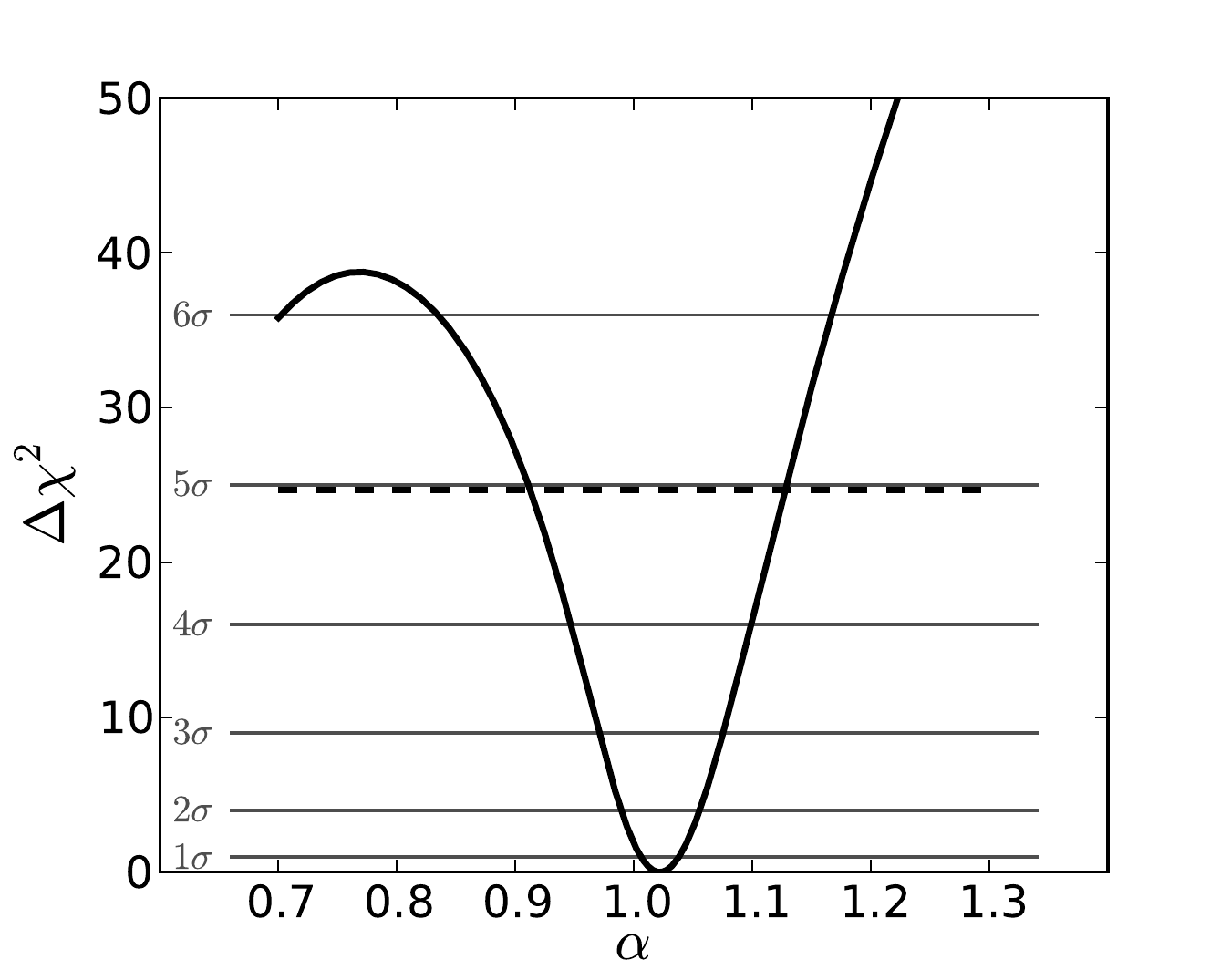}}
  \resizebox{0.95\columnwidth}{!}{\includegraphics{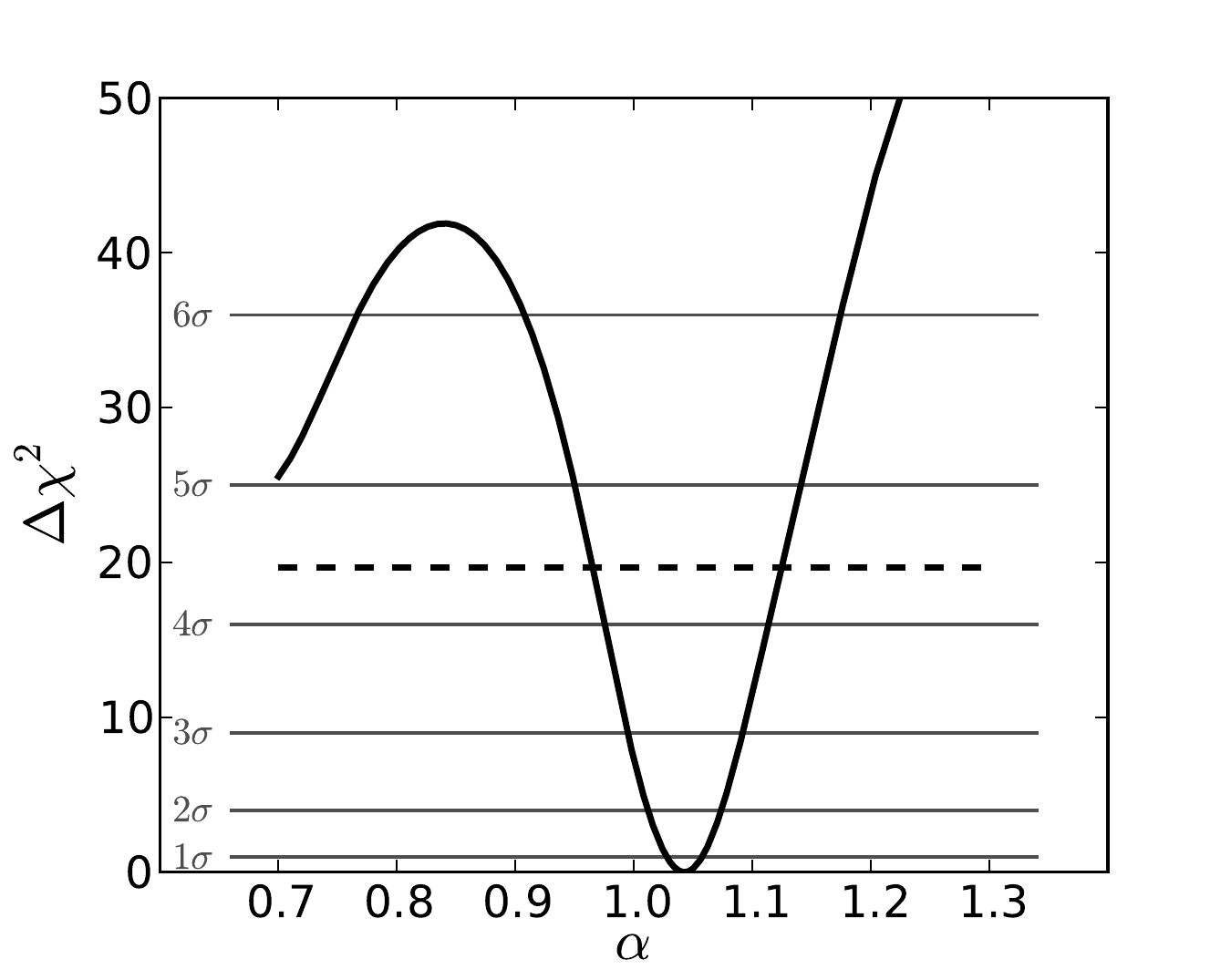}}
  \caption{Significance of the CMASS DR9 BAO feature before (left) and
    after (right) reconstruction as measured from $P(k)$. This figure
    is analogous to Figure \ref{fig:DR7_9chi2} measured from our
    $\xi(r)$ analysis. In our $P(k)$ analysis, before reconstruction
    we detect the BAO in the CMASS DR9 sample at around $5\sigma$
    confidence, similar to our result for $\xi(r)$. After
    reconstruction we see a slight drop in the detection level with
    respect to the pre-reconstruction result. The global maximum is
    found within a valley whose depth is greater than $6\sigma$.}
  \label{fig:chi2_pk}
\end{figure*}

The BAO detection from the CMASS DR9 data is highly significant, with
a $\chi^2$ difference between best-fit models with and without the BAO
component being approximately $5\sigma$ before reconstruction,
dropping slightly to $4.5\sigma$ post-reconstruction. The relatively
small difference between significance before and after reconstruction
matches the difference in $\sigma_\alpha$ discussed previously. The
$\chi^2$ surfaces are shown in Fig.~\ref{fig:chi2_pk} before (left)
and after (right) reconstruction. Low values of $\alpha$ result in the
BAO signal being moved to large scales where the cosmic variance error
increases, which is why these models give comparatively good fits.

From the 600 mocks, pre-reconstruction we recover a mean value of the
dilation parameter of $\langle\alpha\rangle=1.004$, with average error
on any single realisation of $0.029$ and standard error on the mean of
$0.002$. Post-reconstruction this reduces to
$\langle\alpha\rangle=1.003$, with average error on any single
realisation of $0.019$ and standard error on the mean of
$0.001$. There is therefore evidence for a small systematic shift
between the true value ($\alpha=1$) for the mocks and the values
recovered. A discussion of the systematic errors associated with our
measurements is provided in Section~\ref{sec:consensus}. The decrease
in mock rms post-reconstruction demonstrates the positive effect that
reconstruction has on average. Note that when calculating the above
mean recovered errors we excluded two mocks pre-reconstruction for
which $\sigma_\alpha>0.07$, where the BAO feature was not well
recovered.

\begin{figure}
  \centering
  \resizebox{0.95\columnwidth}{!}{\includegraphics{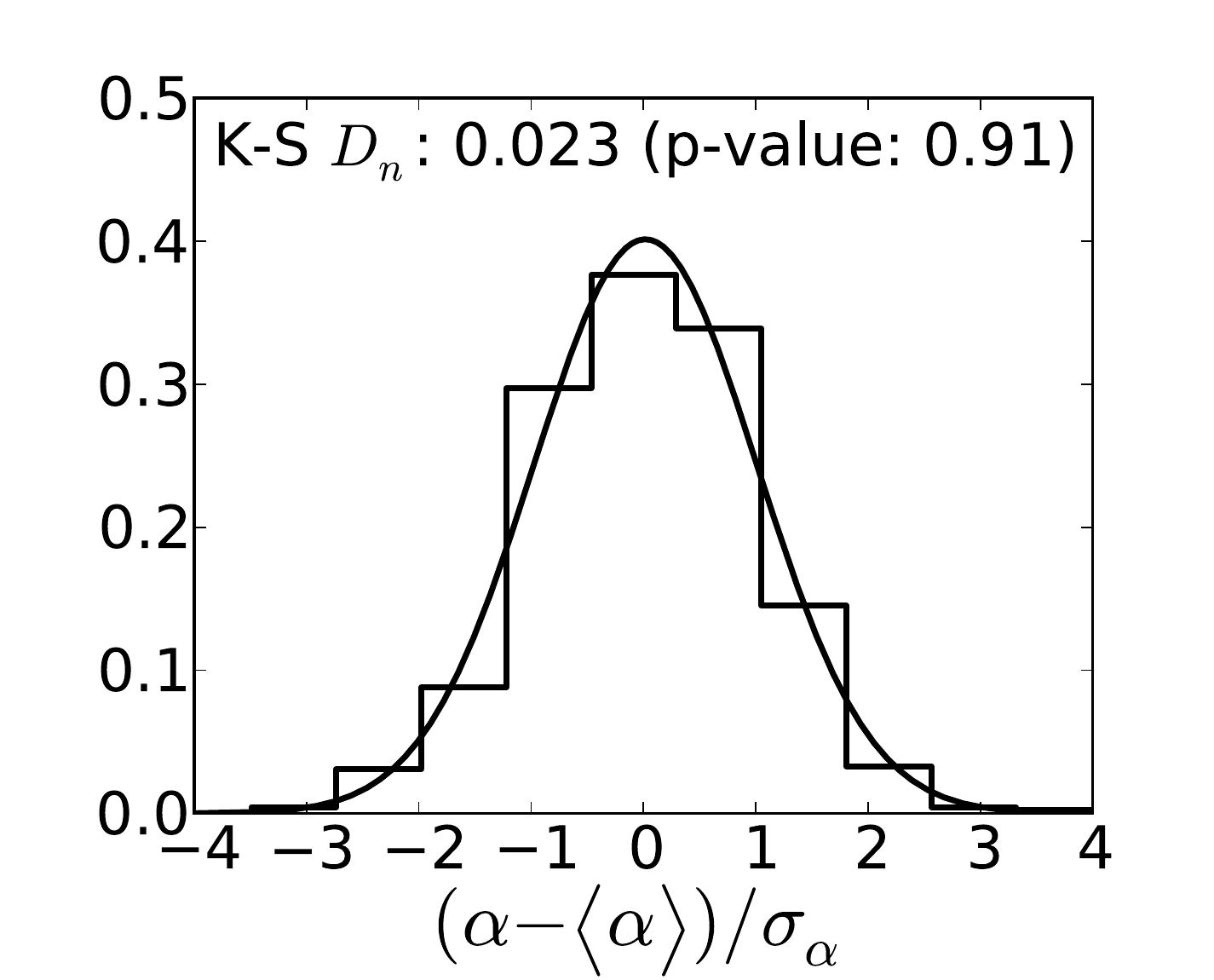}}
  \caption{Histogram of $(\alpha-\langle\alpha\rangle)/\sigma_\alpha$
    measured from $P(k)$ of the post-reconstruction mocks. This figure
    is analogous to Fig.~\ref{fig:snr_xi_rec} obtained from our fits
    to $\xi(r)$. We again see a near-Gaussian distribution as
    indicated by the small K-S value. This indicates that the
    $\sigma_\alpha$ values we measure from the $\chi^2$ distribution
    are reasonable estimates of the error on $\alpha$ measured by
    fitting $P(k)$. Note that round-off accounts for the fact that the
    quoted $D_n$ values are the same in this Figure and
    Figure~\ref{fig:snr_xi_rec}, while the $p$-values differ
    slightly. }
  \label{fig:snr_pk_rec}
\end{figure}

The mean values $\langle\sigma_\alpha\rangle$ match perfectly
with the standard deviation of the recovered $\alpha$ values from
the mocks, which give $\langle(\alpha-1)^2\rangle^{1/2}=0.029$ and
$\langle(\alpha-1)^2\rangle^{1/2}=0.019$ pre- and post-reconstruction,
indicating that the likelihood is extremely well behaved. This is
not the case if the damping parameter $\Sigma_{nl}$ is fixed at an
incorrect value in the model to be fitted to the data: insufficient
damping results in recovered errors that are too small with respect to
the distribution, while over-damping leads to over-prediction of the
errors. As in Section~\ref{sec:results_xi}, we now test the nature of the
distribution of recovered dilation parameters. Fig.~\ref{fig:snr_pk_rec}
shows a histogram of $(\alpha-\langle\alpha\rangle)/\sigma_\alpha$ compared with
a standard Normal distribution. As is clearly evident, the data are
extremely well matched to the Gaussian prediction; this is also
indicated by the result of a K-S test.

\begin{figure}
  \centering
  \resizebox{0.9\columnwidth}{!}{\includegraphics{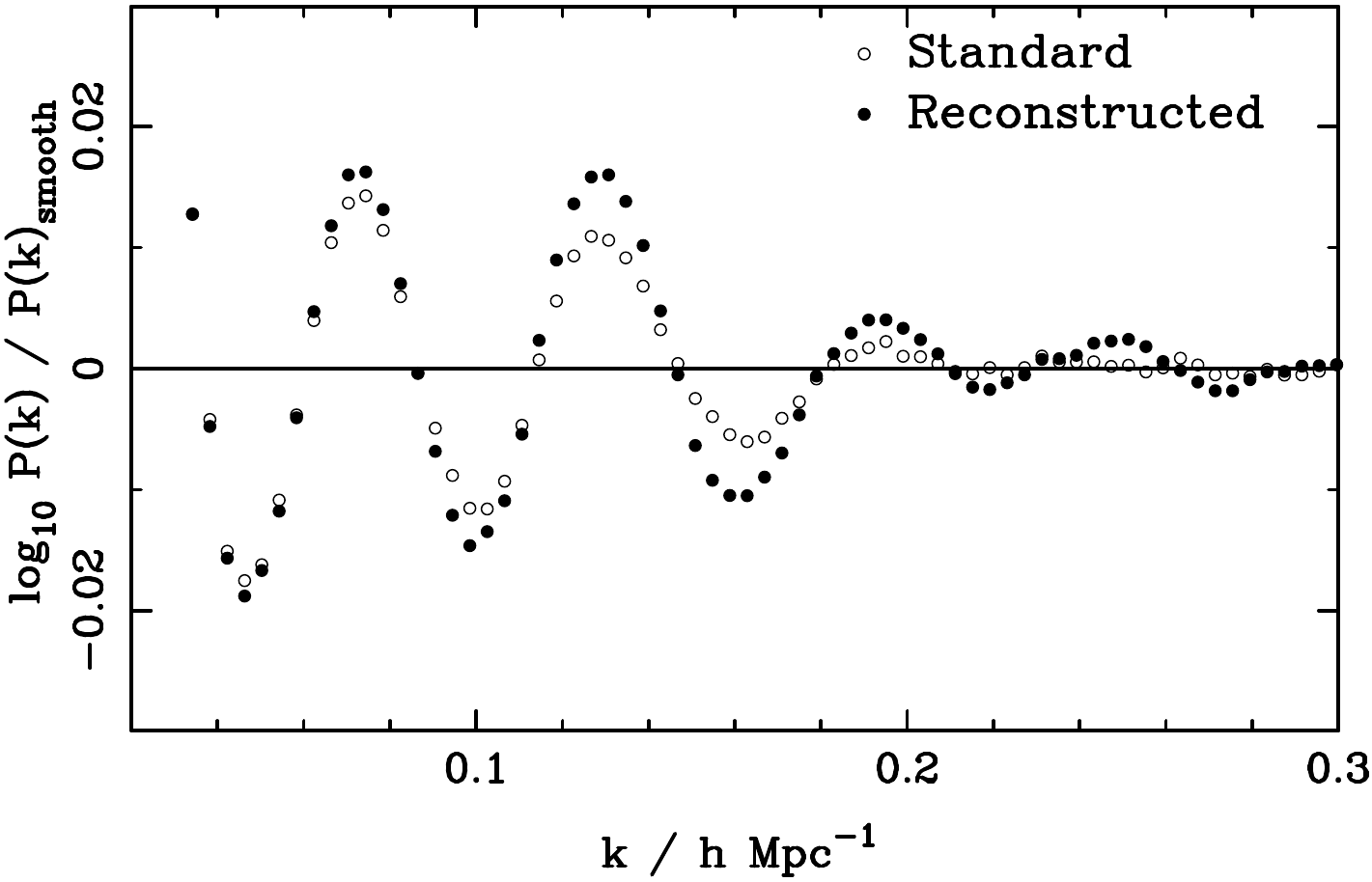}}
  \caption{Average BAO signal calculated by dividing the measured
    power from each of the 600 mocks by the best-fit smooth model
    (solid symbols after reconstruction, open symbols before
    reconstruction). Clearly reconstruction enhances the small-scale
    BAO, where cosmic variance errors are significantly
    reduced. \label{fig:BAO_mocks}}
\end{figure}

Finally in this section we consider the average BAO signal recovered
from the mock catalogues. For each mock, we divide the measured power
spectrum by the smooth component of the best-fit solution convolved
with the survey window function. The average of these values over all
of the mocks is shown in Fig.~\ref{fig:BAO_mocks} both before and
after applying the reconstruction algorithm. The average effect of
reconstruction is evident on small scales, with the BAO feature being
enhanced by this algorithm. Fitting to the mocks without assuming a
prior on $\Sigma_{nl}$ gives average best-fit values of
$\langle\Sigma_{nl}\rangle=8.24\mpcoh$ before reconstruction and
$\langle\Sigma_{nl}\rangle=4.47\mpcoh$ following reconstruction, which
shows the extent of the improvement afforded by this technique. Note
that these are systematically different from the values of
$\Sigma_{nl}$ recovered from the correlation function fits, which
results from the way in which the non-linear shape was fitted leading
to different effective definitions of $\Sigma_{nl}$. In the $P(k)$
fits, a multiplicative correction was used, while for $\xi(r)$, an
additive correction was adopted: the $\xi(r)$ fit required less
damping as he additive correction already acts to damp the importance
of the BAO component, while the multiplicative correction for $P(k)$
afforded by the free shape is itself multiplied by the BAO model, and
thus more damping is required.

\section{The Distance to $z=0.57$}
\label{sec:consensus}
\begin{table}
\caption{Key $\alpha$ measurements from BAO in the CMASS DR9 sample.}
\label{tab:key_alphas}

\begin{tabular}{@{}lccc}

\hline
&
$\alpha$&
$\chi^2/dof$&
$D_V/r_s(z=0.57)$\\

\hline
\multicolumn{4}{c}{Before Reconstruction}\\
\hline

$\xi(r)$&
$1.016 \pm 0.017$&
$30.53/39$&
$13.44 \pm 0.22$\\
\\[-1.5ex]
$P(k)$&
$1.022 \pm 0.017$&
$81.5/59$&
$13.52 \pm 0.22$\\

\hline
\multicolumn{4}{c}{After Reconstruction}\\
\hline

$\xi(r)$&
$1.024 \pm 0.016$&
$34.53/39$&
$13.55 \pm 0.21$\\
\\[-1.5ex]
$P(k)$&
$1.042 \pm 0.016$&
$61.1/59$&
$13.78 \pm 0.21$\\
\\[-1.5ex]
Consensus&
$1.033 \pm 0.017$&
$--$&
$13.67 \pm 0.22$\\
\hline
\end{tabular}

\end{table}

\begin{figure*}
  \centering
  \resizebox{0.75\columnwidth}{!}{\includegraphics{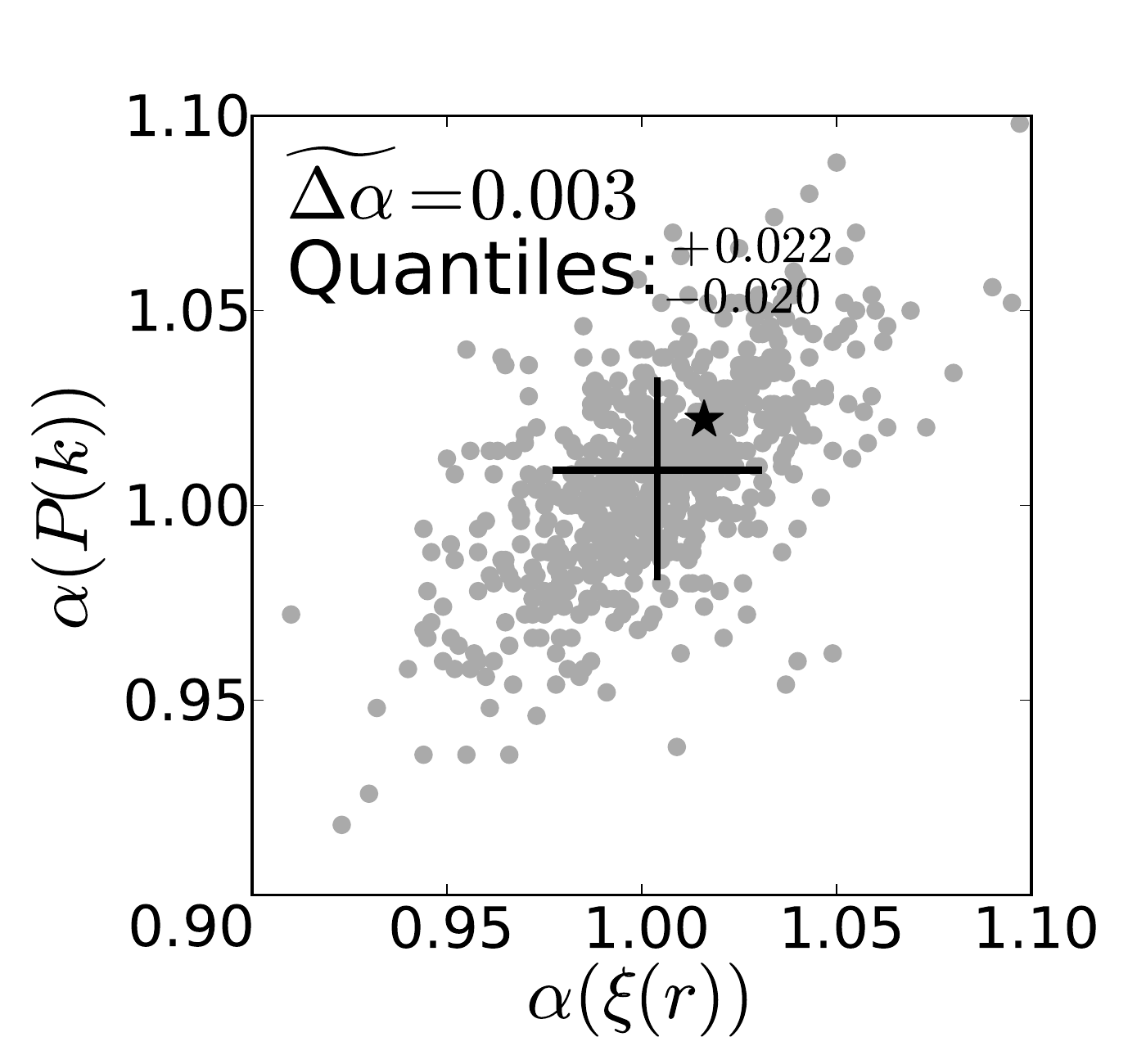}}
  \resizebox{0.75\columnwidth}{!}{\includegraphics{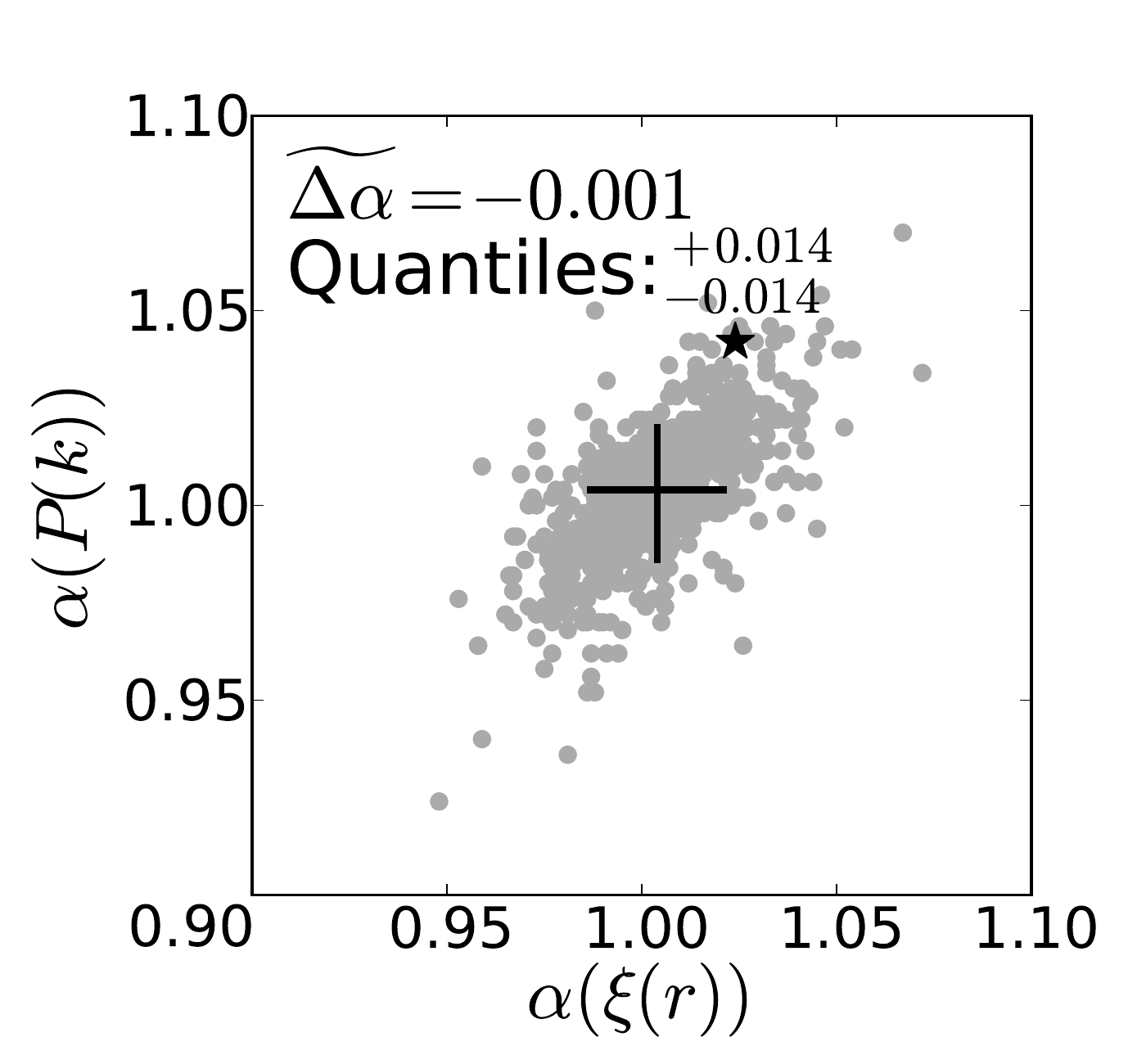}}
  \resizebox{0.75\columnwidth}{!}{\includegraphics{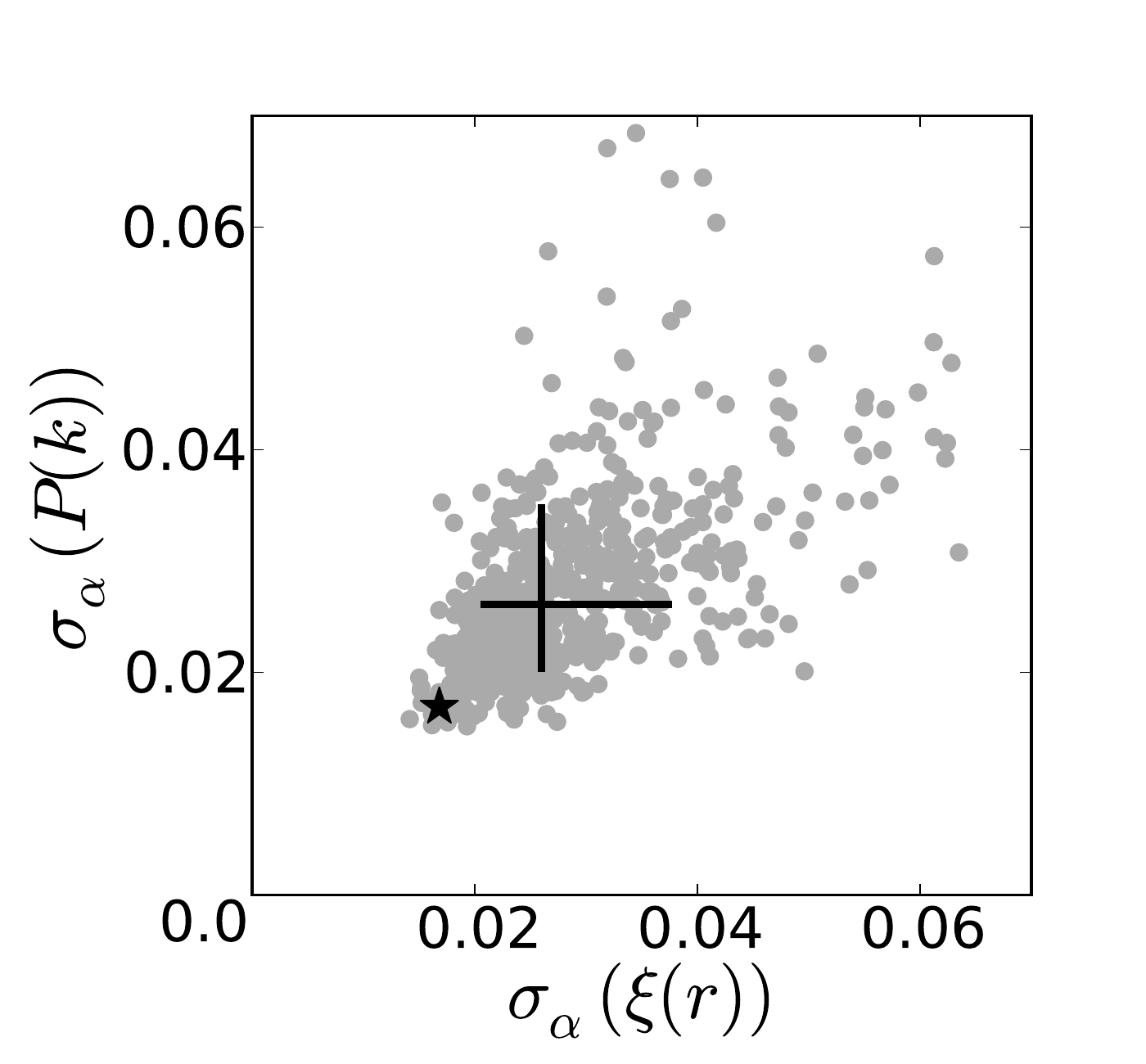}}
  \resizebox{0.75\columnwidth}{!}{\includegraphics{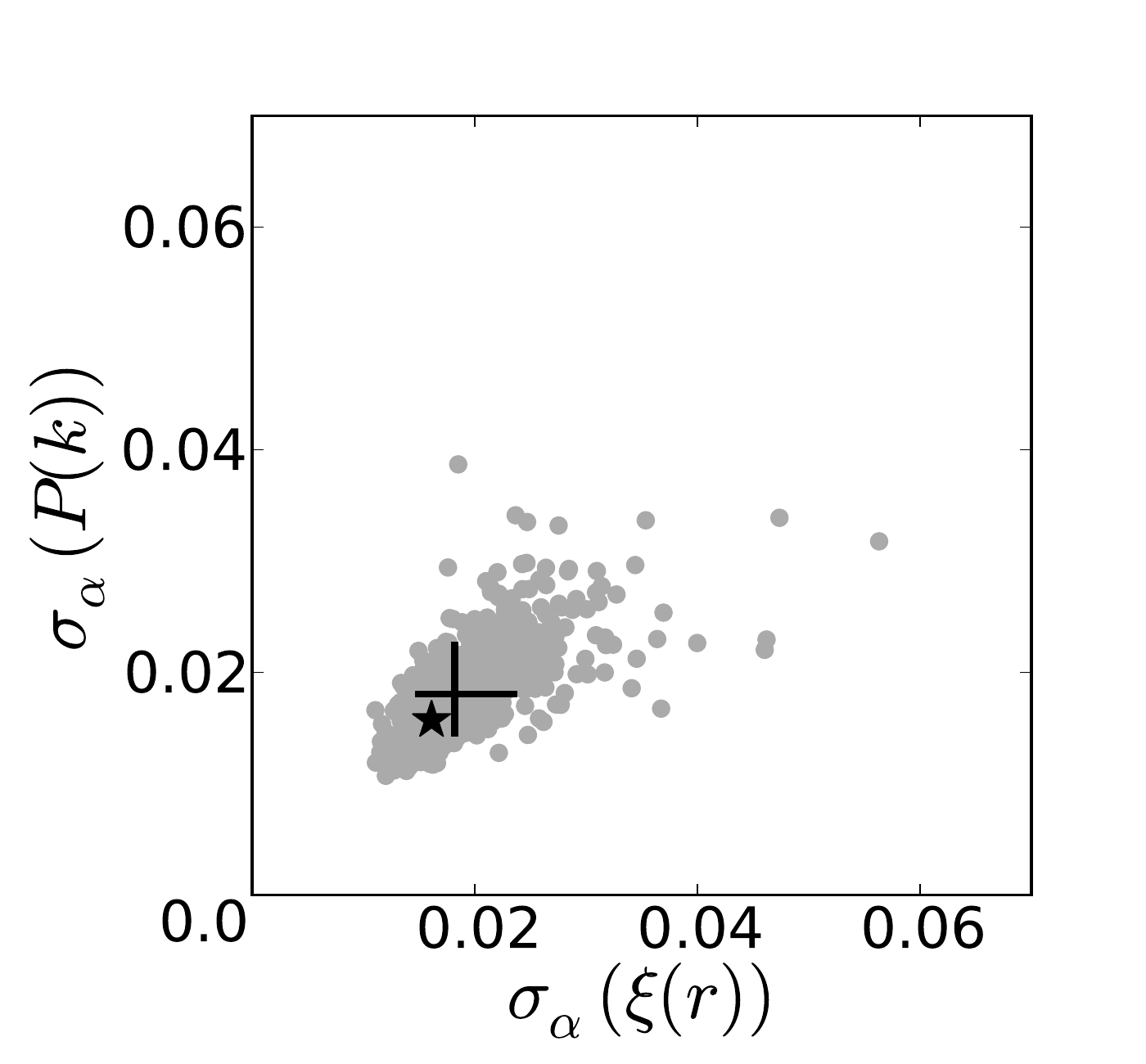}}
  \caption{ Comparison of acoustic scale measurements from $\xi(r)$
    and $P(k)$. The left column shows the pre-reconstruction results
    and the right column shows the post-reconstruction results. The
    top panels show the values of $\alpha$ measured using $\xi(r)$
    versus those measured using $P(k)$; the bottom panels show
    analogous plots for $\sigma_\alpha$. The mock points are shown in
    grey and the CMASS point is overplotted as the black star. The
    black cross marks the median values of $\alpha$ or $\sigma_\alpha$
    along with their quantiles. One can see that there is notable
    scatter between the values of $\alpha$ and $\sigma_\alpha$
    measured from the two different statistics. For example, $\alpha$
    from $\xi$ and $P$ vary by 0.014 after reconstruction.  }
  \label{fig:pxi_comp}
\end{figure*}

\begin{figure*}
  \centering
  \resizebox{0.75\columnwidth}{!}{\includegraphics{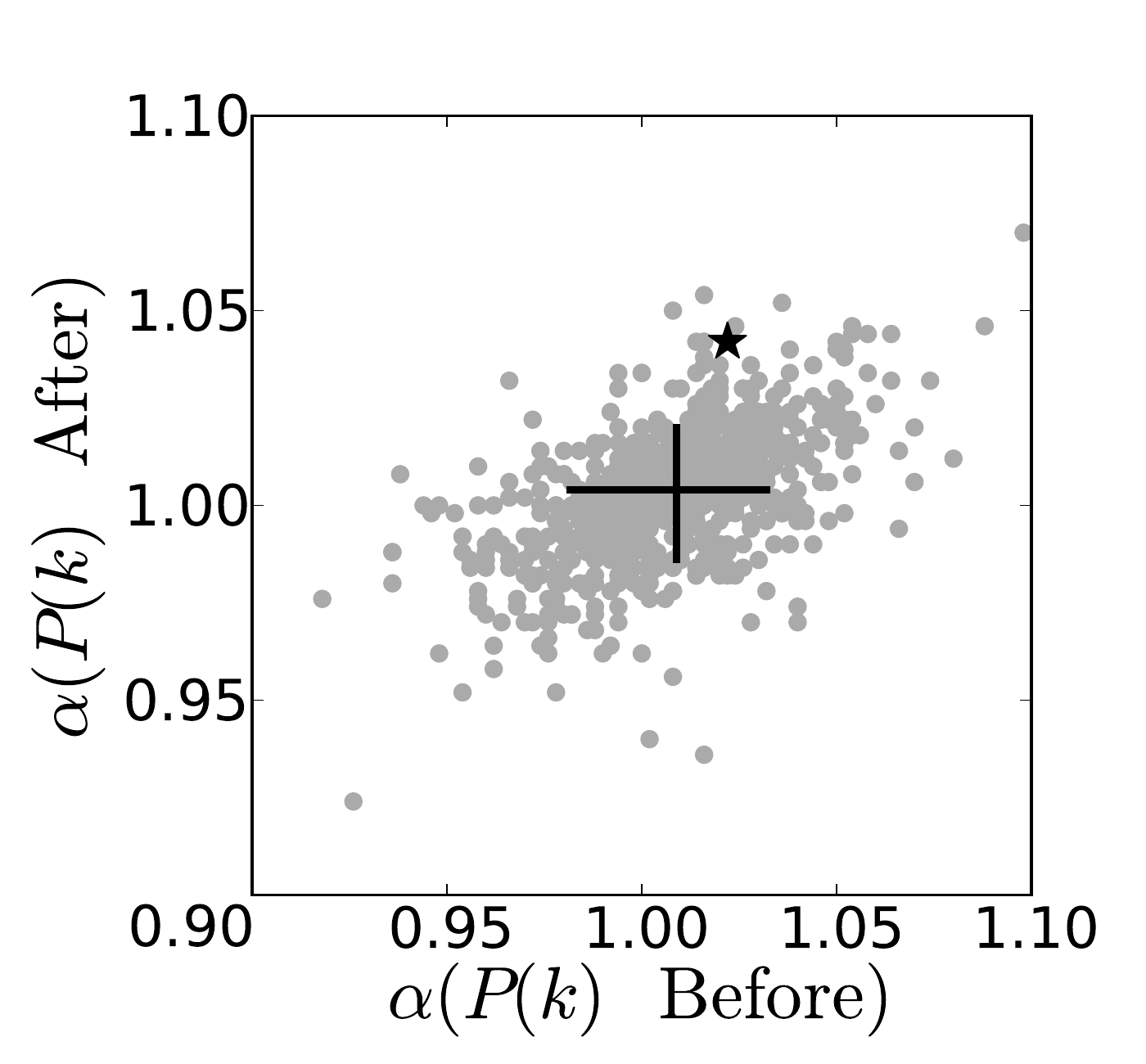}}
  \resizebox{0.75\columnwidth}{!}{\includegraphics{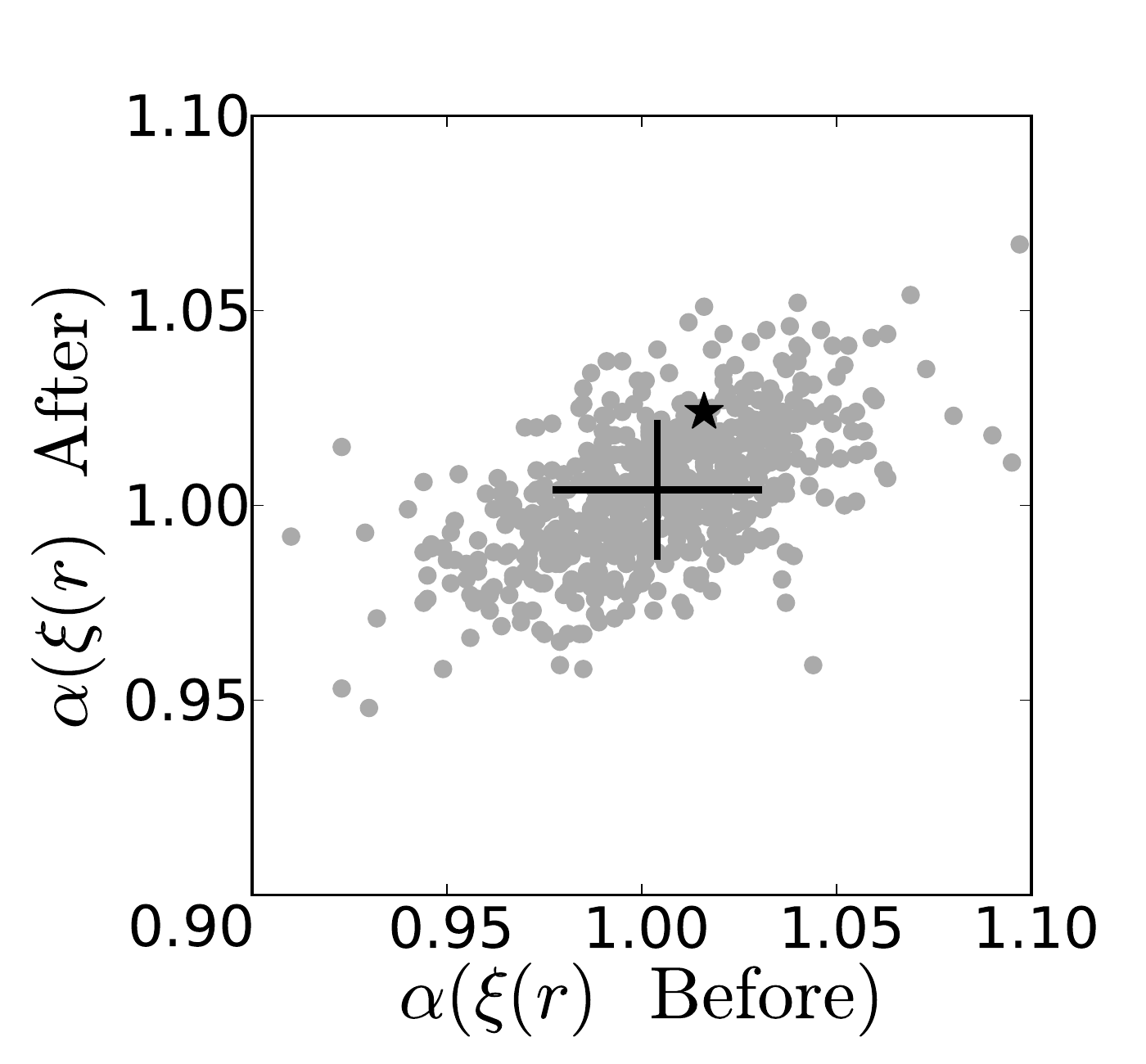}}
  \caption{Comparison of values of $\alpha$ recovered from the mocks
    before and after reconstruction as measured from $\xi(r)$ (left)
    and $P(k)$ (right). As expected, there is a tight correlation
    between the measurements before and after reconstruction, with a
    slopeshowing the reduced scatter after reconstruction. The CMASS
    DR9 measurements (shown by the stars) lie within the locus of
    values recovered from the mocks, and the changes in best-fit
    values seen before and after reconstruction are not unusual.}
  \label{fig:comp}
\end{figure*}

\begin{figure}
  \centering
  \resizebox{0.95\columnwidth}{!}{\includegraphics{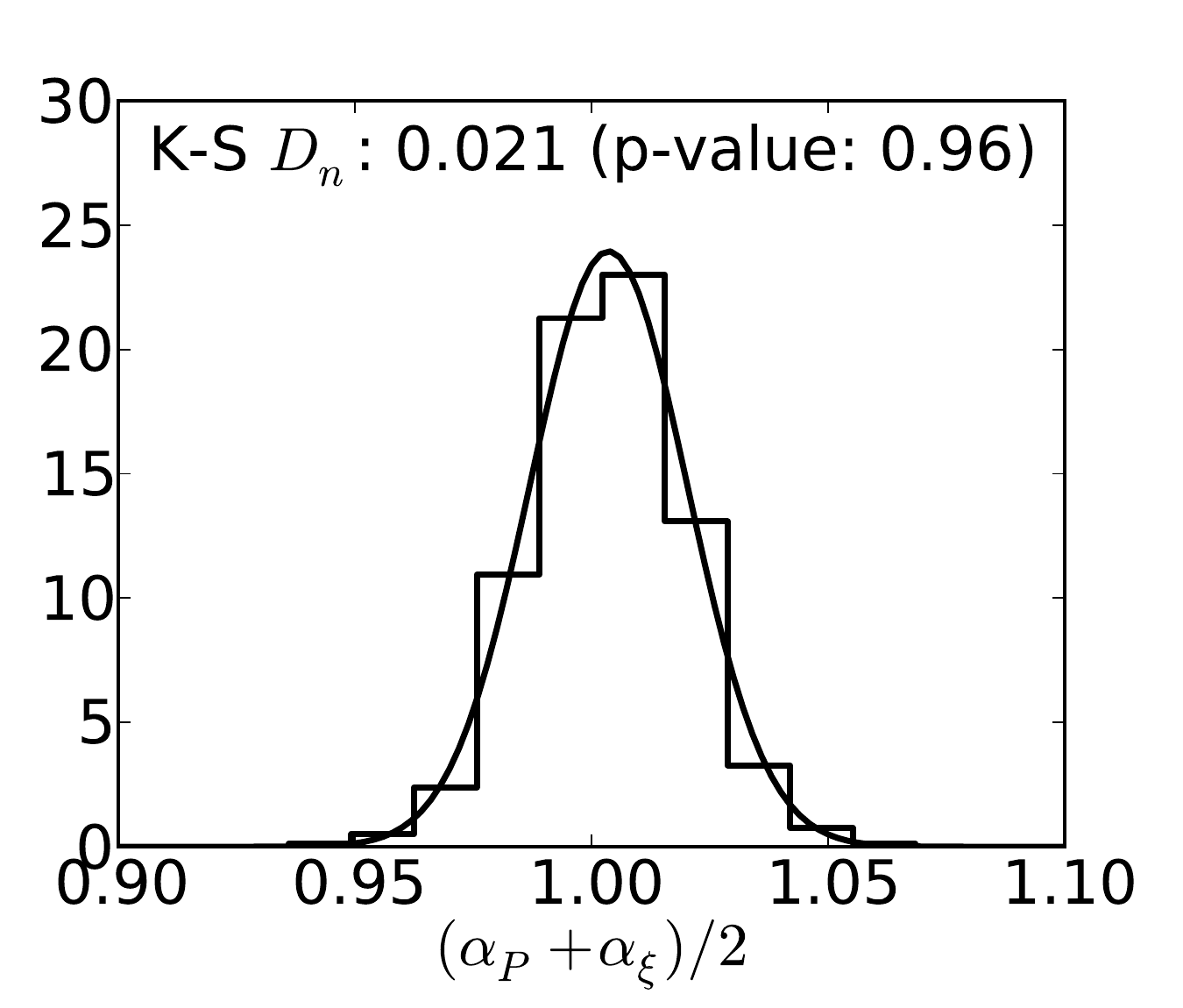}}
  \caption{ Histogram of averaged $\alpha$ values from $\xi(r)$ and
    $P(k)$ as measured from the post-reconstruction mocks. This
    distribution is very close to Gaussian. The near-zero K-S $D_n$
    value and the corresponding p-value indicate we are 96 per cent
    certain that our $\alpha$ values are drawn from a Gaussian. This
    result justifies our using a Gaussian probability distribution for
    our CMASS distance measure based on the standard deviation of this
    distribution.  }
  \label{fig:pkxiavg}
\end{figure}

We now consider how to combine the power spectrum and correlation
function analyses into one estimate for the cosmological distance
scale.  Before reconstruction, we find $\alpha=1.016 \pm 0.017$ from
the correlation function and $\alpha = 1.022 \pm 0.017$ from the power
spectrum.  After reconstruction, we find $\alpha = 1.024\pm0.016$ from
the correlation function and $\alpha = 1.042\pm 0.016$ from the power
spectrum. These measurements are summarised in
Table~\ref{tab:dr9_alphas}.  These are small differences, but they
approach 1$\sigma$ and hence demand a choice to be made.

Importantly, our analysis of the mock catalogs shows that this level
of scatter is not unusual.  Fig. \ref{fig:pxi_comp} compares the
$\alpha$ and errors recovered from $\xi(r)$ and $P(k)$ in our mocks.
While the results are clearly correlated, there is a notable amount of
scatter: about 2.1 per cent before reconstruction and 1.4 per cent
after reconstruction, when one considers the 16--84 per cent quantile.
The observed small difference in our CMASS measurements before
reconstruction is unusually good; the larger difference after
reconstruction is still common, only 1.2$\sigma$. Note that here we
have not discarded any mocks with weak acoustic signals
(i.e. $\sigma_\alpha>0.07$) as we are only comparing how $\xi$ and $P$
results fare against each other, and not examining details of the
BAO. Fig.~\ref{fig:comp} compares $\alpha$ before and after
reconstruction for both estimators. Again we find that the shifts seem
for the CMASS measurements and not unusual given the results from the
mocks.

The mocks indicate that both estimators are reasonably unbiased: the
mean $\alpha$ recovered is shifted by only 0.4 per cent from the input
value, and some of that shift is the actual physical shift of the
acoustic scale due to non-linear gravity and galaxy clustering bias in
the mocks \citep[e.g.,][]{CroSco08,PadWhi09,Seo10,Meh12}.
Furthermore, neither estimator performs notably better.  As shown in
the lower panels of Fig. \ref{fig:pxi_comp}, the errors formed from
$\chi^2$ by the two estimators are comparable and correlate well on a
mock-by-mock basis.

We therefore conclude that both the correlation function and power 
spectrum estimations of the acoustic scale are appropriate and unbiased, 
but that they include the noise from small scales and shot noise 
differently.  Although the power spectrum and correlation function
are Fourier transform partners in an ideal case, in practice this is
not true: both functions are considered only over a limited domain
and with binning.  

Our choice for a consensus distance scale is therefore to average the
two results.  In our cosmological results, we use only the
reconstructed case.  Reconstruction is expected to improve the
acoustic scale shifts from non-linear structure formation and galaxy
clustering bias, and it does decrease the scatter in $\alpha$ in our
mock catalogs.  The actual CMASS data show little change in recovered
error on $\alpha$ under reconstruction, but this is within the range
of behaviours of the mocks and is not an argument for avoiding the
method.

To estimate the error bars on the averaged $\alpha$ estimator, we use
the rms scatter of the results from the mocks for this estimator. Fig.
\ref{fig:pkxiavg} shows the distribution of average $\alpha$ values
from the mocks. The small K-S $D_n$ and p-value of 0.96 indicate that
we are 96 per cent certain our measured $\alpha$ values follow a
Gaussian distribution. Since we expect our DR9 CMASS $\alpha$
measurement to be Gaussian, using the rms of our mock $\alpha$ values
as our CMASS error estimate is valid. The mocks find a scatter of 1.7
per cent on the average $\alpha$, which is a small decrease from the
scatter of 1.8 per cent on $\alpha$ from $\xi(r)$ and 1.9 per cent on
$\alpha$ from $P(k)$.  This scatter is comparable to the 1.6 per cent
error estimated from $\chi^2(\alpha)$ of the CMASS data from the fits
in both $\xi$ and $P$, which would be another reasonable approach to
adopt an error. We note that since the averaging does produce a small
improvement in errors, it is mildly conservative to neglect any
improvement beyond the errors on the individual estimators.

We expect the systematic errors in this measurement to be much smaller
than the statistical errors.  The marginalization over broadband
nuisance terms in the correlation function and over an arbitrary
broadband spline in the power spectrum gives our template fits great
stability to variations in the template or the possibility of
systematic errors in the galaxy catalog.  In Appendix
\ref{sec:robust_tests}, we investigate a wide range of variations in
the fitting methodologies and reconstruction choices, finding
variations in $\alpha$ of 0.2 per cent or less in all physically
reasonable cases.  It is more difficult to test the effects of
systematics in the galaxy catalog, but if we ignore all of the
corrections for the detected angular systematic variations and repeat
the clustering analysis, the recovered $\alpha$ value changes by only
0.1 per cent despite a notable increase in the large-scale power.
This result is not surprising: systematic errors of this form tend to
produce smooth changes in the power spectrum and hence are captured by
the nuisance terms that we remove in our fitting methods.

\citet{Seo08} and \citet{Xeaip} show that mild alterations in the
cosmology used for the template in the fit changes the recovered
$D_V/r_s$ at a negligible level, $\lesssim 0.001$ for variations
consistent with WMAP, when averaged over a number of mock catalogues.
Cosmologies more exotic than the usual cold dark matter families of
course could open up more dramatic changes; in such cases, one should
plan to repeat the fits both to the CMB and BAO data sets.

Our fitting to the mocks does return a value of $\alpha$ that is 0.004
higher than the input value.  As noted above, non-linear structure
formation and galaxy bias do shift the acoustic scale.  \citet{Seo10}
find shifts of order 0.002 from non-linear structure formation, while
\citet{Meh11} finds a similar level from galaxy clustering bias.
Perturbation theory calculations by \citet{PadWhi09} yield similar
results.  However, reconstruction has been found to remove these
shifts, both in periodic box simulations \citep{Seo10,Meh11} and in
SDSS-II mock catalogs \citep{Pad12}.  It is possible that the BOSS DR9
survey geometry is not large and contiguous enough to remove the
shifts in full, but it is also possible that the shift in the real
data might be different than that in the mocks.  As the shift is
small, we have decided not to subtract it from our fitted values and
instead to consider it as a small systematic uncertainty.

More exotic galaxy bias models could in principle add additional
shifts. However, the only physically motivated model known that
does couple to the acoustic scale is that of \citet{Tse10}, in which
relative velocities between the baryons and dark matter at high
redshift modulate the ability of the smallest halos to trap gas.
Whether this modulation will affect the properties of galaxies a
million times more massive is speculative. \citet{Yoo11} discuss
how the imprint on the acoustic scale in galaxy clustering could
be detected and removed using the three-point function, but we have
not yet investigated this in the CMASS sample.

In summary, our consensus value for the acoustic scale fit is $\alpha
= 1.033 \pm 0.017$.  Our estimates of systematic errors are
significantly smaller than the statistical error and are negligible in
quadrature.  Our best value corresponds to a distance constraint of
\begin{equation}\label{eq:bestDV}
D_V(0.57)/r_s = 13.67 \pm 0.22.
\end{equation}
We adopt this as our primary result and use it for all of 
our cosmological interpretations and comparisons to other work. For
easy reference, the key values of $\alpha$ are summarised in Table
\ref{tab:key_alphas}.  For the fiducial sound horizon of 153.19~Mpc,
Eq. (\ref{eq:bestDV}) corresponds to $D_V(0.57) = 2094\pm34$~Mpc.

\section{The BAO distance ladder}
\label{sec:ladder}
\subsection{Comparison to Previous BAO Measurements}

\begin{figure}
  \centering
  \resizebox{0.95\columnwidth}{!}{\includegraphics{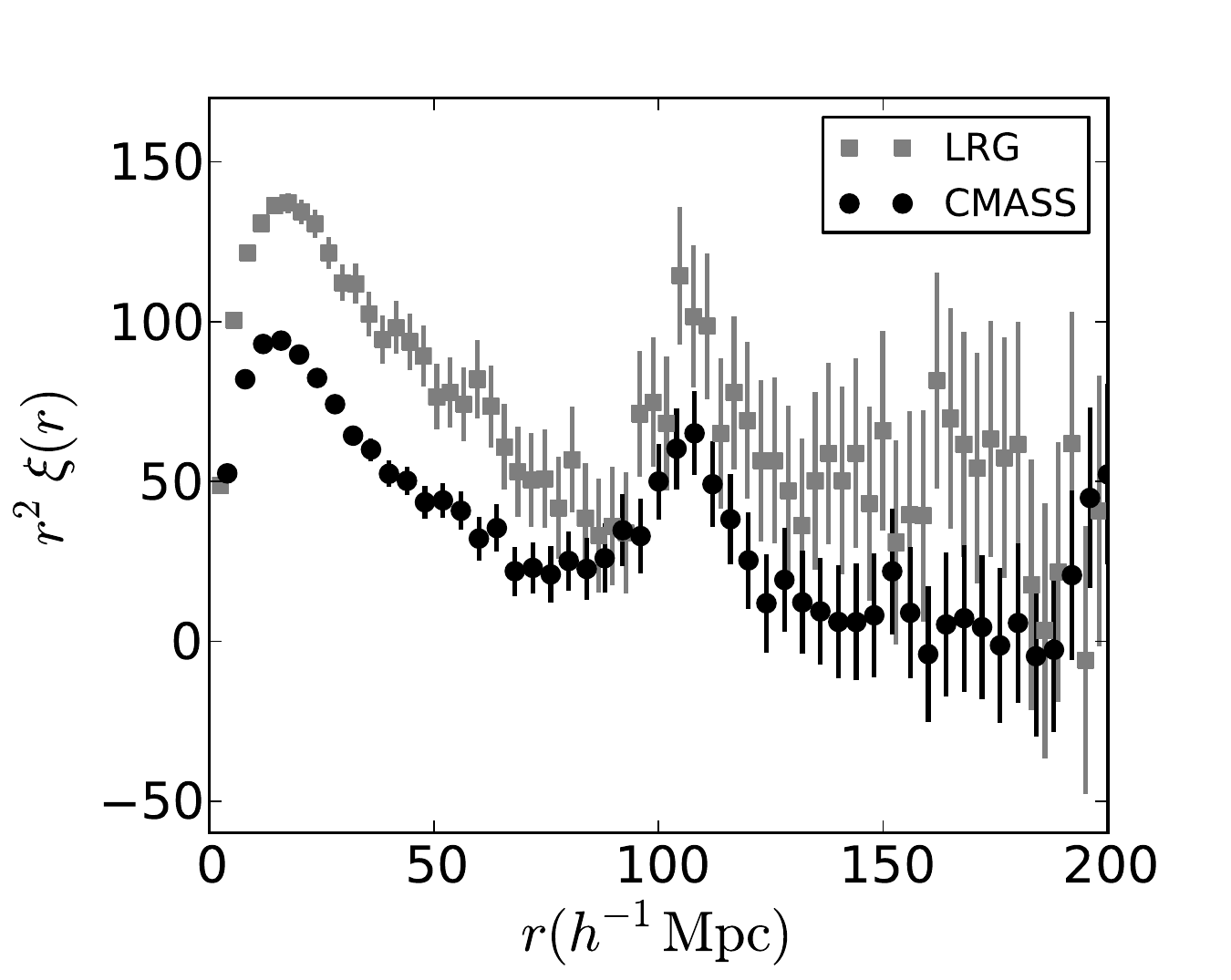}}
  \caption{ The correlation function measured from CMASS data (black
    circles) versus that from SDSS-II LRG data (grey squares) as shown
    in \protect\citet{Pad12}.  The vertical offset is due to the
    difference in galaxy bias between the samples; on average the SDSS-II LRGs
    are more luminous and reside in more massive halos.  These two
    analyses used slightly different fiducial cosmologies; we have
    scaled the SDSS-II LRG points to cosmology of this paper. One can
    clearly see that the acoustic peak is located at the same position
    in both datasets.  As an aside, we note that the difference in the
    size of the errors has several contributions in addition to sample
    size: the CMASS sample has a higher number density and less shot
    noise, the CMASS sample used $4h^{-1}$~Mpc bins, whereas the
    SDSS-II analysis used $3h^{-1}$~Mpc bins, and the linear scaling
    of the vertical axis causes equal fractional errors to appear
    larger in the higher bias sample.
  \label{fig:dr79_rec}
}
\end{figure}

\begin{figure}
  \centering
  \resizebox{0.95\columnwidth}{!}{\includegraphics{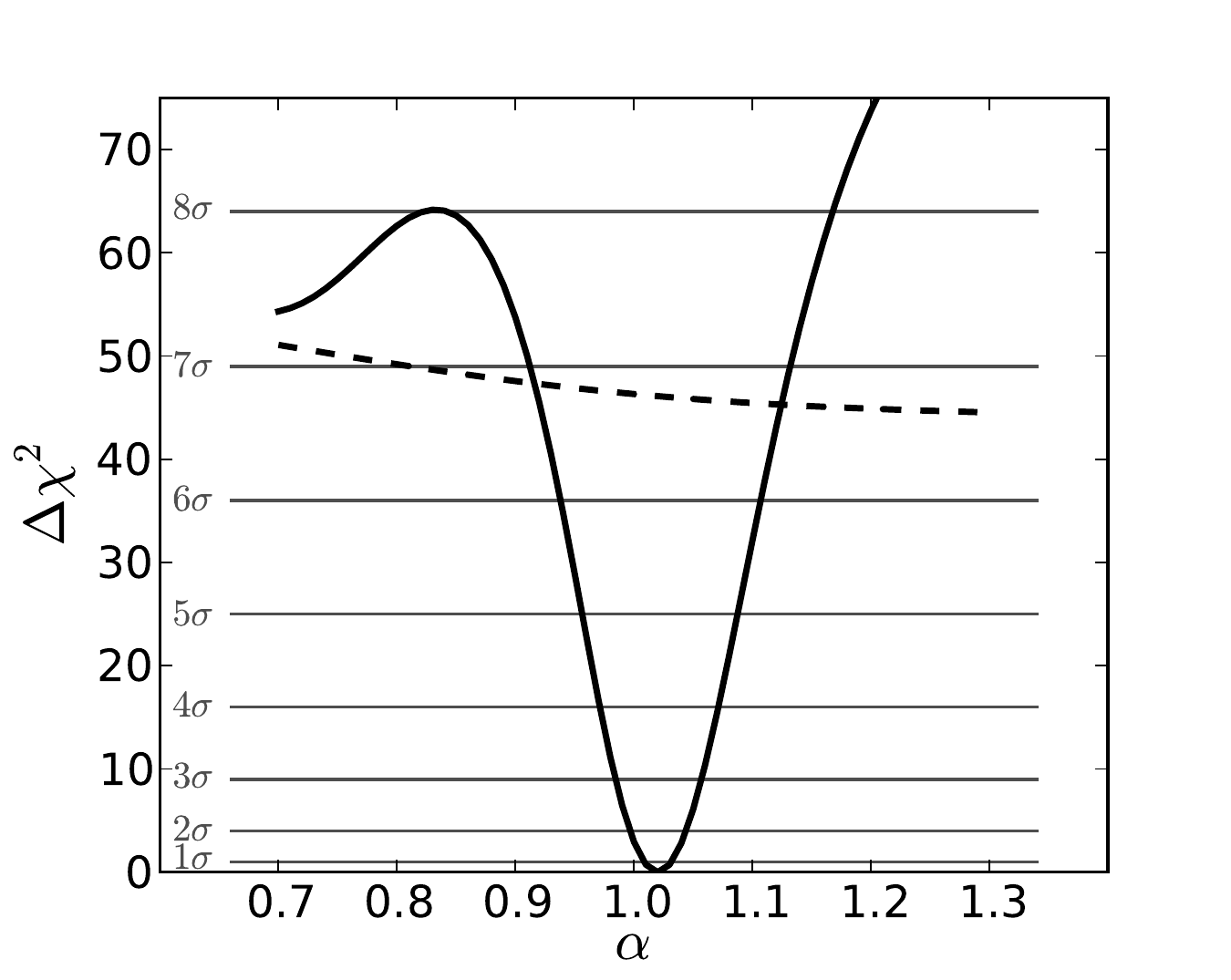}}
  \caption{The total significance of the BAO feature, combining the
    CMASS and SDSS-II LRG results, both after reconstruction. This
    figure is analogous to Fig.~\ref{fig:DR7_9chi2} and indicates that
    in the combined CMASS and LRG data sets, we have detected the
    acoustic peak at greater than $6.5\sigma$, with the local minima
    extending to the $\sim8\sigma$ level.
  \label{fig:BAOsig}
}
\end{figure}

\begin{figure}
  \centering
  \resizebox{0.9\columnwidth}{!}{\includegraphics{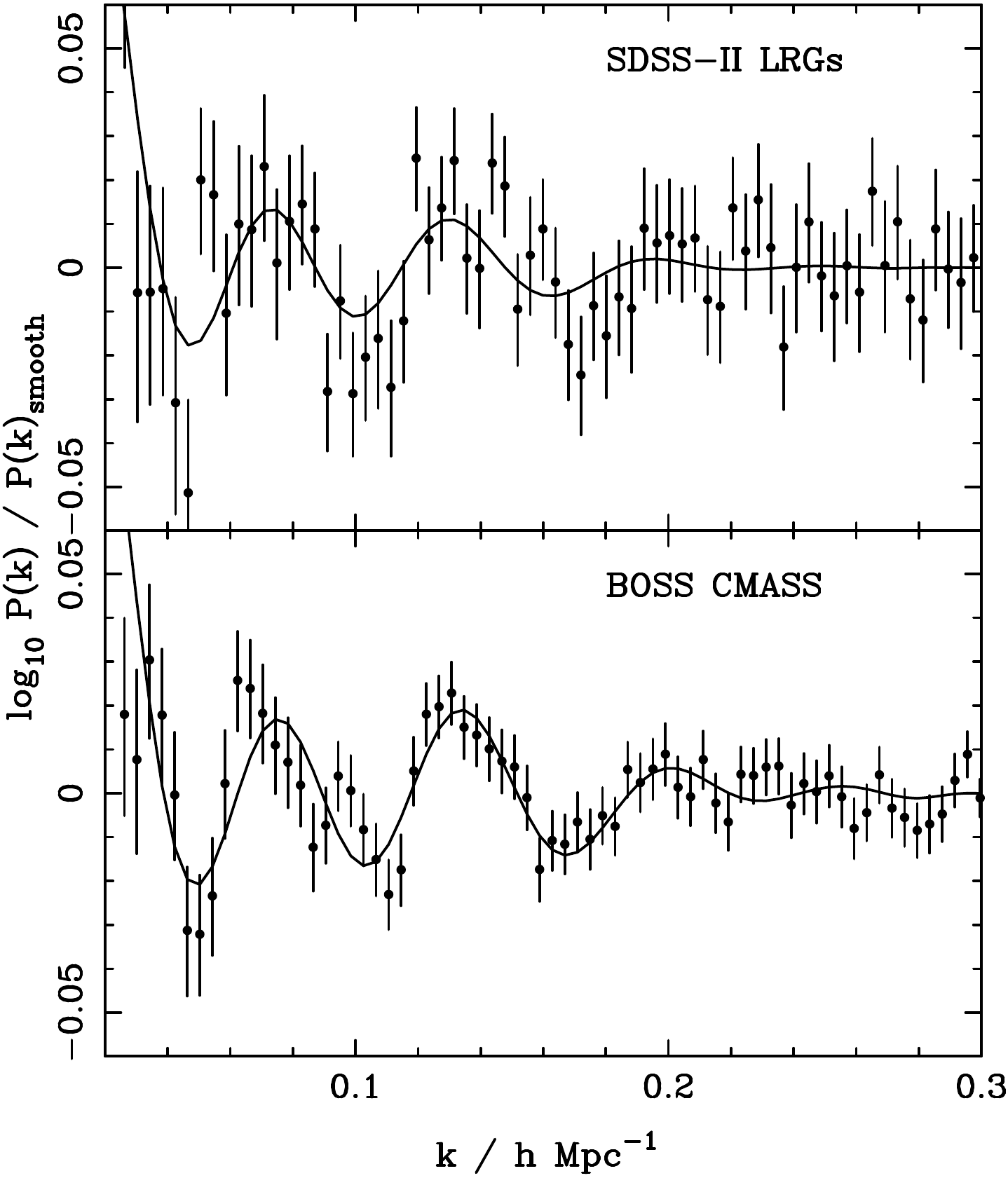}}
  \caption{
  BAO in the power spectrum measured from the reconstructed CMASS
  data (solid circles with $1\sigma$ errors, lower panel) compared
  with unreconstructed BAO recovered from the SDSS-II LRG data
  (solid circles with $1\sigma$ errors, upper panel). Best-fit
  models are shown by the solid lines. The SDSS-II data are based
  on the sample and power spectrum calculated in \citet{Rei10} and
  analysed by \citet{Per10}; it has been shifted to match the
  fiducial cosmology assumed in this paper. Clearly the CMASS errors
  are significantly smaller than those of the SDSS-II data, and we
  also benefit from reconstruction, reducing the the BAO damping
  scale.}
  \label{fig:dr79_pk}
\end{figure}

\begin{figure}
  \centering
  \resizebox{0.95\columnwidth}{!}{\includegraphics{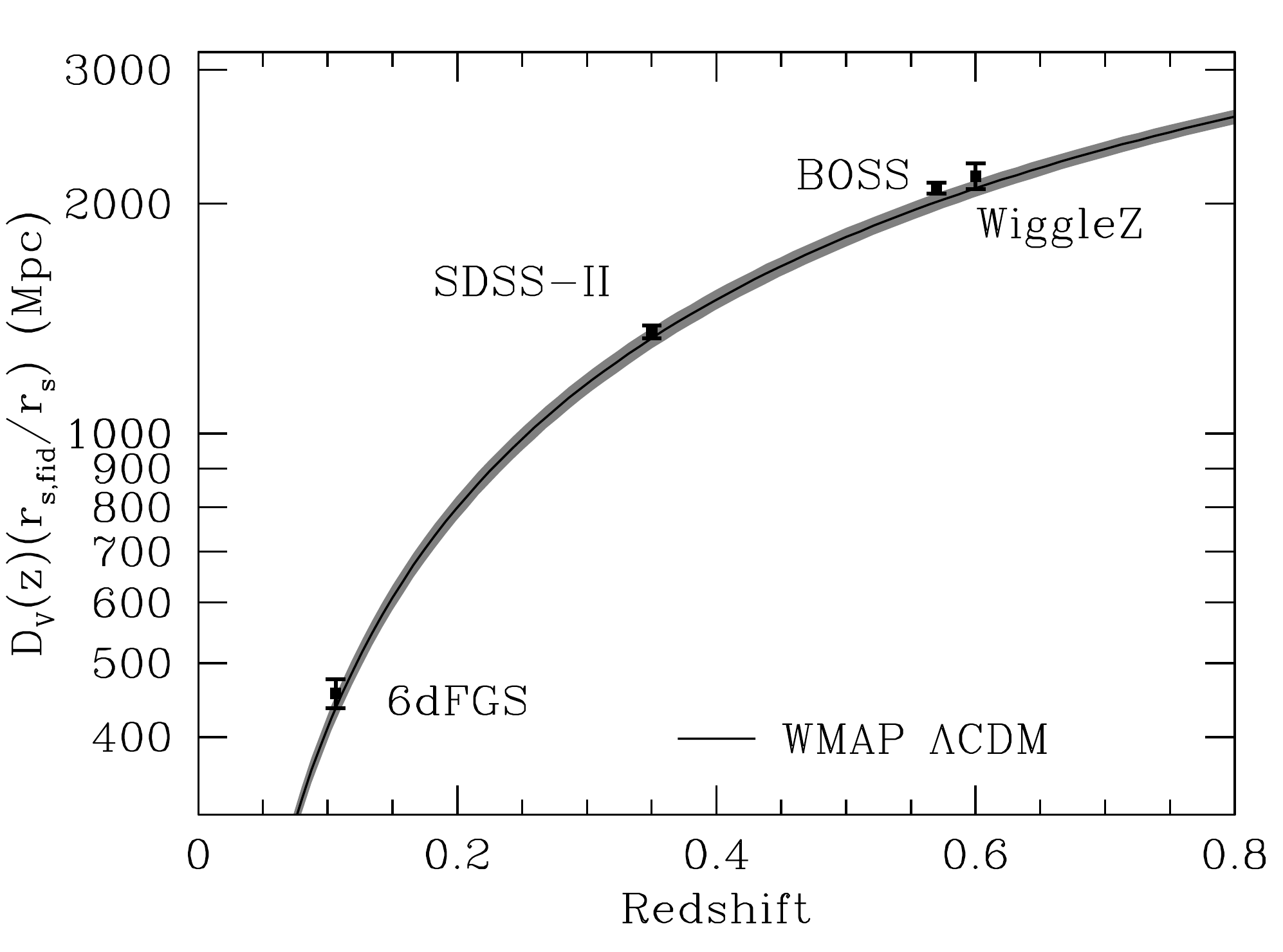}}
  \caption{A plot of the distance-redshift relation from various BAO
    measurements from spectroscopic data sets.  We plot $D_V(z)/r_s$
    times the fiducial $r_s$ to restore a distance.  Included here are
    this CMASS measurement, the 6dF Galaxy Survey measurement at
    $z=0.1$ \citep{Beutler11}, the SDSS-II LRG measurement at $z=0.35$
    \citep{Pad12,Xeaip,Meh12}, and the WiggleZ measurement at $z=0.6$
    \citep{Bla11a}.  The latter is a combination of 3 partially
    covariant data sets.  The grey region is the 1~$\sigma$ prediction
    from WMAP under the assumption of a flat Universe with a
    cosmological constant \citep{komatsu11}.  The agreement between
    the various BAO measurements and this prediction is
    excellent. \label{fig:boss_logDV} }
\end{figure}

\begin{figure}
  \centering
  \resizebox{0.95\columnwidth}{!}{\includegraphics{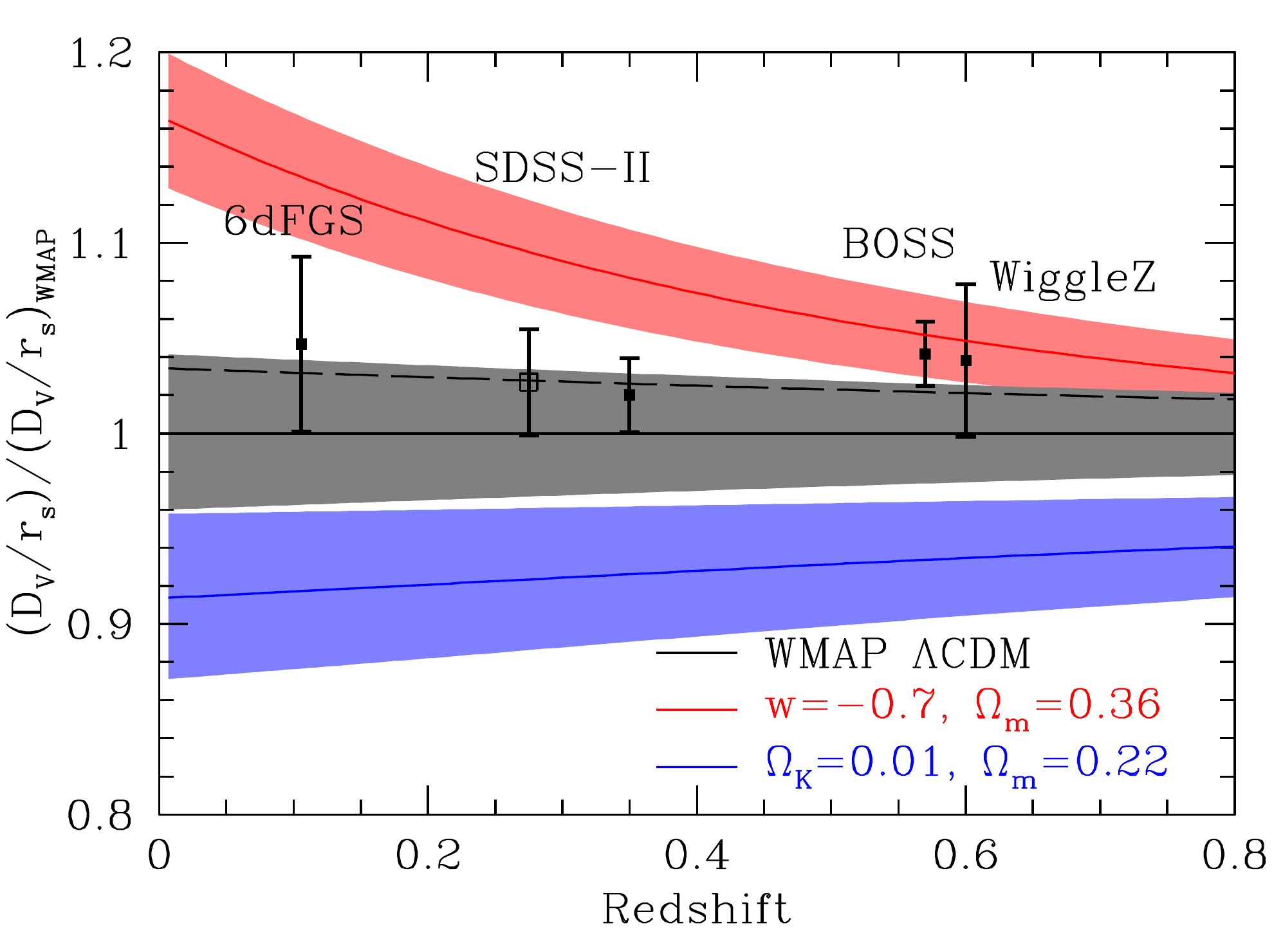}}
  \caption{The BAO distance-redshift relation divided by the best-fit
    flat, $\Lambda$CDM prediction from WMAP ($\Omega_m=0.266$,
    $h=0.708$; note that this is slightly different from the adopted
    fiducial cosmology of this paper).  The grey band indicates the
    1~$\sigma$ prediction range from WMAP \citep{komatsu11}.  In addition to the SDSS-II
    LRG data point from \citet{Pad12}, we also show the result from
    \citet{Per10} using a combination of SDSS-II DR7 LRG and Main
    sample galaxies as well as 2dF Galaxy Redshift Survey data;
    because of the overlap in samples, we use a different symbol.
    The BAO results agree with the best-fit WMAP model at the few
    percent level.  If $\Omega_m h^2$ were 1 $\sigma$ higher than the
    best-fit WMAP value, then the prediction would be the upper edge
    of the grey region, which matches the BAO data very closely.  For
    example, the dashed line is the best-fit CMB+LRG+CMASS flat
    $\Lambda$CDM model from \S~\protect\ref{sec:params}, which clearly
    is a good fit to all data sets.  Also shown are the predicted
    regions from varying the spatial curvature to $\Omega_K=0.01$
    (blue band) or varying the equation of state to $w=-0.7$ (red
    band).
 \label{fig:boss_DVfid}
}
\end{figure}

In the last few years, acoustic scale results have been obtained with
a variety of data sets over a considerable range of redshift.  We now
focus on the comparison between our CMASS DR9 results and past work.  

First, we compare the correlation function at $z=0.57$ from CMASS with
that obtained at $z=0.35$ by the reconstruction analysis of SDSS-II
LRGs presented in \citet{Pad12}.  Fig. \ref{fig:dr79_rec} shows these
two correlation functions as $r^2 \xi(r)$.  The two samples involve
different average masses of galaxies and redshifts, and hence have a
different amplitude of clustering, leading to a vertical offset.  Both
correlation functions use our fiducial $\Lambda$CDM cosmology.  Given
this choice of distance-redshift relation, one can see that the
acoustic peaks are in excellent agreement.

Fig. \ref{fig:BAOsig} shows combined significance of the acoustic peak
detection in $\xi(r)$.  In combining the constraints on CMASS DR9 with
the SDSS-II LRG DR7 data, we neglect the slight overlap in effective
volumes when using these data in cosmological constraints. The LRG
data from\citet{Pad12} cover only the NGC which result in 2496 sq deg
of overlapping area with CMASS over the redshift interval of
$0.43<z<0.47$. We find this is less than 9 per cent of the effective
volume of our CMASS sample, and less than 5 per cent overlap with the
LRG effective volume (fractionally less since the LRG DR7 data cover a
larger area). This is consequently a good but not perfect assumption.
Combining the two correlation functions assuming independence rejects
models without acoustic oscillations at $\Delta\chi^2\approx45$ or
6.7~$\sigma$.  Trying to place the acoustic peak at other nearby
locations and particularly at smaller scales is rejected at
8~$\sigma$.

Fig. \ref{fig:dr79_pk} repeats this comparison with the power spectrum
from the SDSS-II LRG analysis presented in \citet{Rei10} and
\citet{Per10}.  This analysis did not use reconstruction, but one can
see good agreement in the BAO and significant improvement in the error
bars with the CMASS sample.

In Fig. \ref{fig:boss_logDV}, we plot $D_V(z)$ constraints from
measurements of the BAO from various spectroscopic samples.  In
addition to the SDSS-II LRG value at $z=0.35$ \citep{Pad12} and the
CMASS consensus result at $z=0.57$, we also plot the $z=0.1$
constraint from the 6dF Galaxy Survey (6dFGS) \citep{Beutler11} and a
$z=0.6$ constraint from the WiggleZ survey \citep{Bla11a}.  WiggleZ
quotes BAO constraints in 3 redshift bins, but these separate
constraints are weaker and there are significant correlations between
the redshift bins.  We choose here to plot their uncorrelated data
points for $0.2<z<1.0$.  Each data point here is actually a constraint
on $D_V(z)/r_s$, and we have multiplied by our fiducial $r_s$ to get a
distance.

As described further in \citet{Meh12}, the WMAP curve on this graph is
a prediction, not a fit, assuming a flat $\Lambda$CDM cosmology.  For
each value of $\Omega_m h^2$ and $\Omega_b h^2$, one can predict a
sound horizon, and the angular acoustic scale measured by WMAP plus
the assumptions about spatial curvature and dark energy equation of
state then provide a very precise breaking of the degeneracy between
$\Omega_m$ and $H_0$ and hence a unique $D_V(z)/r_s$.  Taking the
1$\sigma$ range of $\Omega_m h^2$ and $\Omega_b h^2$ produces the grey
band in Fig. \ref{fig:boss_logDV}.  There is excellent agreement
between all four BAO measurements and the WMAP $\Lambda$CDM
prediction.

Following \citet{Meh12}, we divide the distance measurements by the
best-fit WMAP prediction ($\Omega_m=0.266$, $h=0.708$) to yield
Fig. \ref{fig:boss_DVfid}.  Focusing first on the data points and the
grey $\Lambda$CDM region, the data points are consistent with the WMAP
prediction, but tend to lie closer to the 1~$\sigma$ upward trend in
WMAP, toward higher $\Omega_m h^2$.  In other words, if the WMAP value
for $\Omega_m h^2$ were 1~$\sigma$ higher, then all of the BAO points
would be in superb consistency with themselves and the CMB under a
flat $\Lambda$CDM cosmology; recall that the swath of grey models is a
nearly one-parameter family so all redshifts move together.  We also
include here the BAO measurement from \citet{Per10}.  Because of the
overlap in sample with the LRG analysis of \citet{Pad12}, we use a
different symbol for this measurement.  The correlation function
analysis from \citet{Kaz10} gives similar agreement. The CMASS BAO
value is in perfect agreement with the WiggleZ measurement
\citep{Bla11a}.  The WiggleZ acoustic scale error is 3.9 per cent
(using their constraint on $A(z)$), so the CMASS DR9 error of 1.7 per
cent represents a 5-fold improvement in the variance.

Also shown in Fig. \ref{fig:boss_DVfid} is how the WMAP prediction
changes as one varies the assumptions about dark energy and spatial
curvature.  For any specific choice of $\Omega_K$ and $w(z)$, WMAP
predicts a narrow region set by the range of $\Omega_m h^2$ and
$\Omega_b h^2$.  Here we present the case of $\Omega_K=0.01$ with a
cosmological constant and a flat Universe with $w=-0.7$; both
produce large differences from the observations.

There have also been BAO measurements at $z\approx 0.55$ using
photometric samples \citep{Pad07,Car11,Seo12}.  These measure the
angular diameter distance $D_A(z)$ rather than $D_V(z)$.
\citet{Car11} measured $(1+z)D_A/r_s=14.7\pm 1.4$ at $z=0.55$ using
the angular correlation function \citep{Crocce11} of SDSS DR7 imaging
data \citep{Abazajian09}.  \citet{Seo12} on the other hand measured
$(1+z)D_A/r_s=14.18\pm 0.63$ at $z=0.54$ using the angular power
spectrum \citep{Ho12} of the SDSS-III DR8 imaging data \citep{DR8}
\footnote{The DR8 measurement used $10\,000$ square degrees of the sky
  that includes the coverage of the CMASS DR9 sample. Therefore the
  overlap in volume between the two samples is approximately 30 per
  cent.}.  Despite the different estimators of two-point statistics
used, both results consistently show a larger distance scale than the
prediction of the WMAP best-fit by $1\sigma$ and $1.4\sigma$,
respectively.  To compare these values with spectroscopic
measurements, we correct the difference between $D_A(z)$ and $D_V(z)$
using the $H(z)$ calculated from the fiducial cosmology, while
translating the percentage error on $D_A(z)$ to be the percentage
error on $D_V(z)$. The deviation from the WMAP prediction will be
reduced due to using the fiducial $H(z)$ during this transformation;
the correction yields $D_V(z=0.55)/r_s=13.6\pm 1.3$ and
$D_V(z=0.54)/r_s=13.22\pm 0.58$, respectively. Extrapolating these
values from $z\approx 0.55$ to $z=0.57$ assuming the fiducial
cosmology gives $D_V(z=0.57)/r_s=14.0\pm 1.4$ for \citet{Car11} and
$D_V(z=0.57)/r_s=13.81\pm 0.61$ for \citet{Seo12}.  Therefore the
photometric BAO measurements show an excellent agreement with
$D_V(z=0.57)/r_s=13.67\pm0.22$ from the CMASS measurement.  It is
clear that these photometric BAO measurements also fall into the
general upward trend relative to the WMAP prediction.

Similarly, there have been spectroscopic BAO measurements that
attempt to separate the line-of-sight and transverse clustering so
as to measure $H(z)$ and $D_A(z)$ separately \citep{Oku08,CW11}.
These measurements are at lower redshift and hence not directly
comparable to our CMASS result.  However, the agreement in the recovered
cosmological parameters is good.

In summary, a precise view of the Hubble diagram from baryon
acoustic oscillations over the range $0.1<z<0.6$ is taking shape.
These measurements appear highly consistent with the standard
cosmological model.

\subsection{Comparison to Supernova and Direct $H_0$ Measurements}
\label{sec:sn}

\begin{figure}
  \centering
  \resizebox{0.9\columnwidth}{!}{\includegraphics{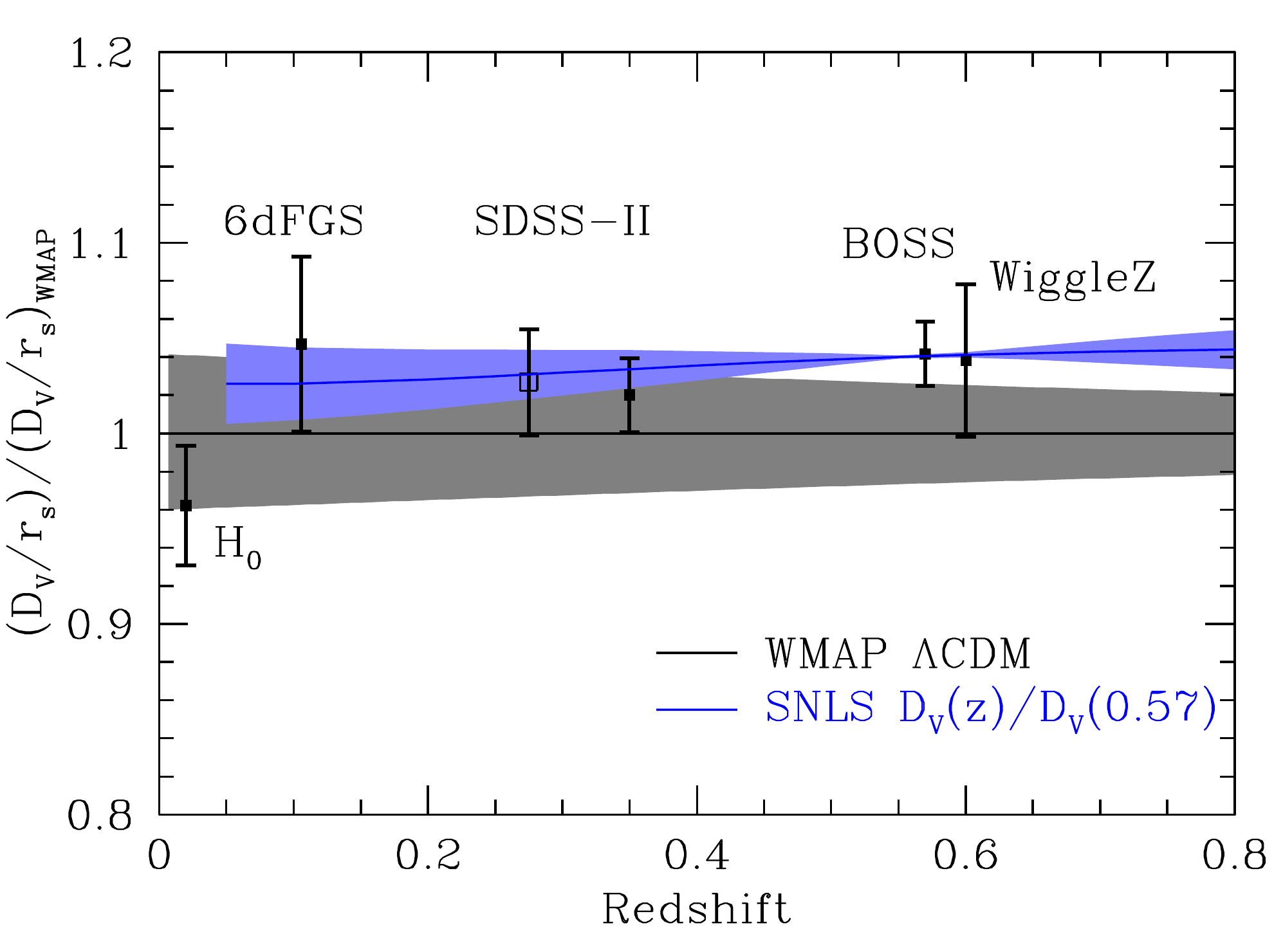}}
  \caption{
The BAO distance-redshift relation divided by the best-fit flat,
$\Lambda$CDM prediction from WMAP (grey band), overlaid with the distance ratio
measurements from the 3-year Supernova Legacy Survey data \citep{conley11} (blue band).
Here, we have fit the SNe data using Markov Chain Monte Carlo in a 
parameter space of $w_0$, $w_a$, and spatial curvature.  From the 
chain, we measure the mean and standard deviation on $D_V(z)/D_V(0.57)$.
We normalise to the CMASS value when plotting on this graph; the supernovae
are providing only a relative distance measure.  There is excellent
agreement between the supernova distance-redshift relation and the measurements
from all BAO experiments.  Also shown is the direct $H_0$ value from
\citet{riess11}; there is mild tension between this measurement and
the BAO and SNe data.
  \label{fig:boss_sn}
}
\end{figure}

We next offer further comparisons to the distance-redshift relation
from Type Ia supernovae and direct $H_0$ measurements.  Type Ia
supernovae can be used to measure relative luminosity distances.  We
use the 3-year Supernova Legacy Survey (SNLS3) results from
\citet{conley11}, including their systematic error treatment.
Comparing the supernova distance-redshift relation to that of the BAO
requires some procedure to bin the individual data points or to fit a
model (see \citealt{Lamp10} for a discussion of possible procedures).
Due to the systematic errors and the absolute distance offset,
combining the supernova data into redshift bins would necessarily
yield correlated results.  Moreover, the supernovae results constrain
the luminosity distance $D_L$, not the $D_V$ measured by the BAO.
$D_V(z)$ requires $H(z)$ as well, which is related to a derivative of
$D_L(z)$, with a mild dependence on spatial curvature.  The need for
this derivative recommends fitting a model instead of binning.

We opt to fit a parametric model.  We run Markov Chain Monte Carlo
(described in the next section) for a model space including spatial
curvature and an equation of state $w(z) = w_0+w_a(1-a)$.  We use CMB
data from WMAP in addition to the SNLS3 data; we opt to include CMB
measurements so that the dependence on spatial curvature remains mild.
CMB data alone would have 3 dimensions of significant degeneracies in
this parameter space; the supernova data will attempt to break these
degeneracies.  We then use the Markov Chain to infer the constraints
on ratios of $D_V(z)$ to $D_V(0.57)$.  This tells us what the
supernova distance-redshift relation predicts for the ratio between
two BAO measurements, subject to the regularisation of the supernova
data implied by the parametric cosmological model that we have chosen.
In effect, we have fit a three-parameter distance-redshift relation to
the supernova data and then used this to infer $D_V(z)$ from the
observed distance moduli.  As a technical note, these results will
differ slightly from those from MCMC that combine BAO and SNe data,
because the BAO data will limit the exploration of the
distance-redshift degeneracy space.

Comparing $z=0.57$ to $z=0.35$, we find that the supernovae measure
$D_V(0.35)/D_V(0.57) = 0.6579\pm0.0063$, a 1.0 per cent inference.
From the CMASS measurement of $D_V(0.57)/r_s = 13.67\pm0.22$, this
predicts $D_V(0.35)/r_s=8.99\pm0.14\pm0.09$, where the first error
arrises from the CMASS error and the second error is due to the error
in the supernova propagation.  This prediction can be compared to the
\citet{Pad12} measurement from SDSS-II LRG of $D_V(0.35)/r_s =
8.88\pm0.17$.  Hence, the ratio of these two BAO measurements agrees
well with the supernova data.

Similarly, at $z=0.10$, we find that the supernovae measure
$D_V(0.10)/D_V(0.57) = 0.2018\pm 0.0038$, a 1.9 per cent inference.
The combination with the CMASS data would then predict 
$D_V(0.10)/r_s = 2.759\pm0.044\pm0.052$, following the notation from
the previous paragraph.  This can be compared to the 6dFGS measurement
of $2.81\pm0.13$, where we have scaled from $z=0.106$ to $z=0.1$.
Again, the ratio of the BAO measurements agrees well with the supernova
distance scale.

We present these results graphically in Fig. \ref{fig:boss_sn}.  Here,
we normalise the $D_V(z)$ from the Markov Chain at $z=0.57$ and
consider the mean and 1~$\sigma$ range explored by the chain.  Of
course, one might have chosen to normalise at another redshift; this
version presents how well the CMASS BAO data can be transferred to
other redshifts.  One can see the excellent agreement with all of the
other BAO results.  One also sees that the supernova relative distance
scale is still more constraining than the BAO relative distance scale,
by a factor of order 2-3.  Of course, the supernovae do not provide an
absolute distance scale; this plot is indicating only their constraint
on the slope of the distance-redshift relation. In the future, we many
wish to combine SNe and BAO distances to further constrain the
reciprocity, or distance-duality, relation which is a generic
prediction of any theory of gravity where photons follow null
geodesics \citep{Bassett04,Lamp10}.

Finally, considering the constraint on the Hubble constant, the
supernovae predict a tight relation between $D_V(0.57)$ and $1/H_0$.
We quote this quantity as $H_0 D_V(0.57)/(0.57c) = 0.844$ with a 2.3
per cent error.  Using this result, the CMASS BAO data with a sound
horizon given by the fiducial cosmological model predicts $H_0 =
68.9$~km/s/Mpc, with 1.7 per cent error from the $z=0.57$ calibration
and 2.3 per cent error from the supernova transfer to $z=0$.  This
value is in mild tension with the direct measurement of $H_0 =
73.8\pm2.4$~km/s/Mpc using the NGC 4258 maser and HST near-IR
observations of Cepheid variable stars \citep{riess11}.
Fig. \ref{fig:boss_sn} plots this measurement, but we remind readers
that the placement of this point assumes the fiducial value of $r_s$,
which creates a 1 per cent uncertainty not included in the errors. We
will quantify this point further in the next section, using a full
Markov Chain.

\section{Cosmological parameters}
\label{sec:params}
\begin{table*}
\caption{\label{tab:datasets} List of the datasets used in the MCMC
chains for measuring cosmological parameters.} \begin{tabular}{lll}
\hline
Dataset Name & Description & References \\
\hline
CMB & WMAP7 data & \cite{komatsu11} \\
LRG & SDSS-II Luminous Red Galaxies & \cite{Pad12} \\
6dFGRS & 6dF Galaxy Redshift Survey sample & \cite{Beutler11} \\
CMASS & SDSS-III Data Release 9 Constant Mass Sample & This paper \\
SN & 3-year Supernova Legacy Survey compilation (SNLS3) & \cite{conley11} \\
H$_0$ & Direct Hubble Constant Measurement & \cite{riess11} \\
\end{tabular}
\end{table*}

\begin{table*}
  \caption{\label{tab:cosmoparams} The first two columns show the cosmological
    model and the data set in each case. The remaining columns show the
    cosmological parameter values estimated from the mean of the posterior distribution.
    The uncertainties, estimated from the second moments of the distribution are in parentheses
    and are the change to the last significant figures. Eg. $0.268(29)$ represents $0.268 \pm 0.029$. 
    Empty values correspond to the cases in which the parameter is
    kept fixed to its fiducial value, i.e. $\Omega_{\rm K}=0, w_0=-1, w_a=0$.
  }

\begin{tabular}{llllllll}
\hline
Cosmological Model & Data Sets$^1$ & $\Omega_{\rm m} h^{2}$ & $\Omega_{\rm m}$ & $H_{0}$ & $\Omega_{\rm K}$ & $w_{0}$ & $w_{a}$ \\
& & & & km/s/Mpc & & & \\
\hline
$\Lambda$CDM & CMB & 0.1341(56) & 0.268(29) & 71.0(26) & \nodata & \nodata & \nodata \\
$\Lambda$CDM & CMB+CMASS & 0.1392(36) & 0.298(17) & 68.4(13) & \nodata & \nodata & \nodata \\
$\Lambda$CDM & CMB+LRG & 0.1362(33) & 0.280(14) & 69.8(12) & \nodata & \nodata & \nodata \\
$\Lambda$CDM & CMB+LRG+CMASS & 0.1384(31) & 0.293(12) & 68.8(10) & \nodata & \nodata & \nodata \\
$\Lambda$CDM & CMB+LRG+CMASS+6dF & 0.1384(31) & 0.293(12) & 68.7(10) & \nodata & \nodata & \nodata \\
$\Lambda$CDM & CMB+LRG+CMASS+SN & 0.1373(30) & 0.287(11) & 69.2(10) & \nodata & \nodata & \nodata \\
$\Lambda$CDM & CMB+LRG+CMASS+SN+6dF & 0.1373(30) & 0.288(11) & 69.1(10) & \nodata & \nodata & \nodata \\
\hline
oCDM & CMB & 0.1344(55) & 0.423(175) & 60.0(123) & -0.039(44) & \nodata & \nodata \\
oCDM & CMB+CMASS & 0.1340(53) & 0.299(16) & 67.0(15) & -0.008(5) & \nodata & \nodata \\
oCDM & CMB+LRG & 0.1333(53) & 0.278(15) & 69.3(16) & -0.004(5) & \nodata & \nodata \\
oCDM & CMB+LRG+CMASS & 0.1336(51) & 0.288(12) & 68.1(11) & -0.006(5) & \nodata & \nodata \\
oCDM & CMB+LRG+CMASS+6dF & 0.1336(50) & 0.288(12) & 68.1(11) & -0.006(5) & \nodata & \nodata \\
oCDM & CMB+LRG+CMASS+SN & 0.1322(51) & 0.284(12) & 68.3(12) & -0.006(5) & \nodata & \nodata \\
oCDM & CMB+LRG+CMASS+SN+6dF & 0.1321(50) & 0.284(12) & 68.2(11) & -0.007(5) & \nodata & \nodata \\
\hline
$w$CDM & CMB & 0.1342(58) & 0.263(105) & 75.4(138) & \nodata & -1.12(41) & \nodata \\
$w$CDM & CMB+CMASS & 0.1358(59) & 0.323(43) & 65.4(60) & \nodata & -0.87(24) & \nodata \\
$w$CDM & CMB+LRG & 0.1349(57) & 0.285(25) & 69.0(39) & \nodata & -0.97(17) & \nodata \\
$w$CDM & CMB+LRG+CMASS & 0.1370(58) & 0.294(27) & 68.6(44) & \nodata & -0.99(21) & \nodata \\
$w$CDM & CMB+LRG+CMASS+6dF & 0.1363(51) & 0.298(20) & 67.8(31) & \nodata & -0.95(15) & \nodata \\
$w$CDM & CMB+LRG+CMASS+SN & 0.1399(37) & 0.280(13) & 70.8(18) & \nodata & -1.09(8) & \nodata \\
$w$CDM & CMB+LRG+CMASS+SN+6dF & 0.1396(37) & 0.282(13) & 70.4(17) & \nodata & -1.08(8) & \nodata \\
\hline
o$w$CDM & CMB+LRG+CMASS & 0.1345(53) & 0.250(42) & 74.1(70) & -0.008(5) & -1.31(34) & \nodata \\
o$w$CDM & CMB+LRG+CMASS+6dF & 0.1334(52) & 0.271(31) & 70.5(43) & -0.007(6) & -1.14(23) & \nodata \\
o$w$CDM & CMB+CMASS+SN & 0.1338(53) & 0.280(17) & 69.2(21) & -0.009(5) & -1.10(8) & \nodata \\
o$w$CDM & CMB+LRG+CMASS+SN & 0.1337(53) & 0.275(14) & 69.8(18) & -0.007(5) & -1.09(8) & \nodata \\
o$w$CDM & CMB+LRG+CMASS+SN+6dF & 0.1333(52) & 0.276(13) & 69.6(17) & -0.008(5) & -1.09(8) & \nodata \\
\hline
$w_0w_a$CDM & CMB+LRG+CMASS & 0.1377(58) & 0.282(52) & 70.7(68) & \nodata & -1.11(51) & 0.18(122)* \\
$w_0w_a$CDM & CMB+LRG+CMASS+6dF & 0.1369(55) & 0.292(41) & 68.9(48) & \nodata & -1.02(42) & 0.44(113)* \\
$w_0w_a$CDM & CMB+CMASS+SN & 0.1389(62) & 0.281(17) &70.3(23) & \nodata & -1.07(16) &-0.85(96)* \\
$w_0w_a$CDM & CMB+LRG+CMASS+SN & 0.1392(59) & 0.280(14) & 70.6(19) & \nodata & -1.08(15) & 0.10(87) \\
$w_0w_a$CDM & CMB+LRG+CMASS+SN+6dF & 0.1385(58) & 0.281(14) & 70.2(17) & \nodata & -1.08(15) & 0.08(81) \\
\hline
o$w_0w_a$CDM & CMB+LRG+CMASS & 0.1347(54) & 0.263(54) & 72.7(79) & -0.009(6) & -1.13(54) & -0.70(139)* \\
o$w_0w_a$CDM & CMB+LRG+CMASS+6dF & 0.1341(53) & 0.284(40) & 69.2(50) & -0.009(7) & -0.93(41) & -0.93(130)* \\
o$w_0w_a$CDM & CMB+CMASS+SN & 0.1344(54) & 0.280(17) & 69.5(21) & -0.012(6) & -0.91(17) & -1.31(102)* \\
o$w_0w_a$CDM & CMB+LRG+CMASS+SN & 0.1348(53) & 0.277(14) & 69.8(18) & -0.012(5) & -0.89(16) & -1.44(93)* \\
o$w_0w_a$CDM & CMB+LRG+CMASS+SN+6dF & 0.1343(52) & 0.278(14) & 69.5(17) & -0.012(5) & -0.88(15) & -1.40(94)* \\
o$w_0w_a$CDM & CMB+LRG+CMASS+SN+H0 & 0.1364(51) & 0.270(12) & 71.1(15) & -0.010(5) & -0.93(16) & -1.46(95)* \\
o$w_0w_a$CDM & CMB+LRG+CMASS+SN+H0+6dF & 0.1359(50) & 0.270(12) & 70.8(14) & -0.010(5) & -0.93(16) & -1.39(96)* \\
\hline
\end{tabular}
\newline
\begin{flushleft} 
$^{1}$ {\footnotesize The CMB+LRG values are taken from \citet{Meh12}.
* {\footnotesize Datasets allow chains to explore parameter space outside the prior of $-3.0 \leq w_a \leq 2.0$.}
}
\end{flushleft}
\end{table*}

\begin{figure}
\centering
\resizebox{0.9\columnwidth}{!}{\includegraphics{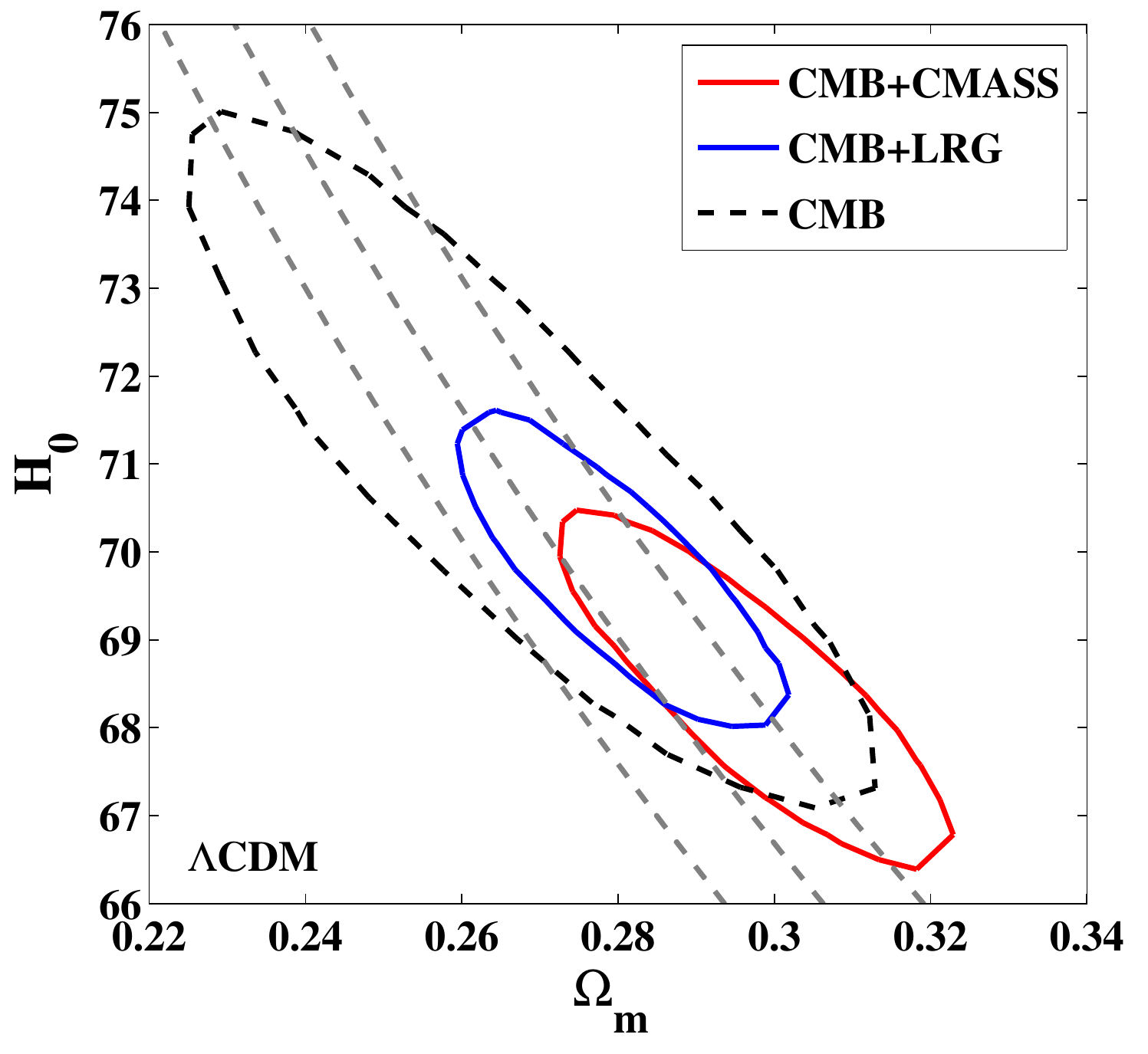}}
\caption{$68$ per cent contours for $H_0$ vs $\Omegam$ in the
  $\Lambda$CDM cosmological model. The CMASS DR9 BAO data improve our
  measurements of $H_0$ and $\Omegam$, and are consistent with the
  SDSS-II LRG measurements. The dashed grey lines are lines of
  constant $\Omegam h^2$ using the WMAP7 values and modulated by
  $1\sigma$ ($\Omegam h^2 =
  0.1334^{+0.0056}_{-0.0055}$).} \label{fig:LCDM}
\end{figure}

\begin{figure}
\centering
\resizebox{0.9\columnwidth}{!}{\includegraphics{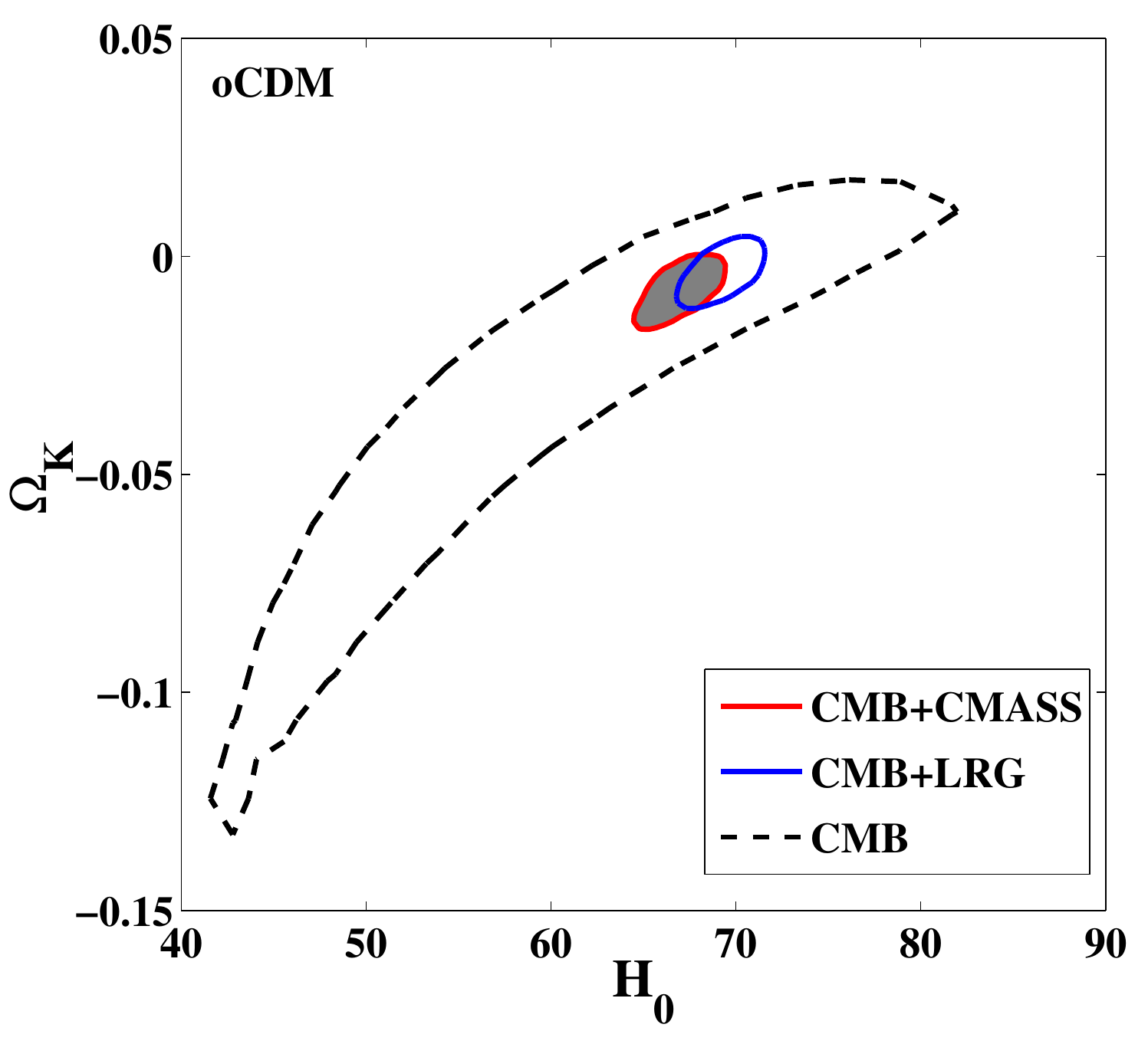}}
\caption{$68$ per cent contours for $H_0$ vs $\Omegak$ in the $o$CDM
  cosmological model. The BAO data break the geometrical degeneracy in
  the CMB, and the CMASS DR9 measurements are consistent with the
  SDSS-II LRG measurements. } \label{fig:oCDM}
\end{figure}

\begin{figure}
\centering
\resizebox{0.9\columnwidth}{!}{\includegraphics{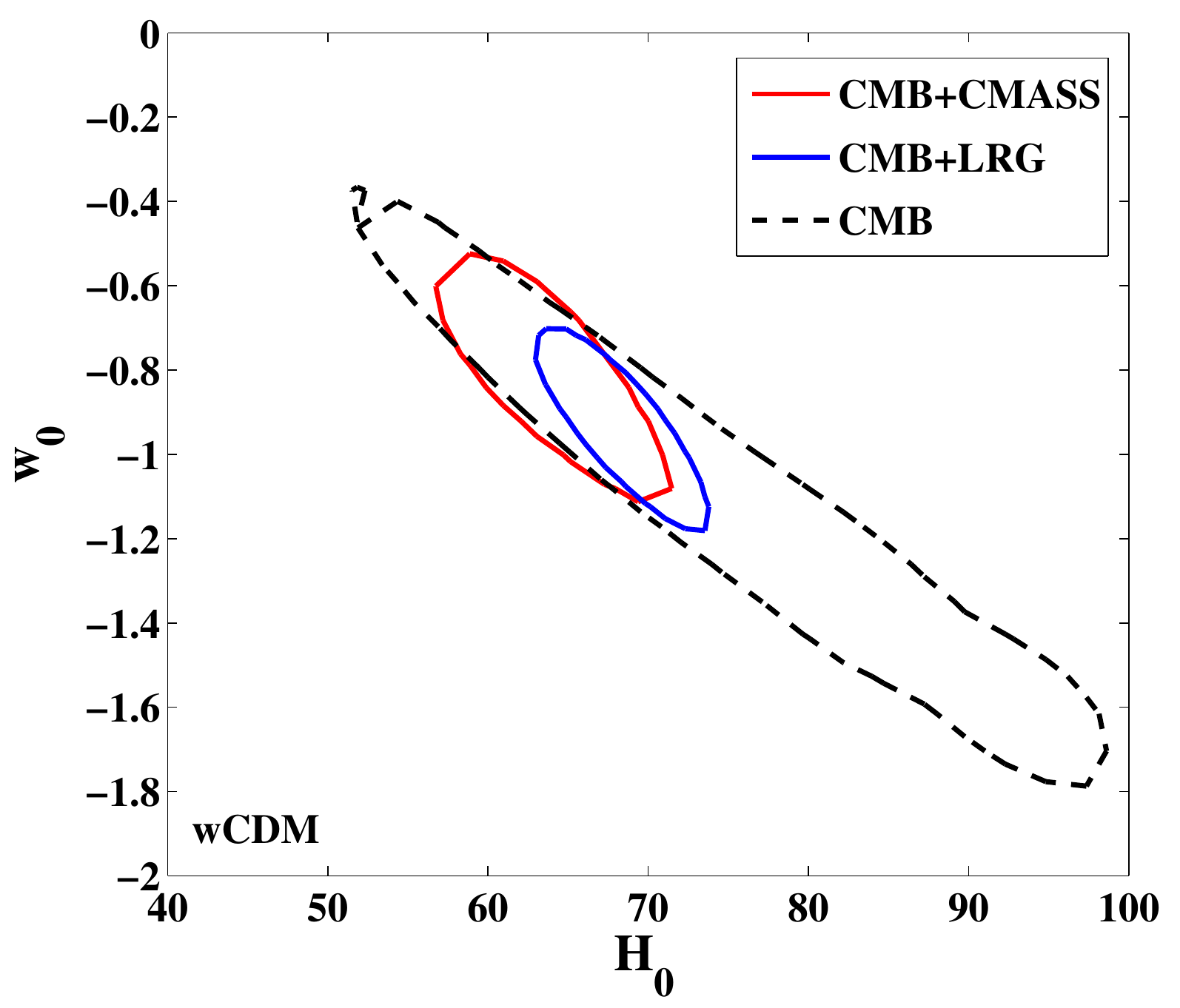}}
\caption{$68$ per cent contours for $H_0$ vs $w$ in the $w$CDM
  cosmological model.  As in Fig.~\ref{fig:oCDM}, addition of the BAO
  data break the degeneracy in the CMB data. The differences in the
  two are due to the different redshift dependence of dark energy and
  curvature.}
\label{fig:wCDM}
\end{figure}

\begin{figure}
\centering
\resizebox{0.9\columnwidth}{!}{\includegraphics{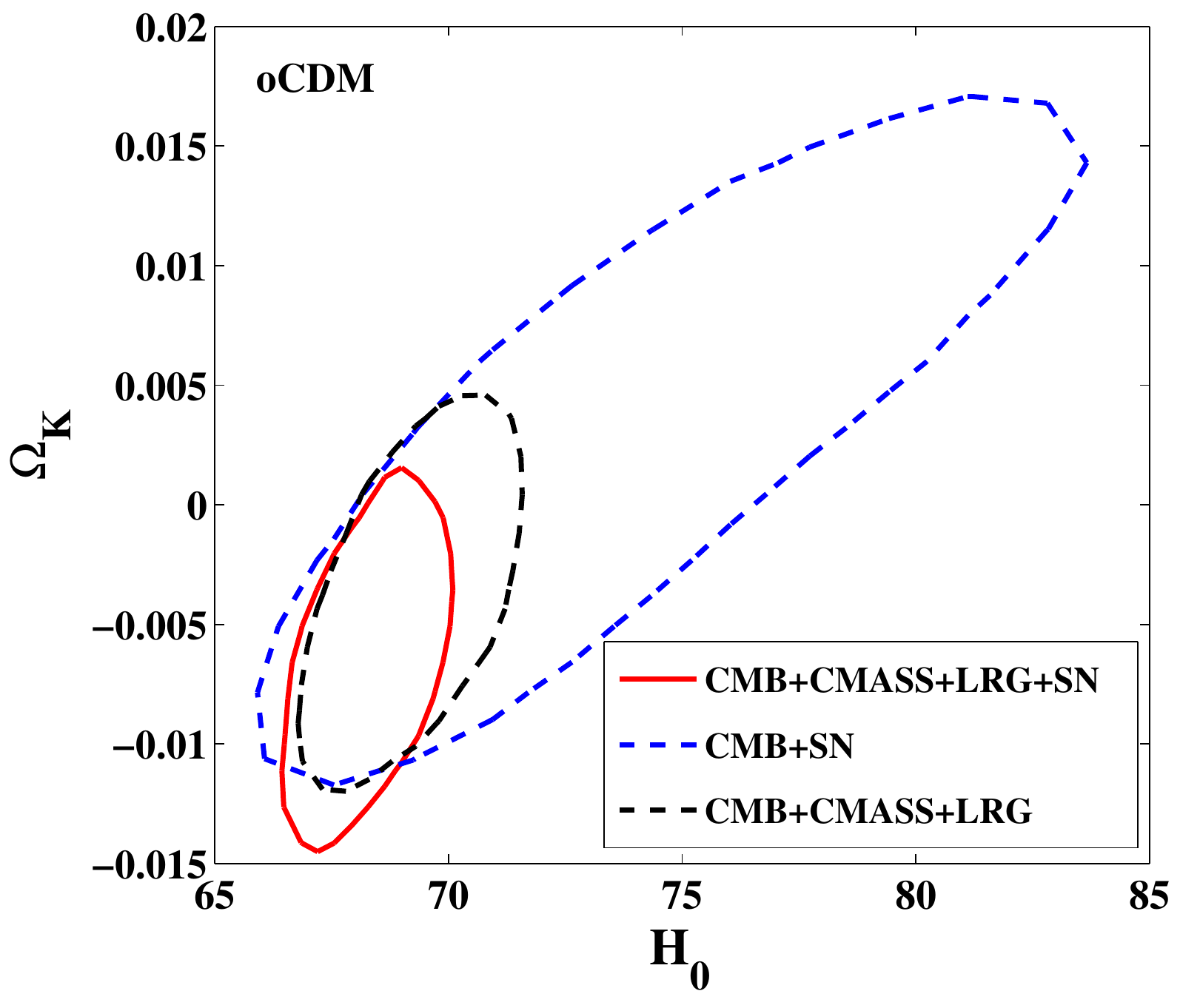}}
\caption{$68$ per cent contours for $H_0$ vs $\Omegak$ in the $o$CDM
  cosmological model comparing different datasets. The SNe data are
  less effective at constraining curvature, given its subdominance at
  low redshifts.} \label{fig:oCDM-Compare}
\end{figure}

\begin{figure}
\centering
\resizebox{0.9\columnwidth}{!}{\includegraphics{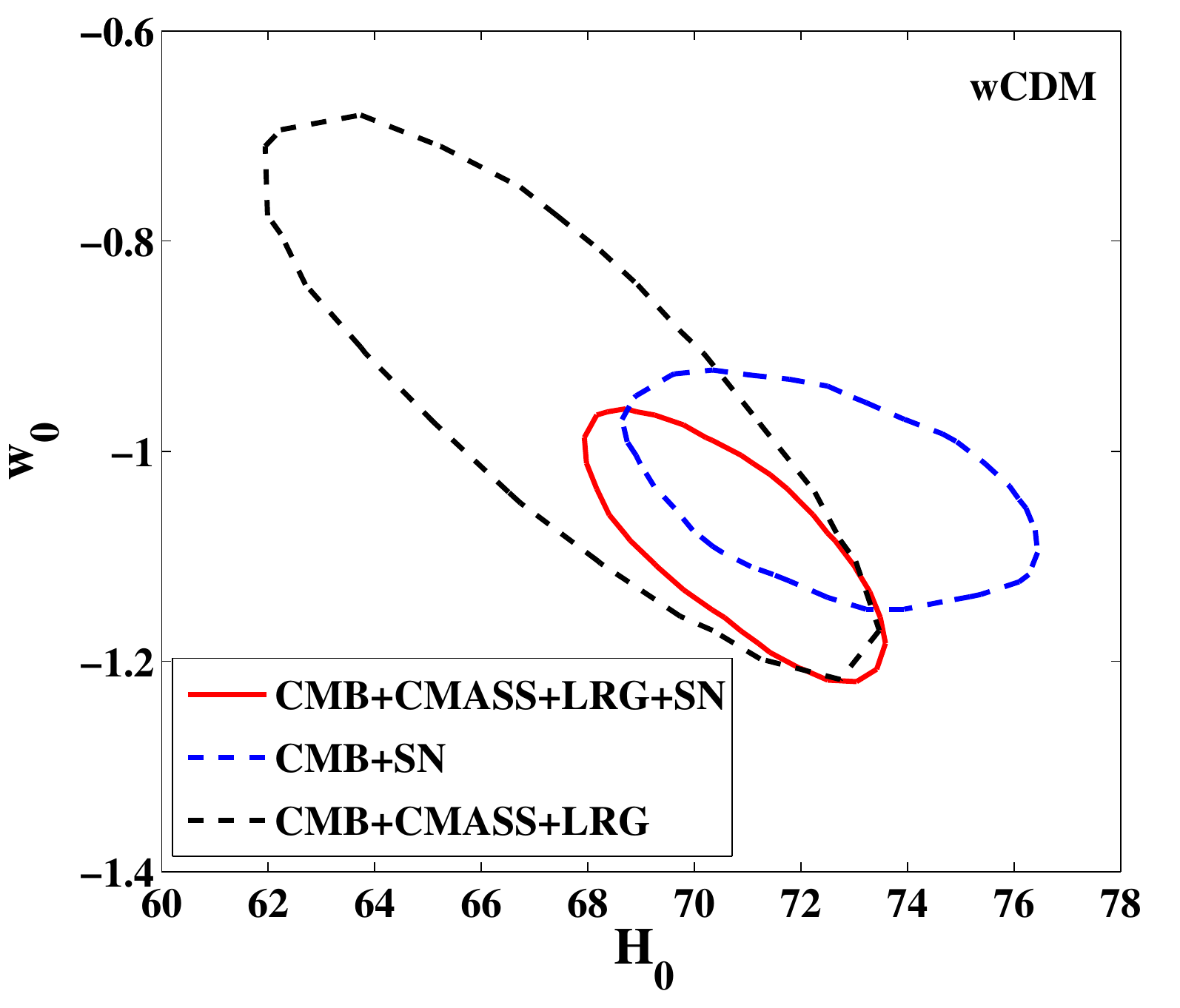}}
\caption{$68$ per cent contours for $H_0$ vs $w$ in the $w$CDM cosmological
model comparing different datasets. Contrast this with
Fig.~\ref{fig:oCDM-Compare}; the smaller redshift lever arm of the BAO data
makes them less sensitive to variations
in the equation of state.} \label{fig:wCDM-Compare}
\end{figure}

\begin{figure}
\centering
\resizebox{0.9\columnwidth}{!}{\includegraphics{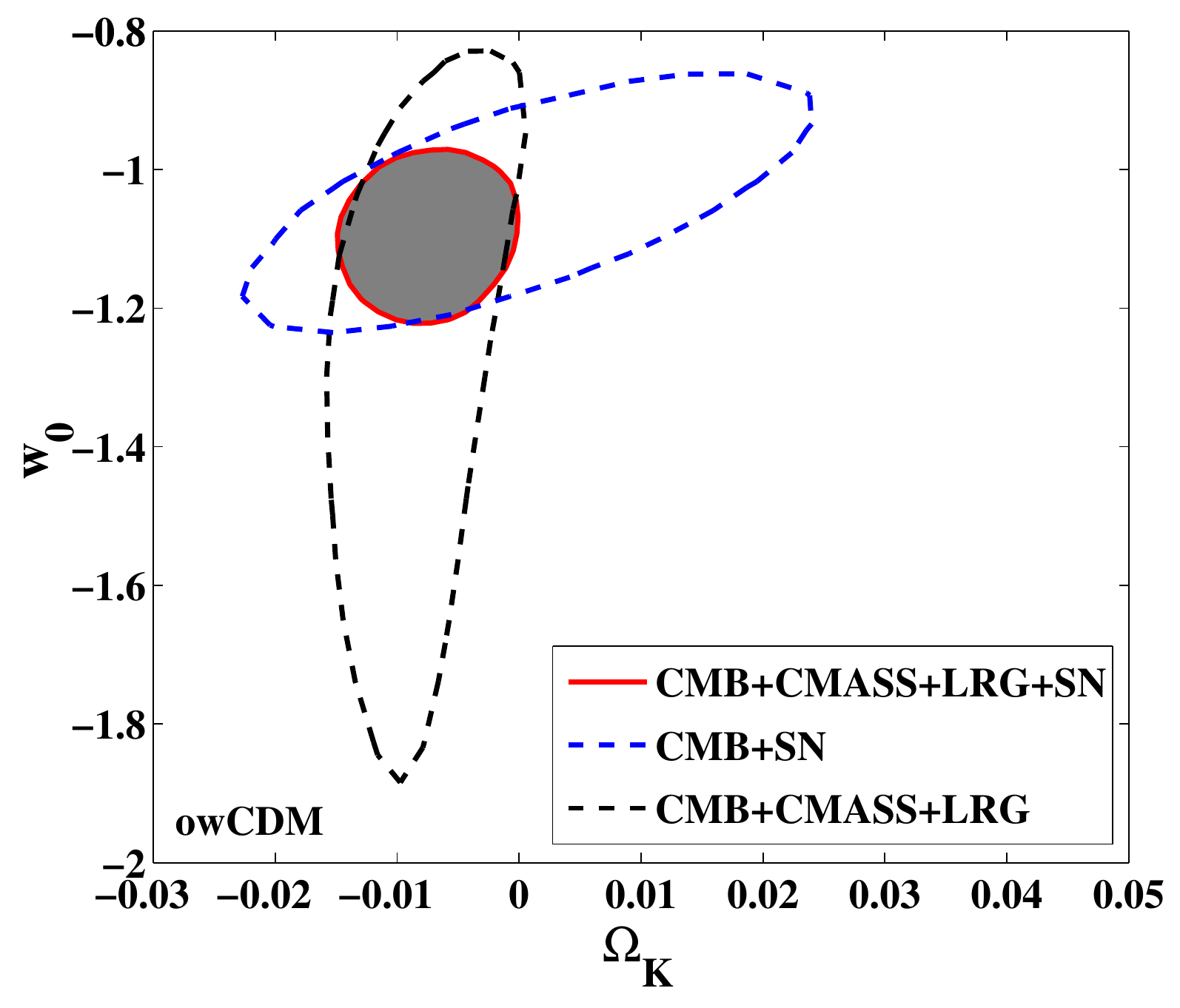}}
\caption{$68$ per cent contours for $w_0$ vs $\Omegak$ in the o$w$CDM cosmological
model for CMB+LRG+CMASS+SN (shaded red), CMB+SN (dashed blue), and
CMB+LRG+CMASS (dashed black) datasets. Note the relative orthogonality of the 
contours -- the BAO data are very effective at constraining curvature, while the SNe data constrain 
the equation of state. Combining the two yield tight constraints both on $\Omega_K$ and $w_0$. } 
\label{fig:owCDM} 
\end{figure}

\begin{figure}
\centering
\resizebox{0.9\columnwidth}{!}{\includegraphics{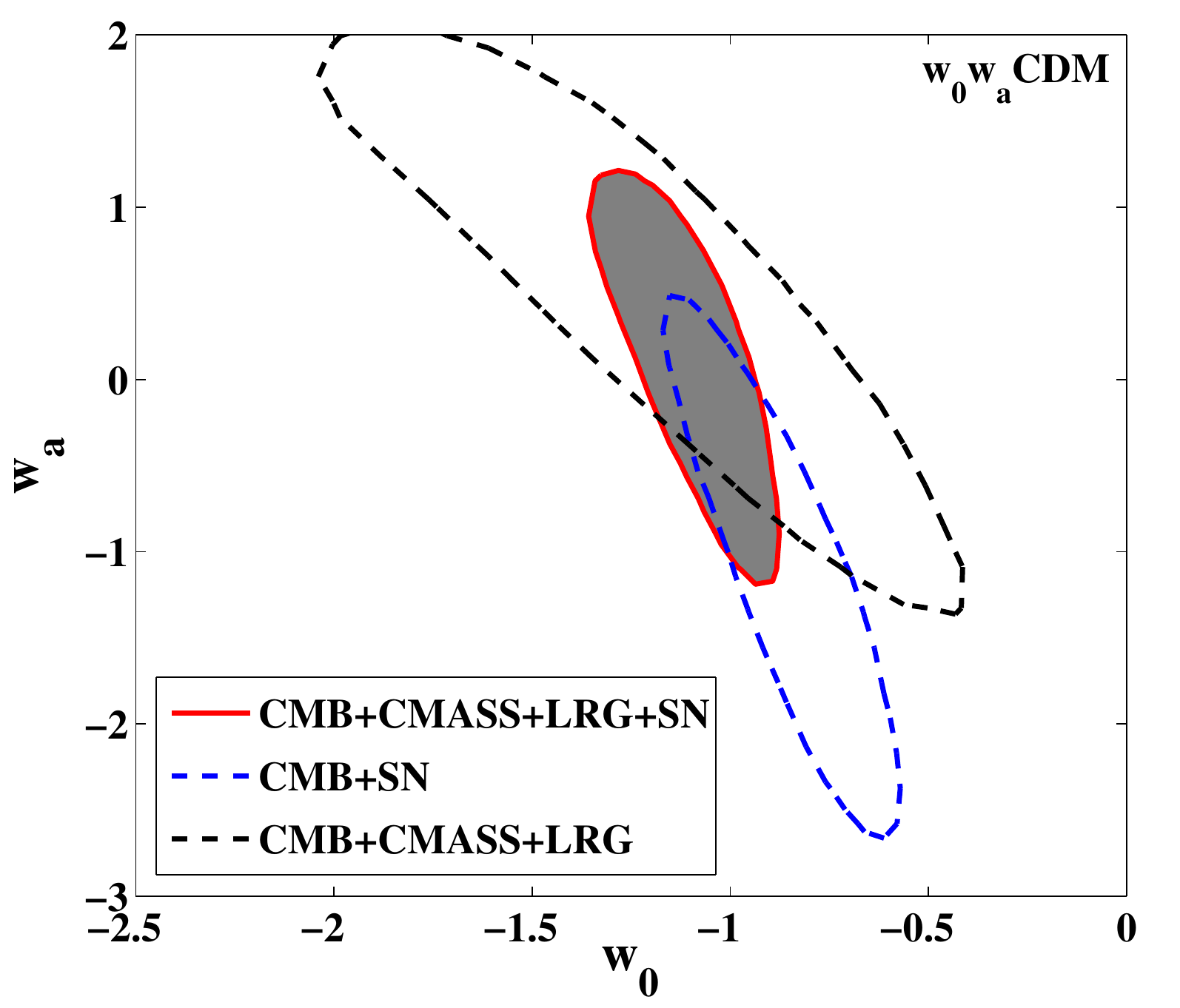}}
\caption{$68$ per cent contours for $w_0$ vs $w_a$ in the $w_0wa$CDM
cosmological model for CMB+LRG+CMASS+SN (shaded red), CMB+SN (dashed blue),
and CMB+LRG+CMASS (dashed black) datasets. We have used a prior on $w_a$ as
follows: $-3.0 \leq w_a \leq 2.0$. Compare the overlaps in this case with Fig.~\ref{fig:owCDM};
the constraints from the BAO and SNe are less complementary.}
\label{fig:w0waCDM}
\end{figure}

\begin{figure}
\centering
\resizebox{0.9\columnwidth}{!}{\includegraphics{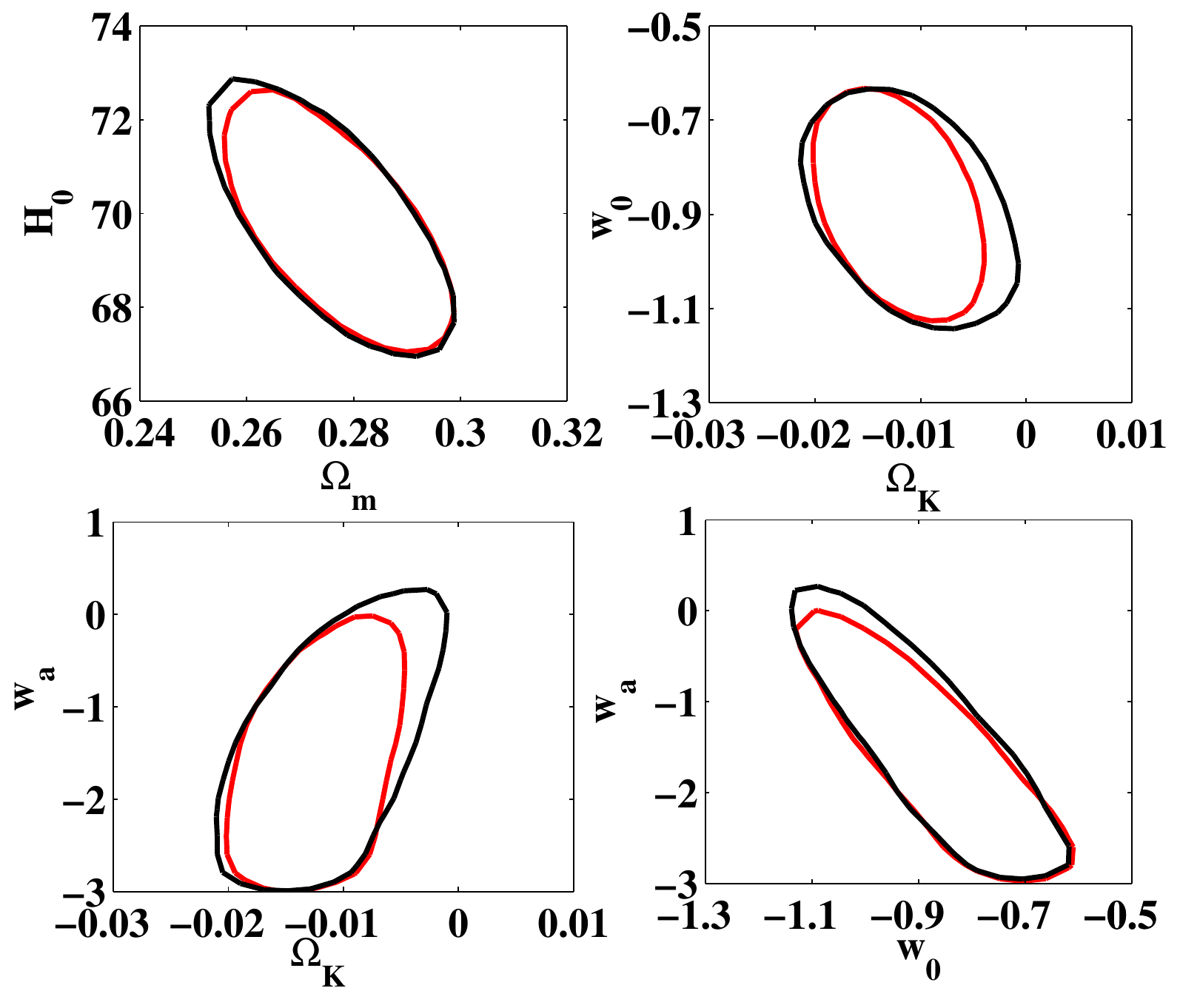}}
\caption{$68$ per cent contours for $H_0$ vs $\Omegam$ (top left), $w_0$ vs
$\Omegak$ (top right), and $w_a$ vs $\Omegak$ (bottom left), and $w_a$ vs $w_0$
(bottom right), in the o$w_0wa$CDM cosmological model for CMB+LRG+CMASS+SN
(solid red) and CMB+LRG+SN (dashed black) datasets. We have used a prior on
$w_a$ as follows: $-3.0 \leq w_a \leq 2.0$.}
\label{fig:ow0waCDM}
\end{figure}

To explore the implications of these results for the values of
cosmological parameters, we consider the standard CDM parametrisation
of the baryon and matter densities $\{ \Omegam, \Omegab \}$, the
primordial power spectrum slope $n_{s}$, the optical depth to
reionization $\tau$, the Hubble constant $H_0$ and the amplitude of
matter clustering $\sigma_{8}$.  We also examine models with a
non-zero curvature $\Omegak$ as well as models where the dark energy
differs from a cosmological constant with an equation of state
parameterised by $w(a) = w_0 + (1-a) w_a$ \citep{Chev01,linder03},
where $a$ is the scale factor.

We follow the methodology in \citet{Meh12}, using the \texttt{CosmoMC}
\citep{lewis02} Markov Chain Monte Carlo sampler to map the posterior
distributions of these parameters.  Our BAO distance constraints are
parameterised as described above as a measurement on $D_V/r_s$ at
$z=0.57$; we augment these with the $z=0.35$ measurement from
\citet{Pad12} as well as the 6dF measurement at $z=0.106$
\citep{Beutler11}. These measurements have very little overlap in
redshift and cover different angular patches, and we treat them
independently. We do not include the WiggleZ measurements
\citep{Bla11a, Bla11b} given the significant overlap with the sample
presented here. However, as discussed in the previous section, the
WiggleZ measurements agree very well with the distances derived in
this work. In addition to these BAO measurements, we include
observations of the temperature and polarization fluctuations in the
cosmic microwave background (CMB) by the WMAP satellite
\citep{komatsu11}, as well as measurements of the expansion history by
the 3-year Supernova Legacy Survey \citep{conley11} and local
measurements of the Hubble constant by \citet{riess11}. We summarise
the data sets used in Table~\ref{tab:datasets}.

We summarise our estimated cosmological parameters and their uncertainties for different assumptions
about the background cosmology in Table~\ref{tab:cosmoparams}. The discussion
below highlights particular cross-sections through this space of models and
parameters, focusing on comparisons between the LRG and CMASS samples as well
as comparisons between the cosmological constraints from the BAO and supernova
data.

The most restricted model we consider (denoted $\Lambda$CDM) is a
$\Lambda$CDM cosmology with no spatial curvature; the dark energy is
assumed to be a cosmological constant with $w = -1$. As is clear from
Fig.~\ref{fig:LCDM}, this model is already highly constrained by the
CMB through a combination of constraints on the physical matter
density $\Omega_m h^2$ and the distance to the last scattering
surface. However, the current WMAP data cannot fully separate
$\Omega_m$ and $h$, leading to an uncertainty in both of these
measurements along the direction of constant $\Omega_m h^n$, where
$n\sim3$ \citep{Per02}. Adding a single low redshift distance
measurement, from either the LRG or CMASS data, significantly reduces
this uncertainty. The similar errors of the two BAO distances lead to
similar constraints on $H_0$: $\pm$1.2 kms$^{-1}$Mpc$^{-1}$ for the
LRG sample and $\pm$1.3 kms$^{-1}$Mpc$^{-1}$ for the CMASS sample (a
1.7 per cent measurement). Combining these reduces this error to
$\pm$1.0 kms$^{-1}$Mpc$^{-1}$ (a 1.4 per cent measurement), a
reduction by $\sim \sqrt{2}$ (Table~\ref{tab:cosmoparams}).

Allowing the curvature or $w_0$ (for a constant equation of state) to
vary (denoted $o$CDM and $w$CDM respectively) opens up a degeneracy in
the $\Omegak / w_0 - H_0$ plane when only the CMB data are considered
(Figs.~\ref{fig:oCDM} and \ref{fig:wCDM}). This degeneracy is broken
by the introduction of a single distance measurement, as one might
have expected from Fig.~\ref{fig:boss_DVfid}.  The larger degeneracy
in $\Omegak - H_0$ and the subsequently tighter constraints from the
BAO measurements compared to $w_0 - H_0$ results from the different
redshifts at which curvature and dark energy become important. The BAO
and CMB measurements are connected through the sound horizon and
curvature has the dominant effect on this lever arm. This effect is
visually apparent in Fig.~\ref{fig:boss_DVfid}, where the effect of
curvature is mostly an offset in the distance-redshift relation (over
the redshifts for which we are sensitive), while changing $w_0$
results in a non-trivial change to the shape of the distance redshift
relation. This result also explains the difference in the improvement
when the LRG and CMASS samples are combined. The two samples do not
have a wide enough lever arm to improve the constraints on $H_0$ in
the $w$CDM case. By contrast, for the $o$CDM case, the errors in
$H_0$ drop by $\sim 25$ per cent from the LRG only case.

For both these cosmological models, one can also compare the constraints from the SN data with
those from the BAO data as shown in Figs.~\ref{fig:oCDM-Compare} and \ref{fig:wCDM-Compare}.
The qualitative difference between the SN and BAO distance ladders is that while the SN data are a regular 
distance ladder, building out from low redshift to high redshift, the BAO are 
an ``inverse'' distance ladder, calibrated at the CMB and extending down to low 
redshift. The SN therefore only weakly constrain the curvature (Fig.~\ref{fig:oCDM-Compare}) and
are more sensitive to $w_0$, with the reverse being true for BAO. This
effect is reflected
in Figs.~\ref{fig:oCDM-Compare} and \ref{fig:wCDM-Compare}. 
The constraints on curvature are significantly improved by
the BAO data, and they do not improve significantly upon the addition of the SN data. 
For the $w$CDM case, while the BAO measurements have lower constraining power, 
their different redshift dependence gives them a different degeneracy direction to the
SN, resulting in improved constraints. 
These trends are repeated when we consider two parameter models of the expansion history:
$ow$CDM (Fig.~\ref{fig:owCDM}) and $w_0 w_a$CDM (Fig.~\ref{fig:w0waCDM}), with the error ellipses being more
orthogonal when the curvature is allowed to vary.

None of the individual probes are currently sufficiently sensitive to
constrain the combination of $\Omegak$, $w_0$ and $w_a$. We therefore
combine the SN and BAO data to obtain constraints on these models
(Fig.~\ref{fig:ow0waCDM}). This cosmological model is also the one
recommended by the Dark Energy Task Force \citep{albrecht06} as the
baseline to compare different dark energy experiments. They recommend
using the inverse of the area of the 95 per cent error ellipse in the
$w_0 - w_a$ plane as a ``Figure of Merit'' (FoM) for the
experiment. Our results (CMB+LRG+SN+CMASS) yield a FoM of 14.4,
compared to a FoM of 11.5 (CMB+LRG+SN) reported by \cite{Meh12}; the
improvement driven by the inclusion of the higher precision BOSS
measurement is clear.

\begin{figure}
\centering
\resizebox{0.9\columnwidth}{!}{\includegraphics{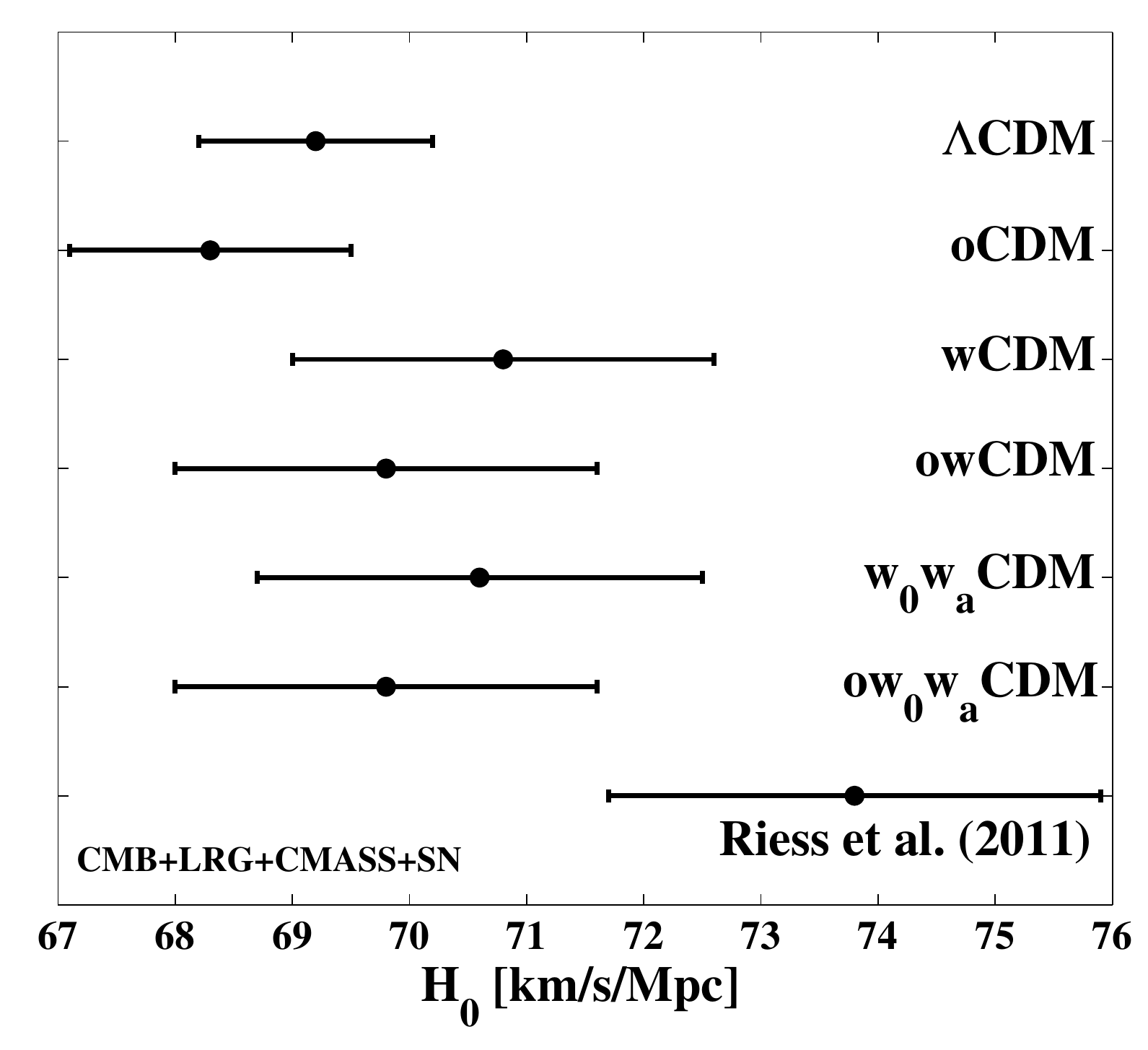}}
\caption{Measured values for $H_0$ using the CMB+LRG+CMASS+SN dataset
  under various cosmological models. This figure shows that we get
  consistent results for $H_0$, which is slightly smaller than the
  direct measurement by \citet{riess11}.} \label{fig:H0-Compare}
\end{figure}

Finally, as was discussed extensively in \citet{Meh12}, the
combination of the SN and BAO distances allows one to transfer the CMB
distance scale down to the local Universe and constrain $H_0$.
Fig.~\ref{fig:H0-Compare} demonstrates that the resulting value of
$H_0$ is robust to changes in assumed cosmological model.
While the difference between the 
inferred value of $H_0$ from these data and the direct measurement of 
\citet{riess11} is not statistically significant in
these data ($\sim 1\sigma$), they may be brought into better agreement by adding an
additional relativistic energy density component equivalent to $4.26
\pm 0.56$ neutrino species, instead of the canonical $3.04$ (see
e.g. \citealt{Meh12} for more details on the mechanism).  Improvements
in both data sets in the future will elucidate if the introduction of
new physics is warranted or if the explanation is more mundane.

\section{Discussion}
\label{sec:discuss}
We have presented the first constraints on cosmology and the distance
scale from the Data Release 9 CMASS galaxy sample of the Baryon
Oscillation Spectroscopic Survey.  Our results are based on accurate
3D positions of $264\,283$ massive galaxies covering 3275
sq.~deg.~with an effective redshift $z=0.57$.  This is the largest
sample of the Universe ever surveyed at this high density and the
derived BAO constraints are the most accurate determination of the
distance scale within the crucial redshift range where the expansion
of the Universe begins to accelerate.

The large survey volume and high sampling density of the CMASS
galaxies allow the detection of the acoustic oscillations predicted by
theories of the early Universe at very high significance
($>5\,\sigma$).  The acoustic signature is seen both as a clear peak
in the correlation function and a series of ``wiggles'' in the power
spectrum.  The measures are highly consistent, and we use both
statistics in our final results.  We determine the statistical
significance of our measurements using a large number of mock catalogs
based on second order Lagrangian perturbation theory \citep{Man12},
and test the covariance matrices so derived with two analytic methods.
Our analysis of the mock catalogs shows that our measurements and
their errors are not at all unusual, and would be expected given our
sampling if the underlying cosmology were of the $\Lambda$CDM family.
Applying reconstruction \citep{Eis07a} to the data does not
significantly improve our measurement of the acoustic signature, which
is to be expected based on comparison to mock catalogs since the
pre-reconstruction error in the CMASS DR9 data is smaller than for a
``typical'' realisation.

We obtain a distance measurement from the power spectrum and
correlation function by fitting the acoustic feature to an
appropriately scaled template, while marginalising over variations in
the broad-band shape.  Our results are very robust to the procedure
employed to marginalize over broad-band power, and indeed the
configuration- and Fourier-space fits provide consistent constraints
even though the template form and procedure are quite different.  The
scale parameter, $\alpha$, relates $D_V/r_s$ to the value in a
fiducial cosmology.  Since we use angle-averaged statistics in this
work, the relevant distance measure is
$D_V=\left[cz(1+z)^2D_A^2/H\right]^{1/3}$, and it is measured relative
to the sound horizon, $r_s$.  Since the correlation function and power
spectrum include noise from small-scales and shot-noise differently,
we average the two determinations to obtain our consensus result on
the distance to $z=0.57$, $D_V/r_s = 13.67\pm 0.22$, where we use the
scatter in the mock catalogs as an estimate of the error ($1.7$ per
cent) on this average.

\citet{Rei12} and \citet{Sanchez12} use the correlation function over
a wide range of scales to constrain cosmological parameters.  We find
excellent agreement between their results and the pure-BAO measurement
described here, despite slightly different choices of binning, fit
range, etc.  This demonstrates that the distance information is
dominated by the sharp acoustic feature rather than the broad-band
power (which we have explicitly marginalised over in our analysis).

The BOSS result can be combined with other BAO measurements to form an
``inverse distance ladder'' which tightly constrains the expansion
rate from $z\simeq 0.1$ to $z\sim 0.6$.  The acoustic signature
measured in BOSS is in excellent agreement with earlier SDSS results
\citep{Per10,Pad12}, and the distance to $z\simeq 0.6$ is in almost
perfect agreement with that inferred by WiggleZ \citep{Bla11a}.  In
general the independent BAO results are all consistent with the same
underlying (flat, $\Lambda$CDM) cosmology.  Even with only a fraction
of the survey completed, the BOSS constraint is already the tightest
distance constraint in the ladder ($1.7$ per cent), with an errorbar
2.3 times smaller at $z\simeq 0.6$ than the combined, earlier WiggleZ
measurements \citep{Bla11a}.  The BAO distance ladder suggests a
slightly larger distance scale than the best-fit to the 7-year WMAP
data, lying closer to the $1\,\sigma$ upper limit in WMAP toward
higher $\Omega_m h^2$.  With this slightly higher value of $\Omega_m
h^2$, the BAO measurements are in superb agreement with each other and
the CMB within the context of a flat $\Lambda$CDM cosmology.  While
SNe do not provide an absolute distance, the relative distance scale
inferred from SNLS SNe data is in good agreement with that inferred
from BAO.

BOSS continues to amass data, and we expect these constraints to tighten
significantly, as data will be collected through mid-2014.

\section{Acknowledgements}

WJP is grateful for support from the UK Science and Technology
Facilities Research Council, the Leverhulme Trust, and the European
Research Council.
AJR is grateful to the UK Science and Technology Facilities Council
for financial support through the grant ST/I001204/1.
CGS acknowledges funding from project AYA2010-21766-C03-02 of the Spanish Ministry of Science and
Innovation (MICINN).
DJE, XX, and KM were supported by NSF grant AST-0707725 and NASA grant NNX07AH11G.
SH and DJS are supported by the US Department of Energy’s Office of High Energy Physics (DE-AC02-05CH11231). 
NP and AJC are partially supported by NASA grant NNX11AF43G.
FP acknowledges support from the Spanish MICINN’s Consolider grant MultiDark CSD2009-00064.
MAS was supported by NSF grant AST-0707266.
MECS was supported by the NSF under Award No. AST-0901965.

Funding for SDSS-III has been provided by the Alfred P. Sloan
Foundation, the Participating Institutions, the National Science
Foundation, and the U.S. Department of Energy Office of Science. The
SDSS-III web site is http://www.sdss3.org/.

SDSS-III is managed by the Astrophysical Research Consortium for the
Participating Institutions of the SDSS-III Collaboration including the
University of Arizona,
the Brazilian Participation Group,
Brookhaven National Laboratory,
University of Cambridge,
Carnegie Mellon University,
University of Florida,
the French Participation Group,
the German Participation Group,
Harvard University,
the Instituto de Astrofisica de Canarias,
the Michigan State/Notre Dame/JINA Participation Group,
Johns Hopkins University,
Lawrence Berkeley National Laboratory,
Max Planck Institute for Astrophysics,
Max Planck Institute for Extraterrestrial Physics,
New Mexico State University,
New York University,
Ohio State University,
Pennsylvania State University,
University of Portsmouth,
Princeton University,
the Spanish Participation Group,
University of Tokyo,
University of Utah,
Vanderbilt University,
University of Virginia,
University of Washington,
and Yale University.

We acknowledge the use of the Legacy Archive for Microwave Background
Data Analysis (LAMBDA). Support for LAMBDA is provided by the NASA
Office of Space Science.

The reconstruction computations were supported by facilities and staff
of the Yale University Faculty of Arts and Sciences High Performance
Computing Center.  The MCMC computations in this paper were run on the
Odyssey cluster supported by the FAS Science Division Research
Computing Group at Harvard University.

Catalogue construction, power spectrum calculations, and fitting made
use of the facilities and staff of the UK Sciama High Performance
Computing cluster supported by the ICG, SEPNet and the University of
Portsmouth, and the COSMOS/Universe supercomputer, a UK-CCC facility
supported by HEFCE and STFC in cooperation with CGI/Intel.

\appendix

\section{Comparisons of the NGC and SGC}
    \label{sec:northsouth}
    In this paper, we analyse the full sample of the CMASS DR9 galaxy catalog 
combining the Northern Galactic Cap (NGC) and Southern Galactic Cap (SGC). 
We justify this choice in this section, since we find no significant 
differences in the clustering beyond acceptable statistical fluctuations. 

The DR9 BOSS footprint contains two disjoint regions: a 2635\,deg$^2$
region in the NGC and a 709\,deg$^2$ region in the SGC.  The SGC
imaging, on average, is at coordinates with larger Galactic extinction
and was taken under conditions with higher airmass and sky background,
compared to the NGC. However, \citet{Ross12} find that none of these
factors has a measurable systematic effect on the clustering of DR9
CMASS galaxies. Comparing the NGC and SGC, \citet{Ross12} found that
the projected number density is 3.2 per cent larger in the SGC.
\cite{Sch11} found offsets in photometry between the NGC and SGC data
(due to a combination of calibration offsets in the SDSS imaging data
and systematic errors in extinction corrections). Applying these
offsets to our target selection criteria only removes objects in the
SGC, so we can apply this correction to our catalogue. Doing so
reduces the difference in projected number density to 0.2 per cent,
well within the 2 per cent standard deviation found in the mocks. The
$n(z)$ in the two hemispheres also appear different. Constructing a
covariance matrix based on the mock $n(z)$, \citet{Ross12} found that
6 per cent of consistent samples exhibit larger NGC to SGC variations
in the $n(z)$.  This fraction was increased to 11 per cent after
applying the \cite{Sch11} offsets to the sample selection. Our
approach is to treat the two regions as having separate selection
functions, and to optimally combine the pair counts from each sample
in order to obtain $\xi(s)$ measurements for the full sample.

\citet{Ross12} found that the clustering in the NGC and SGC is
generally consistent, to within 2$\sigma$. Indeed, we find
$\chi^2=56.6$ (with 44 data points fitted) when comparing the two
regions' $\xi(s)$ measurements in the range $28<s<200\mpcoh$. The
greatest differences are found close to the BAO scale, and
\citet{Ross12} found no treatment of the data (e.g., applying the
\citealt{Sch11} offsets to the sample selection or applying
alternative weighting schemes) that could ameliorate this
tension. Before reconstruction, we find $\alpha = 1.000 \pm 0.018$ in
the NGC and $\alpha = 1.091 \pm 0.030$ in the SGC; the BAO position
differs by 2.6$\sigma$. After reconstruction, we measure $\alpha =
1.012 \pm 0.018$ in the NGC and $\alpha = 1.067 \pm 0.035$ in the SGC.
Hence, after reconstruction, the BAO positions differ by only
1.4~$\sigma$.

In addition, through analysing our mock catalogues, we find no
difference between the ensemble properties of the NGC and the
SGC. From the mock catalogues, we recover a mean
$\langle{\alpha}\rangle = 1.005$ with average error on any single
realisation of $0.031$ and a median $\tilde{\alpha} = 1.006$ with
quantiles $^{+0.026}_{-0.029}$ in the NGC before reconstruction. In
the SGC we find a mean $\langle{\alpha}\rangle = 0.999$ with average
error in any single realisation of $0.047$ and a median
$\tilde{\alpha} = 0.999$ with quantiles $^{+0.042}_{-0.045}$. After
reconstruction, we recover a mean $\langle{\alpha}\rangle = 1.004$
with average error on any single realisation of $0.020$ and a median
$\tilde{\alpha} = 1.004$ with quantiles $^{+0.020}_{-0.019}$ in the
NGC. In the SGC we find a mean $\langle{\alpha}\rangle = 1.006$ with
average error on any single realisation of $0.041$ and a median
$\tilde{\alpha} = 1.006$ with quantiles $^{+0.039}_{-0.038}$. The mean
$\langle{\alpha}\rangle$ and median $\tilde{\alpha}$ values of the NGC
and SGC mocks are consistent with each other before and after
reconstruction. These results also suggest that reconstruction not
only improves the precision of BAO position measurements, but it also
makes the likelihood distributions more Gaussian. Accepting that a
difference greater than 1.4~$\sigma$ happens 16 per cent of the time
and that \citet{Ross12} find no evidence that the SGC $\xi(s)$
measurements have significant systematic uncertainties affecting the
BAO scale, we conclude that there is no systematic difference between
the two hemispheres and therefore base our BAO measurements on the
combined sample.

\section{Robustness Tests}
\label{sec:robust_tests}
  \subsection{Robustness to Reconstruction Parameters}
    \label{sec:robust_recon}
    \begin{figure}
  \centering 
  \includegraphics[width=0.9 \linewidth]{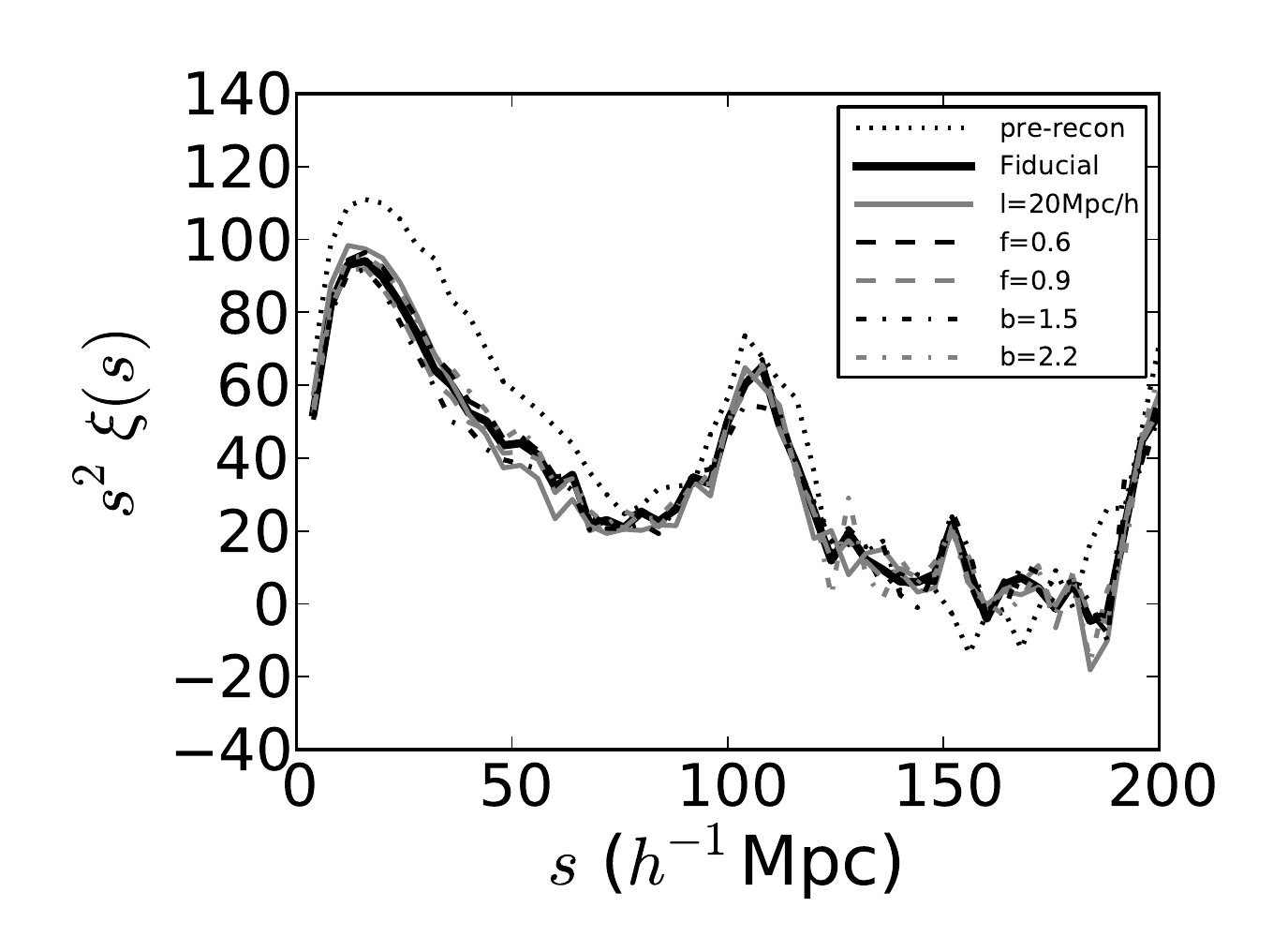}
  \caption{ Correlation function of CMASS DR9 galaxies after
    reconstruction, with different curves corresponding to different
    input parameters of the reconstruction code. There is good
    agreement between the fiducial choice of the parameters (black
    solid line) and other choices.  For each case, we replace either
    the assumed value of the bias $b$ (red and yellow dot-dash lines),
    the growth rate $f$ (blue and green dashed lines), or the
    smoothing length of the density field (cyan solid line). For
    reference, the dotted line represents the CMASS DR9 correlation
    function before reconstruction. }
  \label{fig:robust_recon_data}
\end{figure}

In this section, we test the sensitivity of reconstruction to our
fiducial values of the bias $b$, the growth rate $f$, and the
smoothing length $l$ (used to remove shot noise in the computation of
the density field).  In order to do this, we run reconstruction in a
sample of 100 PTHalos mocks assuming the standard values for all
except one of these parameters. In particular, we tested the effect of
assuming a bias of $b=1.5$ and $b=2.2$ (i.e. 20 per cent below and
above the average recovered bias from the mocks, which was adopted as
our standard value), and also the effect of assuming a growth rate of
$f=0.6$ and $f=0.9$ (again, 20 per cent below and above our standard
value). We also tested the effect of choosing a more conservative
smoothing length of $l=20h^{-1}$ Mpc. As both the correlation function
and power spectrum fits give consistent results for our standard
reconstruction run (see Section~\ref{sec:consensus}), for simplicity
in this Appendix we only report results measuring the acoustic scale
from the correlation function. 

The results are shown at the end of Table~\ref{tab:mock_alphas}. On
average, different choices for the values of the reconstruction
parameters do not bias our distance scale measurements by more than
$\avg{\Delta\alpha} \lesssim 0.3$ per cent with respect to the
measurements using the fiducial values. However, even these small
biases are not significant considering their errors.  We conclude that
our measurements of the distance scale are not sensitive to changes in
the fiducial values of the reconstruction parameters over a wide range
of values, and in good agreement with the fiducial case.

We also study the dependence on reconstruction parameters of the CMASS
DR9 measurements of the distance scale from the correlation function
after reconstruction. We select the same cases we studied above for
the case of the PTHalos mocks, and run reconstruction on CMASS DR9
galaxies for each choice of the parameters. The results are shown in
Table~\ref{tab:dr9_alphas}. We find that in all cases the distance
scale measurements are consistent with the results from the fiducial
case. It is worth noting that the choice of a galaxy bias 20 per cent
smaller ($b=1.5$) than the fiducial case drives the measurement and
the errors above the rest of the cases. The reason becomes evident in
Figure~\ref{fig:robust_recon_data}, which shows the correlation
functions after reconstruction for different values of the
reconstruction parameters. The shape of the correlation function
around the BAO peak has been distorted by this particular choice of
bias, whereas all other choices show results more similar to the
fiducial case. This is an indication that our estimates for the galaxy
bias and the cosmological parameters cannot be completely arbitrary if
we want to reconstruct the density field accurately.  However, for
reasonable values for these parameters, we do not find a large
sensitivity of our measurements to these parameters, and find our
results to be consistent with the fiducial case.

  \subsection{Robustness of Fitting Algorithm for $\xi(r)$}
    \label{sec:robust_fit}





\begin{table*}
  \caption{Fitting results for various models, found by varying the
    fiducial fitting model and reconstruction parameters on the
    ensemble distance scale measurement from the mocks (i.e. the mean
    $\langle\alpha\rangle$ and the median $\tilde{\alpha}$) as well as
    the difference in the observed distance scale with respect to the
    fiducial model on a mock-by-mock basis ($\Delta\alpha$). The
    results for the fiducial model, and for different broadband $A(r)$
    fitting functions ($poly0$, $poly2$, $poly4$), fitting ranges, and
    non-linear damping $\Sigma_{nl}$ of the acoustic scale, are shown
    for the correlation function before and after reconstruction. We
    also present the results of fitting with a different covariance matrix (ML) derived based on the technique in \citet{Xeaip}. For our reconstruction tests, we present the effects of changing the fiducial galaxy bias by $+20$ per cent and $-20$ per cent ($b=1.5$ and $b=2.2$), the fiducial growth rate by $+20$ per cent and $-20$ per cent ($f=0.6$ and $f=0.9$), and the smoothing length to $20\mpcoh$, a more conservative choice than our fiducial smoothing of $15\mpcoh$.}
\label{tab:mock_alphas}

\begin{tabular}{@{}lccccccccc}

\hline
Model&
$\langle\alpha\rangle$&
rms&
$\tilde{\alpha}$&
Quantiles&
$\langle\Delta\alpha\rangle^{1,2}$&
rms&
$\widetilde{\Delta\alpha}$&
Quantiles&
$\langle\chi^2\rangle/dof$\\

\hline
\multicolumn{10}{c}{Before Reconstruction}\\
\hline

Fiducial $[f]$ &
1.004&
0.027&
1.004&
$^{+0.026}_{-0.026}$&
--&
--&
--&
--&
39.60/39\\
\\[-1.5ex]
Fit with $poly0$. &
0.999&
0.026&
1.000&
$^{+0.024}_{-0.024}$&
-0.005&
0.009&
-0.004&
$^{+0.007}_{-0.008}$&
42.93/42\\
\\[-1.5ex]
Fit with $poly2$. &
1.001&
0.027&
1.002&
$^{+0.025}_{-0.025}$&
-0.002&
0.004&
-0.002&
$^{+0.003}_{-0.003}$&
41.24/40\\
\\[-1.5ex]
Fit with $poly4$. &
1.004&
0.027&
1.004&
$^{+0.025}_{-0.025}$&
0.000&
0.001&
-0.000&
$^{+0.001}_{-0.001}$&
38.27/38\\
\\[-1.5ex]
Fit between $20<r<200\mpcoh$. &
1.001&
0.028&
1.003&
$^{+0.025}_{-0.028}$&
-0.002&
0.006&
-0.002&
$^{+0.004}_{-0.004}$&
41.78/41\\
\\[-1.5ex]
Fit between $50<r<200\mpcoh$. &
1.005&
0.027&
1.005&
$^{+0.025}_{-0.026}$&
0.001&
0.003&
0.001&
$^{+0.003}_{-0.002}$&
34.20/34\\
\\[-1.5ex]
Fit with $\snl \rightarrow 0$. &
1.000&
0.030&
0.999&
$^{+0.028}_{-0.026}$&
-0.004&
0.015&
-0.005&
$^{+0.012}_{-0.011}$&
41.60/39\\
\\[-1.5ex]
Fit with $\snl \rightarrow \snl-2$. &
1.002&
0.028&
1.003&
$^{+0.026}_{-0.025}$&
-0.001&
0.005&
-0.002&
$^{+0.004}_{-0.004}$&
39.72/39\\
\\[-1.5ex]
Fit with $\snl \rightarrow \snl+2$. &
1.005&
0.028&
1.005&
$^{+0.026}_{-0.027}$&
0.001&
0.005&
0.001&
$^{+0.004}_{-0.004}$&
40.11/39\\
\\[-1.5ex]
Fit using ML covariance matrix. &
1.003&
0.029&
1.005&
$^{+0.023}_{-0.028}$&
-0.001&
0.008&
-0.000&
$^{+0.006}_{-0.007}$&
40.07/39\\
\hline
\multicolumn{10}{c}{After Reconstruction}\\
\hline
Fiducial $[f]$ &
1.004&
0.018&
1.004&
$^{+0.017}_{-0.018}$&
--&
--&
--&
--&
40.95/39\\
\\[-1.5ex]
Fit with $poly0$. &
1.002&
0.018&
1.002&
$^{+0.017}_{-0.018}$&
-0.002&
0.004&
-0.002&
$^{+0.003}_{-0.004}$&
45.15/42\\
\\[-1.5ex]
Fit with $poly2$. &
1.004&
0.018&
1.003&
$^{+0.017}_{-0.017}$&
-0.001&
0.001&
-0.001&
$^{+0.001}_{-0.001}$&
42.53/40\\
\\[-1.5ex]
Fit with $poly4$. &
1.004&
0.018&
1.004&
$^{+0.017}_{-0.017}$&
-0.000&
0.000&
-0.000&
$^{+0.000}_{-0.000}$&
39.94/38\\
\\[-1.5ex]
Fit between $20<r<200\mpcoh$. &
1.010&
0.017&
1.010&
$^{+0.017}_{-0.017}$&
0.006&
0.003&
0.005&
$^{+0.003}_{-0.003}$&
47.38/41\\
\\[-1.5ex]
Fit between $50<r<200\mpcoh$. &
1.004&
0.018&
1.003&
$^{+0.017}_{-0.018}$&
-0.001&
0.002&
-0.001&
$^{+0.002}_{-0.002}$&
34.55/34\\
\\[-1.5ex]
Fit with $\snl \rightarrow 0$. &
1.003&
0.019&
1.003&
$^{+0.017}_{-0.018}$&
-0.001&
0.003&
-0.001&
$^{+0.003}_{-0.003}$&
40.87/39\\
\\[-1.5ex]
Fit with $\snl \rightarrow \snl-2$. &
1.003&
0.018&
1.004&
$^{+0.017}_{-0.018}$&
-0.001&
0.002&
-0.001&
$^{+0.002}_{-0.002}$&
40.84/39\\
\\[-1.5ex]
Fit with $\snl \rightarrow \snl+2$. &
1.006&
0.018&
1.006&
$^{+0.016}_{-0.018}$&
0.001&
0.003&
0.001&
$^{+0.002}_{-0.002}$&
41.62/39\\
\\[-1.5ex]
Fit using ML covariance matrix. &
1.004&
0.019&
1.003&
$^{+0.019}_{-0.018}$&
-0.000&
0.004&
-0.000&
$^{+0.005}_{-0.004}$&
41.02/39\\
\hline
Fit to recon. with $b \rightarrow 1.5$. &
1.004&
0.019&
1.004&
$^{+0.016}_{-0.022}$&
0.000&
0.006&
0.000&
$^{+0.006}_{-0.006}$&
42.56/39\\
\\[-1.5ex]
Fit to recon. with $b \rightarrow 2.2$. &
1.003&
0.019&
1.005&
$^{+0.015}_{-0.023}$&
-0.000&
0.006&
-0.001&
$^{+0.006}_{-0.005}$&
41.01/39\\
\\[-1.5ex]
Fit to recon. with $f \rightarrow 0.6$. &
1.003&
0.018&
1.004&
$^{+0.017}_{-0.021}$&
-0.000&
0.002&
-0.000&
$^{+0.001}_{-0.002}$&
40.50/39\\
\\[-1.5ex]
Fit to recon. with $f \rightarrow 0.9$. &
1.004&
0.018&
1.005&
$^{+0.015}_{-0.022}$&
0.001&
0.002&
0.001&
$^{+0.001}_{-0.002}$&
41.09/39\\
\\[-1.5ex]
Fit to recon. with $l \rightarrow 20\mpcoh$. &
1.006&
0.019&
1.008&
$^{+0.016}_{-0.023}$&
0.003&
0.007&
0.002&
$^{+0.005}_{-0.005}$&
45.04/39\\
\hline
\end{tabular}

\medskip
$^{1}$ $\Delta \alpha = \alpha_{[i]} - \alpha_{[f]}$, where $i$ is the model indicated in the first column.\\
$^{2}$ Note that the error on the mean $\Delta\alpha$ is $\sqrt{N}$
smaller than the rms from the mocks quoted in the table, where $N$ is
the number of mocks. These much smaller numbers would indicate that
there is a significant detection of the change in the mean as we
change fitting model or reconstruction parameters; however, such a
small change would not be significantly detected in each mock given
the dispersion.

\end{table*}

We test the robustness of our correlation function fitting model by
slightly varying the fiducial model parameters and then re-performing
the fits to see if we recover consistent values of the acoustic scale
$\alpha$. These tests are performed on the mocks as well as the CMASS
DR9 data. Recall that the fiducial model takes on the form given in
Equations (\ref{eqn:fform}) and (\ref{eqn:aform}), where we have taken
$\snl=8\mpcoh$ before reconstruction and $\snl=4\mpcoh$ after
reconstruction. In addition, we specify a fiducial fitting range of
$28<r<200\mpcoh$ and use the sample covariance matrix. Hence, the
fiducial model parameters we alter in performing these tests are the
order of $A(r)$, the value of $\snl$, the fitting range, and the
covariance matrix used. In modifying the form of $A(r)$, $poly0$
corresponds to $A(r)=0$, $poly2$ corresponds to a 2-parameter $A(r) =
a_1/r^2 + a_2/r$, and $poly4$ corresponds to a 4-parameter $A(r) =
a_1/r^2 + a_2/r + a_3 + a_4 r$.

\begin{figure}
  \centering
  \resizebox{0.49\columnwidth}{!}{\includegraphics{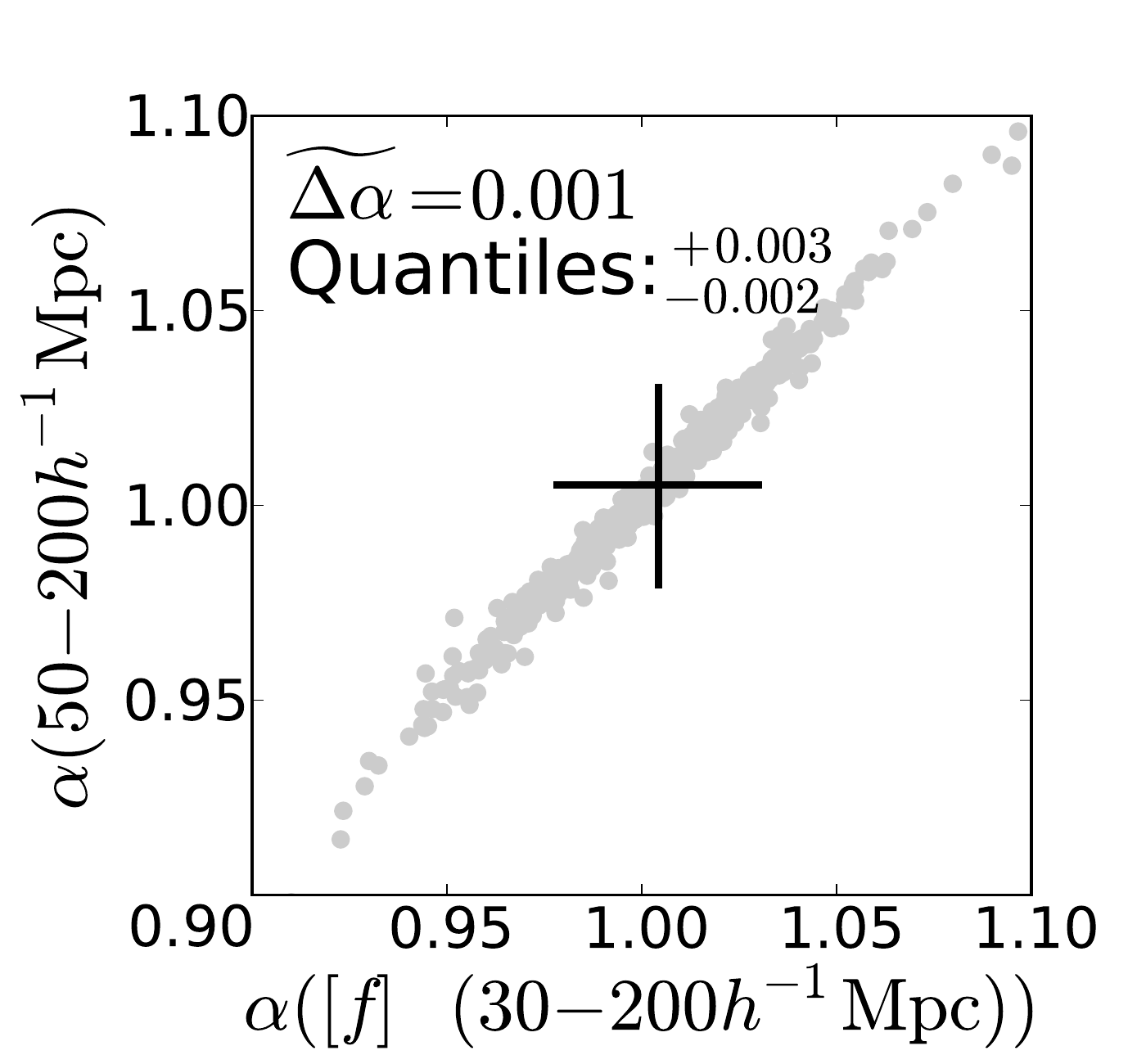}}
  \resizebox{0.49\columnwidth}{!}{\includegraphics{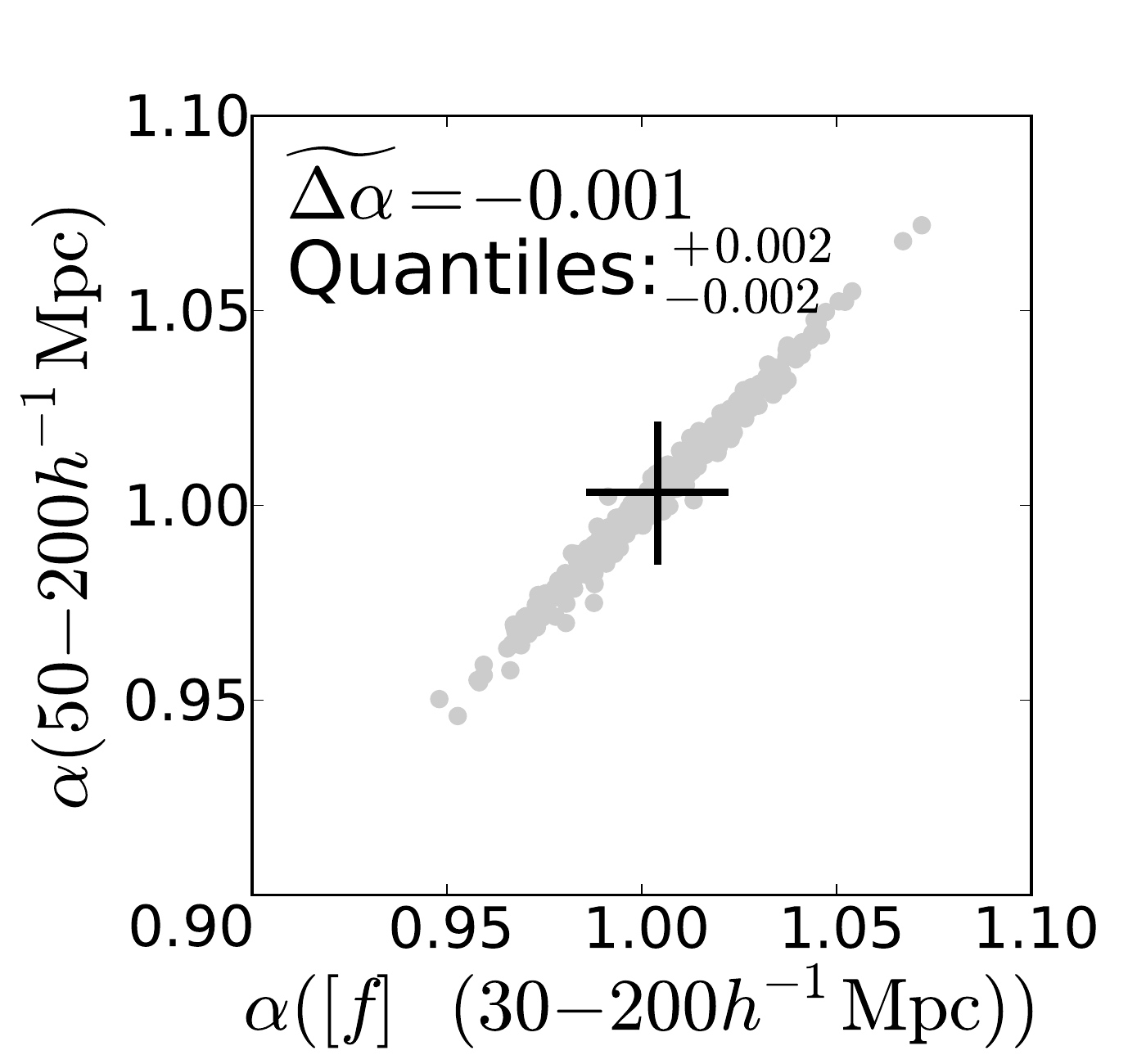}}
  \resizebox{0.49\columnwidth}{!}{\includegraphics{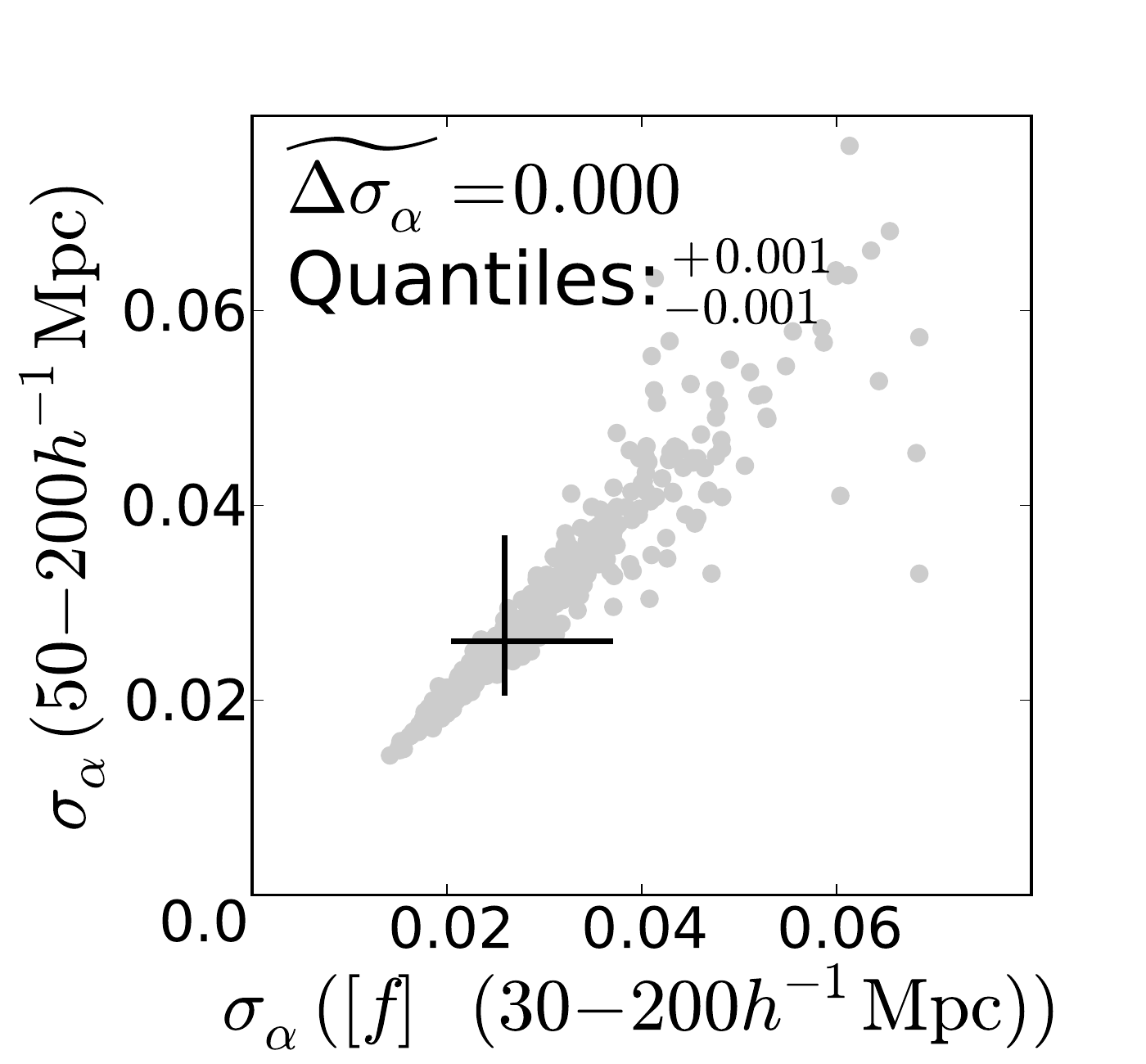}}
  \resizebox{0.49\columnwidth}{!}{\includegraphics{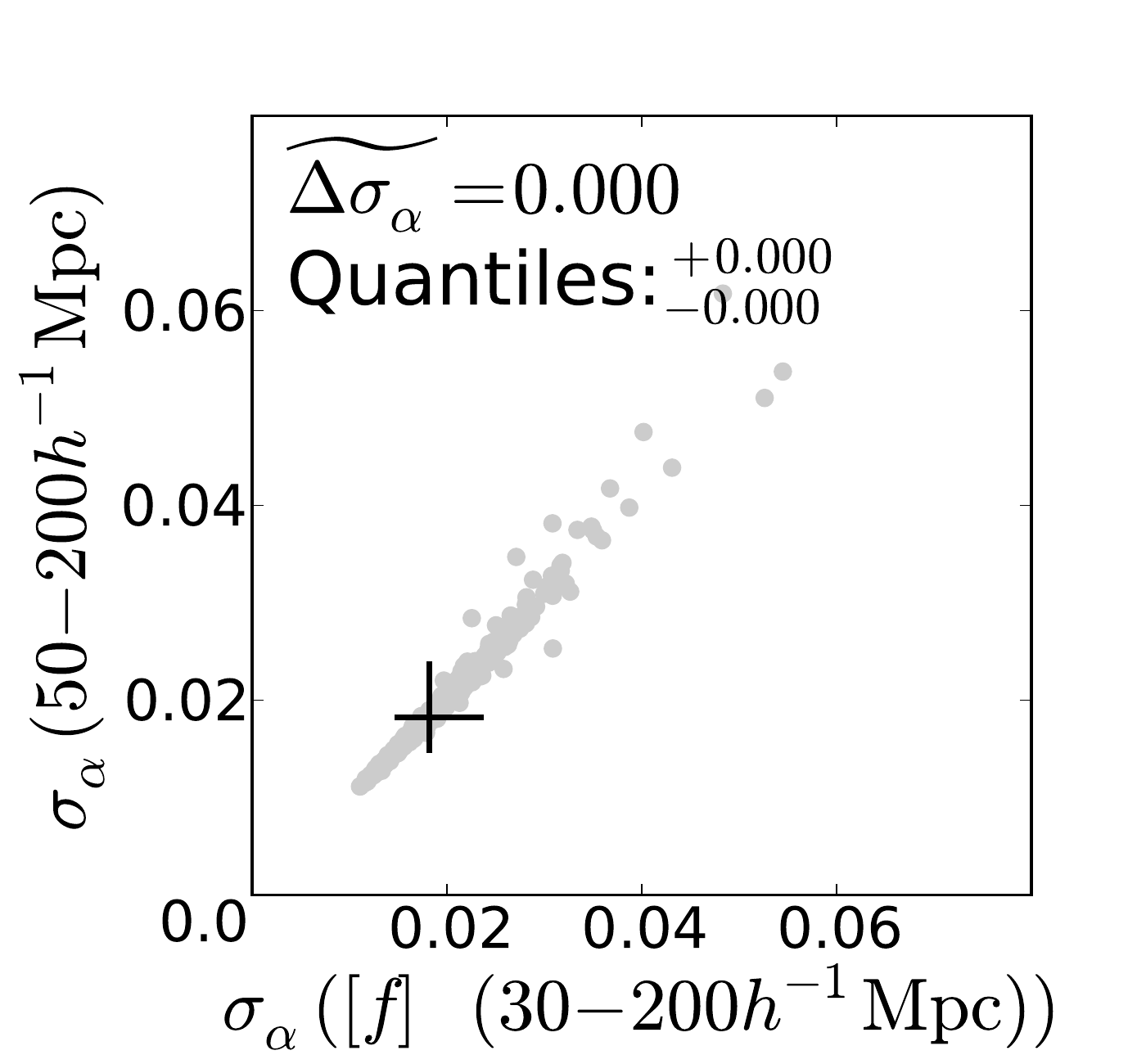}}
  \caption{ \label{fig:comps} Comparison of results obtained by
    fitting the mocks using the fiducial model, Equations
    (\ref{eqn:fform}) and (\ref{eqn:aform}) with a $28<r<200\mpcoh$
    fitting range, and those obtained using a $50<r<200\mpcoh$ fitting
    range. The pre-reconstruction results are shown in the left column
    and the post-reconstruction results are shown in the right
    column. The top panels show comparisons of $\alpha$ values and the
    bottom panels show comparisons of $\sigma_\alpha$ values. The
    black cross marks the median values of $\alpha$ or $\sigma_\alpha$
    along with their quantiles.  Nearly perfect 1:1 correlations exist
    between the $\alpha$ values measured using the fiducial fitting
    range and those measured using a slightly smaller fitting range
    both before and after reconstruction. The agreement between
    $\sigma_\alpha$ values is already good before reconstruction,
    however after reconstruction, the scatter is reduced to nearly
    zero. The results of robustness tests against other fit parameters
    are listed in Table \ref{tab:mock_alphas}. These imply that our
    fiducial model is robust against small changes in model parameters
    and hence should return unbiased measurements of $\alpha$.  }
\end{figure}

The fiducial and tweaked model fit results for 600 mocks are shown in
Table \ref{tab:mock_alphas}.  We remove mock results with poorly
measured values of $\alpha$ since a BAO feature was not clearly
identified ($\sigma_\alpha> 7$ per cent).  Nearly perfect 1:1
correlations between the values of $\alpha$ are measured from the
mocks as shown in Fig.~\ref{fig:comps}. The top two panels of
Fig.~\ref{fig:comps} show the $\alpha$ values measured using the
fiducial model plotted against the $\alpha$ values measured using a
smaller fitting range ($50<r<200\mpcoh$) both before (left) and after
(right) reconstruction. The bottom two panels show the corresponding
plots for $\sigma_\alpha$. Similar 1:1 correlations are seen for most
of the other ``tweaked'' models, implying that our fiducial model
returns unbiased measurements of the acoustic scale. The only cases
that have larger scatter in the correlations are the
pre-reconstruction $poly0$ and $\snl=0\mpcoh$ cases which is not
surprising. The prior implies that before reconstruction, there is
non-negligible broadband smooth signal that may bias our measurement
of the acoustic scale and hence a non-zero form for $A(r)$ is required
to marginalize over this contribution. The latter implies that using a
BAO model that does not account for the effects of non-linear
evolution, which are clearly evident before reconstruction, will also
bias the measurement of $\alpha$. After reconstruction, the scatter in
these cases is greatly reduced as reconstruction partially undoes
large-scale redshift space distortions and non-linear structure
growth.

Similar results for the CMASS DR9 data are shown in
Table~\ref{tab:dr9_alphas}. In general, our choice of model parameters
does not affect the outcome of the fits. A few cases measure slightly
larger or smaller values of $\alpha$, but all fall well within the
1$\sigma$ errorbars.

\begin{table}
  \caption{Fitting results for various models. Here we explore the
    effects of varying the fiducial fitting model and reconstruction
    parameters on our measurements of the distance scale from CMASS
    DR9. The results for the fiducial model, for different broadband
    $A(r)$ fitting functions ($poly0$, $poly2$, $poly4$), fitting
    ranges, and non-linear damping $\Sigma_{nl}$ of the acoustic scale
    are shown for the correlation function before and after
    reconstruction. We also present the results of fitting with a
    different covariance matrix (ML) derived based on the technique in
    \citet{Xeaip}. For our reconstruction tests, we present the
    effects of changing the fiducial galaxy bias by $+20$ per cent and
    $-20$ per cent ($b=1.5$ and $b=2.2$), the fiducial growth rate by
    $+20$ per cent and $-20$ per cent ($f=0.6$ and $f=0.9$), and the
    smoothing length to $20\mpcoh$, which is a more conservative choice than our fiducial smoothing of $15\mpcoh$.}
\label{tab:dr9_alphas}

\begin{tabular}{lcc}
\hline
Model&$\alpha$&$\chi^2$ \\
\hline

\multicolumn{3}{c}{Before Reconstruction}\\
\hline
Fiducial $[f]$ &
$1.016 \pm 0.017$&
30.53/39\\
\\[-1.5ex]
Fit with $poly0$. &
$1.018 \pm 0.020$&
40.84/42\\
\\[-1.5ex]
Fit with $poly2$. &
$1.017 \pm 0.016$&
30.74/40\\
\\[-1.5ex]
Fit with $poly4$. &
$1.016 \pm 0.017$&
30.33/38\\
\\[-1.5ex]
Fit between $20<r<200\mpcoh$. &
$1.020 \pm 0.017$&
32.47/41\\
\\[-1.5ex]
Fit between $50<r<200\mpcoh$. &
$1.018 \pm 0.018$&
22.99/34\\
\\[-1.5ex]
Fit with $\snl \rightarrow 0$. &
$1.005 \pm 0.013$&
30.84/39\\
\\[-1.5ex]
Fit with $\snl \rightarrow \snl-2$. &
$1.012 \pm 0.015$&
29.93/39\\
\\[-1.5ex]
Fit with $\snl \rightarrow \snl+2$. &
$1.019 \pm 0.019$&
32.02/39\\
\\[-1.5ex]
Fit using ML covariance matrix. &
$1.022 \pm 0.018$&
30.64/39\\
\hline
\multicolumn{3}{c}{After Reconstruction}\\
\hline
Fiducial $[f]$ &
$1.024 \pm 0.016$&
34.53/39\\
\\[-1.5ex]
Fit with $poly0$. &
$1.026 \pm 0.017$&
41.82/42\\
\\[-1.5ex]
Fit with $poly2$. &
$1.025 \pm 0.015$&
36.12/40\\
\\[-1.5ex]
Fit with $poly4$. &
$1.024 \pm 0.017$&
33.29/38\\
\\[-1.5ex]
Fit between $20<r<200\mpcoh$. &
$1.031 \pm 0.018$&
47.31/41\\
\\[-1.5ex]
Fit between $50<r<200\mpcoh$. &
$1.022 \pm 0.016$&
25.94/34\\
\\[-1.5ex]
Fit with $\snl \rightarrow 0$. &
$1.019 \pm 0.015$&
34.18/39\\
\\[-1.5ex]
Fit with $\snl \rightarrow \snl-2$. &
$1.020 \pm 0.015$&
34.27/39\\
\\[-1.5ex]
Fit with $\snl \rightarrow \snl+2$. &
$1.029 \pm 0.017$&
35.10/39\\
\\[-1.5ex]
Fit using ML covariance matrix. &
$1.022 \pm 0.017$&
34.30/39\\
\hline
Fit to recon. with $b \rightarrow 1.5$. &
$1.033 \pm 0.020$&
42.97/39\\
\\[-1.5ex]
Fit to recon. with $b \rightarrow 2.2$. &
$1.021 \pm 0.015$&
46.89/39\\
\\[-1.5ex]
Fit to recon. with $f \rightarrow 0.6$. &
$1.024 \pm 0.015$&
33.19/39\\
\\[-1.5ex]
Fit to recon. with $f \rightarrow 0.9$. &
$1.025 \pm 0.017$&
36.53/39\\
\\[-1.5ex]
Fit to recon. with $l \rightarrow 20\mpcoh$. &
$1.026 \pm 0.015$&
43.79/39\\
\hline
\end{tabular}
\end{table}

\begin{figure*}
  \centering
  \begin{tabular}{cc}
    \includegraphics[width=\columnwidth]{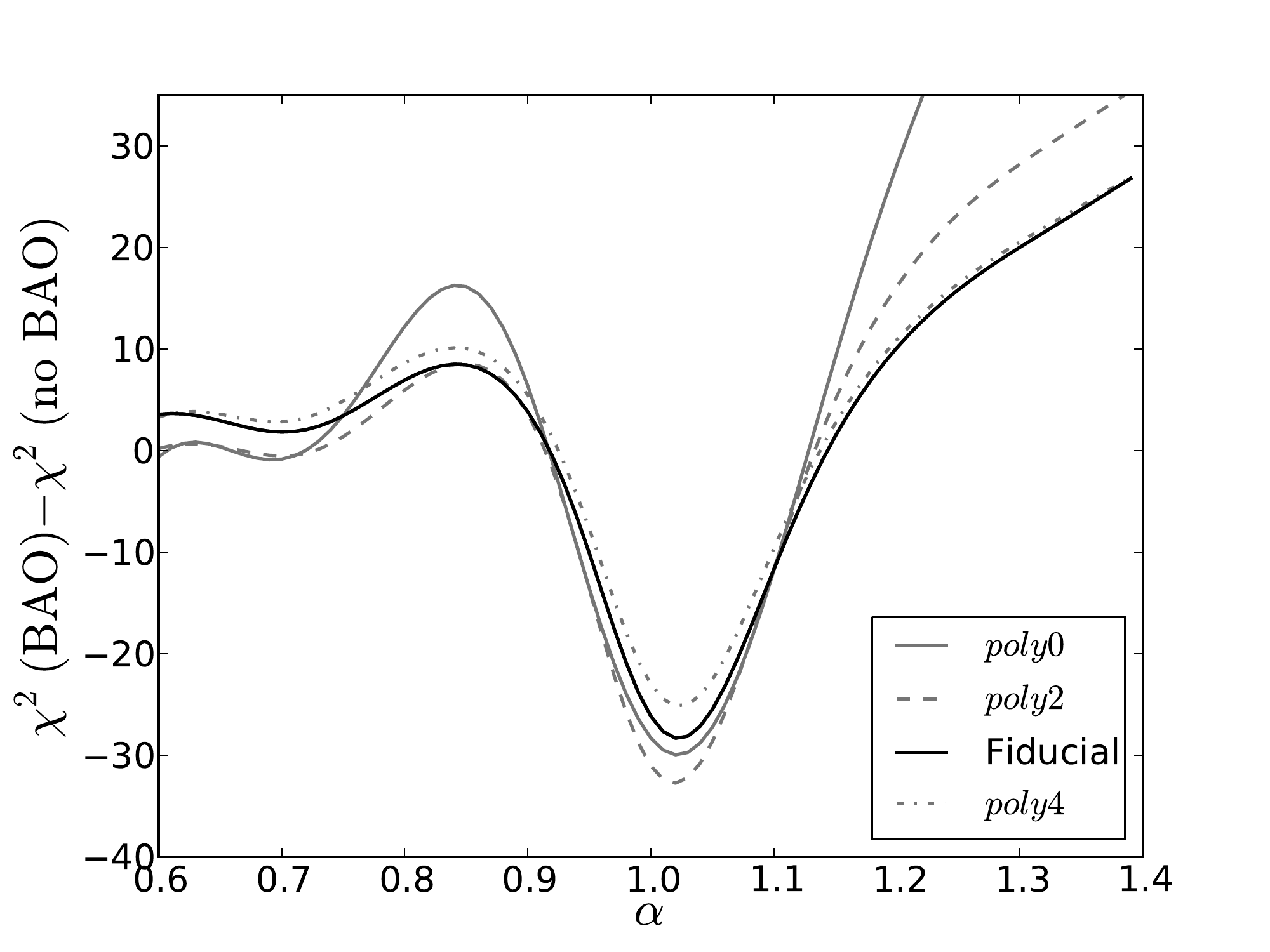}
    \includegraphics[width=\columnwidth]{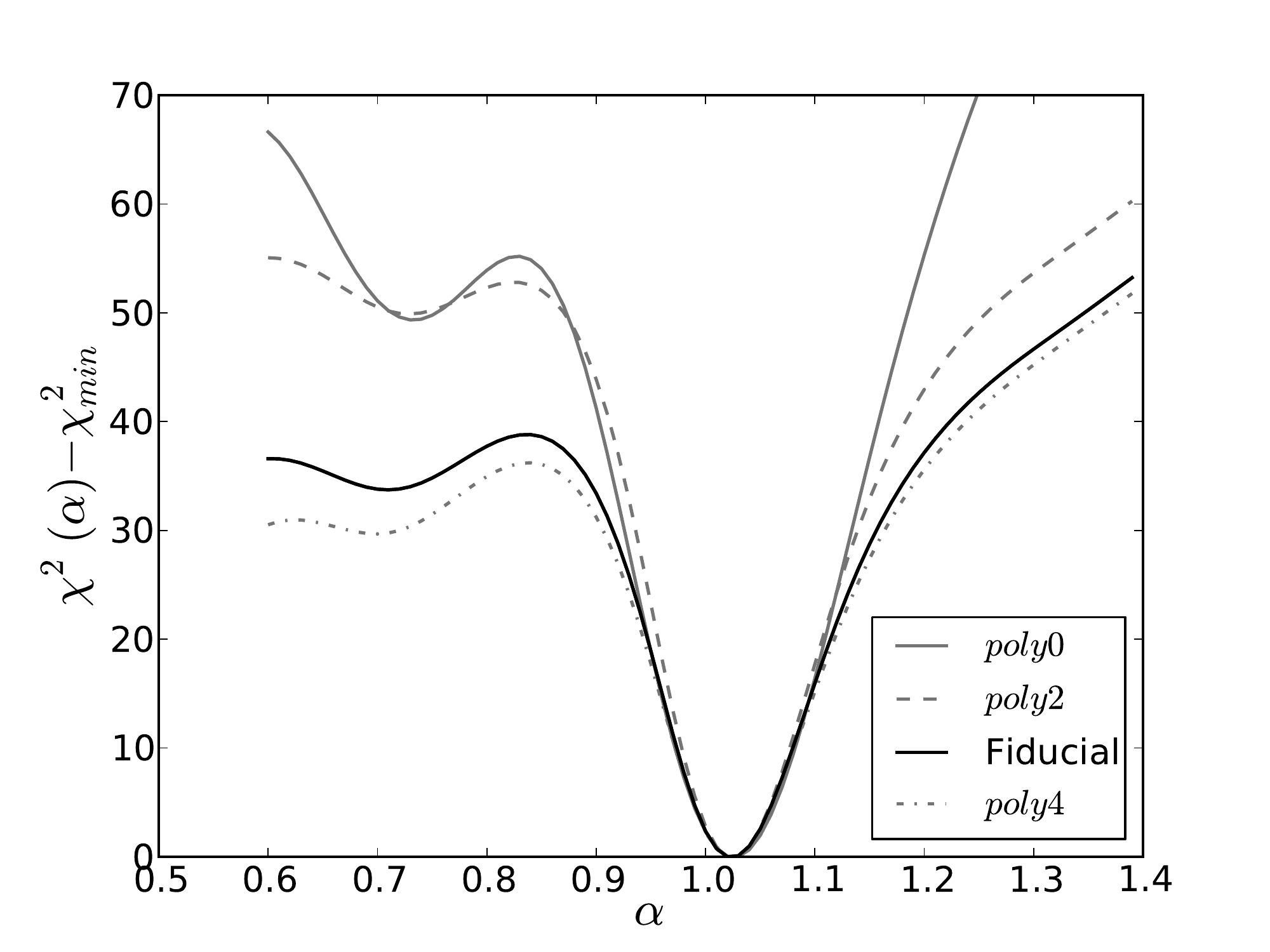}
  \end{tabular}
  \caption{ $\Delta\chi^2$ for CMASS DR9 using various forms of
    $A(r)$. These plots are analogous to Fig.~\ref{fig:DR7_9chi2},
    except we have split the two tests of BAO significance into separate
    panels. The left panel shows how robustly we have detected the BAO
    in the CMASS DR9 sample and the right panel shows how confident we
    are that we have measured the correct acoustic scale. In the left
    panel, we have plotted the difference in $\chi^2$ between 2 fits
    to the data, one using a model containing BAO and one using a
    model without BAO. We see that this $\Delta\chi^2$ is consistently
    around $-30$ for all forms of $A(r)$ indicating that the amount of
    flexibility in the broadband marginalization (i.e. the number of
    nuisance parameters in $A(r)$) does not have a significant impact
    on how well we detect the BAO in the CMASS DR9 sample. In the
    right panel, we have plotted the $\Delta\chi^2$ of the minimum as
    a function of $\alpha$. The various forms of $A(r)$ all identify
    the same best-fit value of $\alpha$ and this best-fit is at a
    $\Delta\chi^2$ well below the plateau in the curve. However, it
    appears that lower orders of $A(r)$ allow more confident measures
    of $\alpha$, possibly due to the increased flexibility in higher
    order forms to fit noise. Regardless, we have at least a $6\sigma$
    measurement of best-fit $\alpha$ in all cases which is
    robust. \label{fig:archi}}
\end{figure*}

We also investigate our measurements of BAO significance with respect
to the form of $A(r)$. The results are shown in Fig.~\ref{fig:archi}
after reconstruction. The right panel shows the difference in $\chi^2$
between a fit to the data using a model containing BAO and a fit to
the data using a model without BAO. These curves demonstrate how well
we have detected the BAO in the CMASS DR9 data. The solid black curves
correspond to subtracting the solid line from the dashed line in
Fig.~\ref{fig:DR7_9chi2}. The other lines correspond to various other
forms of $A(r)$, some with more and some with less nuisance
parameters. Here, the more negative $\Delta\chi^2$ is, the more a
model containing BAO is preferred. Allowing more or less flexibility
in the broadband marginalization as parameterised by $A(r)$ does not
change the fact that a model containing BAO is favoured and we have a
robust detection of the BAO in the CMASS DR9 data. The actual
confidence level changes slightly between the different $A(r)$ forms;
however, the variation is small and consistently falls between
$5-6\sigma$.

The right panel shows the $\Delta\chi^2$ values from the minimum (or
best-fit value) and demonstrates how well we have measured the
acoustic scale. The solid black curve is identical to the solid line
in Fig.~\ref{fig:DR7_9chi2}. The other curves correspond to various
other forms of $A(r)$. In all cases, the minima lie at the same value
of $\alpha$ with the plateaus lying at significant $\Delta\chi^2$
above the minima. Although $\Delta\chi^2$ shows significant variation
between the $A(r)$ forms, we see at least a $6\sigma$
($\Delta\chi^2\sim36$) preference for the best-fit value of
$\alpha$. It appears that a lower order or less flexible form for
$A(r)$ may return $\alpha$ at a higher confidence, which indicates that
higher order $A(r)$ may afford the model with enough flexibility to
start fitting noise.

We have tested the robustness of the fitting methods by making
Gaussian realisations of the correlation function from the full
covariance matrix and a template for the correlation function. We have
considered the model defined by Eq.~\ref{eqn:fform}, and computed the
best fit parameters for every simulation. The sample of recovered
parameter values that we recovered has a Gaussian distribution, as
expected. We also checked that our simulations are truly Gaussian by
computing the $\chi^2$ estimator for each simulation at the ``true
model'' (with $\alpha=1$, $B=1$, and $A(r)= 0$), and verifying that it
follows a $\chi^2$-distribution with $\nu = 50$ degrees of freedom,
which is the number of bins in $r$ used in this test.  We have found
that, while in the NGC the inferred errors from the fits agree very
well with the width of the distribution of $\alpha$, in the SGC the
measured errors tend to be slightly under-estimated (by about 0.25
$\sigma_\alpha$).


  \subsection{Robustness of Fitting Algorithm for $P(k)$}
    \label{sec:robust_pk}
    We have tested that our model for the power spectrum, calculated as
described in Section~\ref{sec:fitpk}, provides an adequate match to
the power spectra of the mocks. In fact the data plotted in
Figs.~\ref{fig:DR9pk} \&~\ref{fig:BAO_mocks} already show this result
to some extent as we plot the deviation between the measured power
spectra and the smooth model: The consistency between the data plotted
and the expected BAO model shows that any residual differences
between data and model are of significantly lower order than the BAO
signal.

\begin{figure}
  \centering
  \resizebox{0.7\columnwidth}{!}{\includegraphics{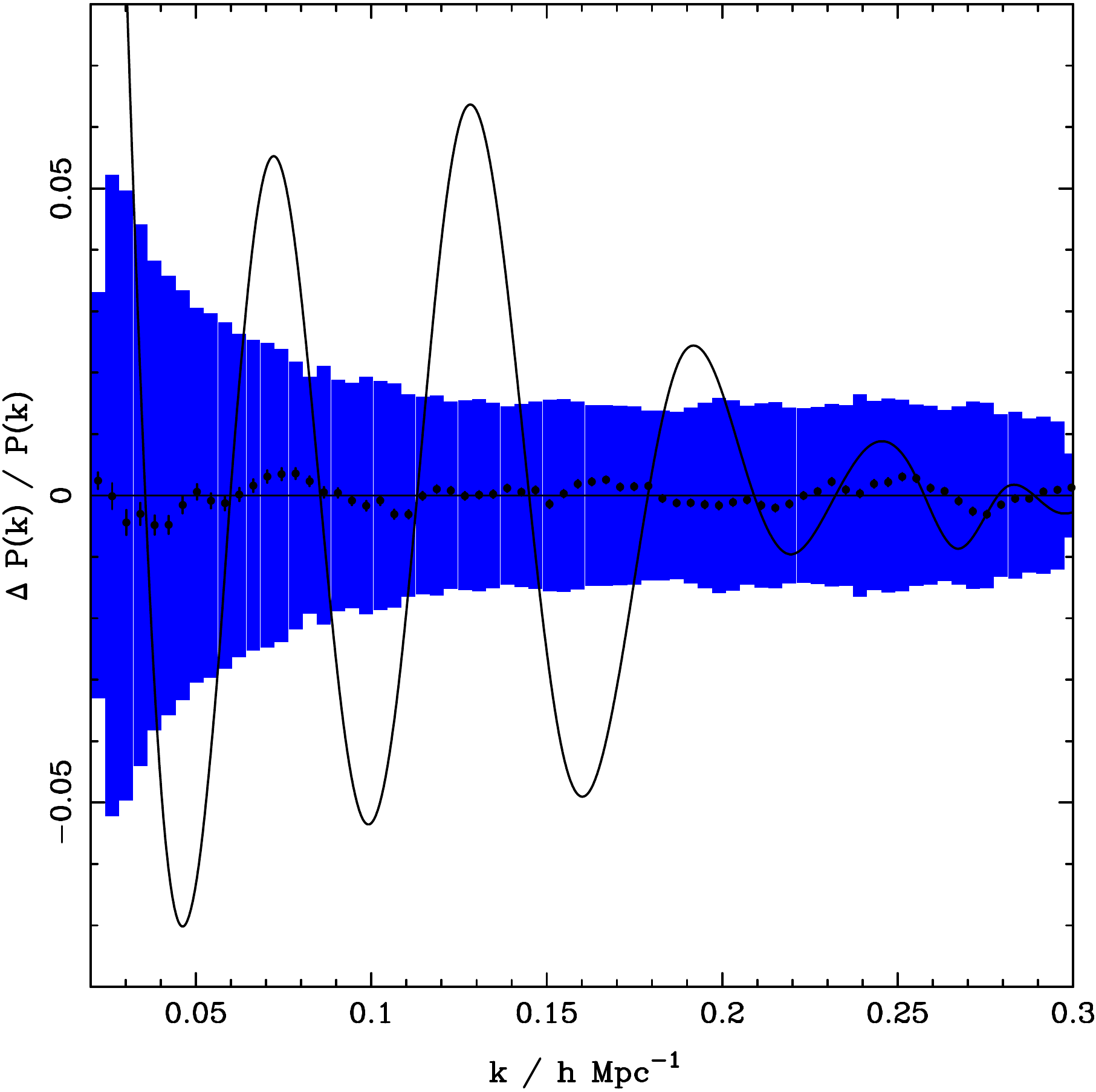}}
  \caption{Average residual recovered from fitting to the power
    spectra derived from the 600 mock catalogues after reconstruction
    (solid circles with $1\sigma$ errors). The shaded region shows the
    expected error for any single fit, while the solid line shows our
    fiducial BAO model. Clearly there is no evidence for any large
    deviations between the model and data, which might have indicated
    that the spline was unable to match the input power
    spectrum. \label{fig:pk_fit_residuals}}
\end{figure}
To test the goodness-of-fit further, Fig.~\ref{fig:pk_fit_residuals}
displays the average residual recovered after fitting to the 600 power
spectra derived from the mock catalogs
\begin{equation}
  \langle P(k_i) - P^{\rm fit}(k_i) \rangle = \frac{1}{600}\sum_{\rm mocks}
    \left[P(k_i) - P^{\rm fit}(k_i)\right],
\end{equation}
where $P(k_i)$ are the measured band powers, and $P^{\rm fit}(k_i)$ is
the best-fit model as defined in Eq.~\ref{eq:defwindow}. As can be
seen in Fig.~\ref{fig:pk_fit_residuals}, the average residual is well
below the scales of both the BAO and the difference expected for any
single fit (shown by the shaded region in
Fig.~\ref{fig:pk_fit_residuals}), so there is no evidence of a
systematic inability to fit the shape of the power spectrum over the
fitted $k$-range.

\begin{figure}
  \centering
  \resizebox{0.7\columnwidth}{!}{\includegraphics{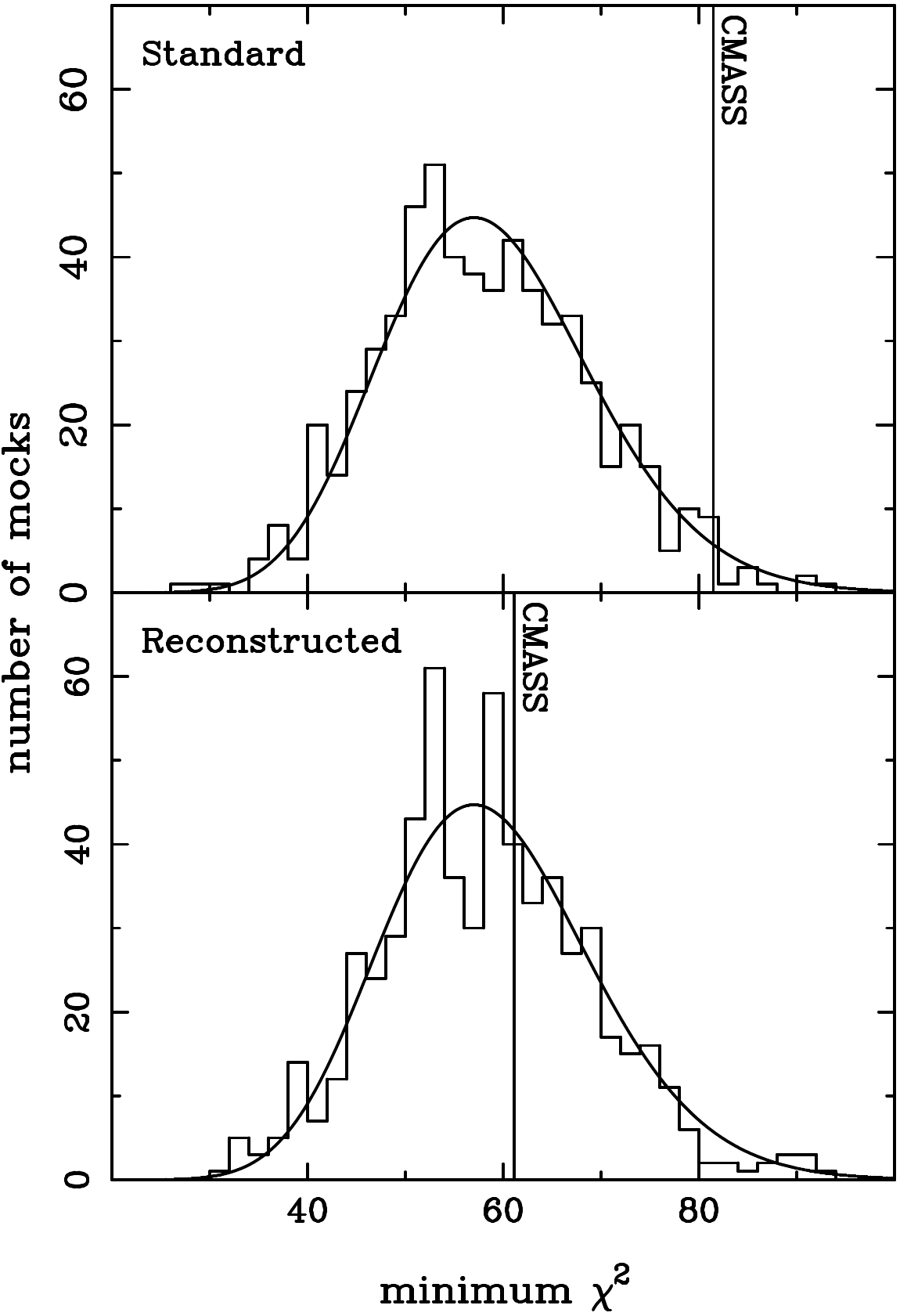}}
  \caption{Histograms of recovered minimum $\chi^2$ values from fits
    to power spectra measured from the mock catalogs:
    pre-reconstruction (upper panel) and post-reconstruction (lower
    panel). The smooth solid line gives the expected distribution of
    $\chi^2$ values for 59 degrees of freedom, and the vertical lines
    show the values recovered from fits to the CMASS DR9
    data. \label{fig:pk_fit_chisq}}
\end{figure}
Fig.~\ref{fig:pk_fit_chisq} shows histograms of the recovered best-fit
$\chi^2$ values from the fits to the 600 mock catalogs, before (upper
panel) and after (lower panel) reconstruction. These values match the
expected distribution of $\chi^2$ values for a fit with 59
degrees-of-freedom, which is also shown in this plot. This agreement
gives us confidence that the fit is behaving as expected for the power
spectra derived from the mock catalogues. If the model was unable to
adequately fit the mocks, we should expect the recovered $\chi^2$
minima to be significantly offset from the expected
distribution. i.e. if the model failed to adequately fit the shape of
the recovered power spectrum, then we would find systematically worse
$\chi^2$ values compared with those expected. The distribution
actually agrees remarkably well with that expected, which gives us
confidence that the model described in Section~\ref{sec:fitpk} is
adequate for these data.

The $\chi^2$ values from the fits to the data fall within the
distribution of values from the mocks although, for the
pre-reconstruction measurement, only 18/600 mocks give a worse
$\chi^2$ value. However, we know from the analysis presented in
Section~\ref{sec:results_pk} that the pre-reconstruction catalogue
also gives a smaller-than-average error, so this result is perhaps not
that surprising. In conclusion we find no evidence that the
fitting method applied to the power spectra is not adequate for
recovering the BAO scale.

\label{lastpage}

\end{document}